\documentclass[11pt,preprint]{aastex6} 
\usepackage{hyperref}
\hypersetup{pdfauthor={Song et al.}, pdftitle={Ly$\alpha$ halos}}
\usepackage{natbib}
\usepackage{url}
\usepackage{xcolor}
\usepackage{graphicx}
\usepackage{amsmath}
\usepackage{multirow}
\newcommand{\cs}{\scriptsize}

\newcommand{\Lyaa}{Ly$\alpha$ }
\newcommand{\Haa}{H$\alpha$ }
\newcommand{\rscalee}{$rs_\textrm{\tiny \ion{H}{1}}$ }
\newcommand{\rpeakk}{$r_\textrm{\cs peak}$ }
\newcommand{\Vpeakk}{$V_\textrm{\cs peak}$ }
\newcommand{\delVV}{$\Delta V$ }
\newcommand{\rshaloo}{$rs_\textrm{\cs halo}$ }
\newcommand{\rscontt}{$rs_\textrm{\cs cont}$ }
\newcommand{\rmaxx}{$r_\textrm{\cs max}$ }
\newcommand{\rshalorscontt}{$rs_\textrm{\cs halo}/rs_\textrm{\cs cont}$ }
\newcommand{\rscalerscontt}{$rs_\textrm{\tiny \ion{H}{1}}/rs_\textrm{\cs cont}$ }
\newcommand{\rpeakrscontt}{$r_\textrm{\cs peak}/rs_\textrm{\cs cont}$ }

\newcommand{\Lya}{Ly$\alpha$}

\newcommand{\rscale}{$rs_\textrm{\tiny \ion{H}{1}}$}
\newcommand{\rpeak}{$r_\textrm{\cs peak}$}
\newcommand{\Vpeak}{$V_\textrm{\cs peak}$}
\newcommand{\delV}{$\Delta V$}
\newcommand{\rshalo}{$rs_\textrm{\cs halo}$}
\newcommand{\rscont}{$rs_\textrm{\cs cont}$}
\newcommand{\rmax}{$r_\textrm{\cs max}$}
\newcommand{\rshalorscont}{$rs_\textrm{\cs halo}/rs_\textrm{\cs cont}$}
\newcommand{\rscalerscont}{$rs_\textrm{\tiny \ion{H}{1}}/rs_\textrm{\cs cont}$}

\slugcomment{Last updated: \today}
\shorttitle{Ly$\alpha$ halos}
\shortauthors{Song et al}
\begin{document}

\title{Ly$\alpha$ radiative transfer: 
	Modeling spectrum and surface brightness profile of Ly$\alpha$-emitting galaxies at $z=$3--6}

\author{Hyunmi Song\altaffilmark{1,2}, Kwang-il Seon\altaffilmark{2,3}, and Ho Seong Hwang\altaffilmark{2}}
\altaffiltext{1}{Department of Astronomy, Yonsei University, 50 Yonsei-ro, Seodaemun-gu, Seoul 03722, Republic of Korea}
\altaffiltext{2}{Korea Astronomy \& Space Science Institute, Daedockdae-ro 776, Yuseong-gu, Daejeon 34055, Republic of Korea}
\altaffiltext{3}{Astronomy and Space Science Major, University of Science and Technology, Daejeon 34113, Republic of Korea}

%%%%%%%%%%%%%%%%%%%%%%%%%%%%%%%%%%%%%%%%%%%%%%%%%%%%%%%%%%%%%%%%%%%%%%%%%%%%%%%%%%%%%%%%%%%%%%%%%%%
\begin{abstract}
We perform \Lyaa radiative transfer calculations
	for reproducing \Lyaa properties of star-forming galaxies at high redshifts.
We model a galaxy as a halo in which the density distributions of 
	\Lyaa sources and \ion{H}{1} plus dust medium are described with exponential functions.
We also consider an outflow of the medium
	that represents a momentum-driven wind in a gravitational potential well.
We successfully reproduce both the spectra and the surface brightness profiles
	of eight star-forming galaxies at $z=$3--6 observed with MUSE
	using this outflowing halo model with \Lyaa scattering.
The best-fit model parameters (i.e., the outflowing velocity and optical depth) for these galaxies
	are consistent with other studies.
We examine the impacts of individual model parameters and input spectrum
	on emerging spectrum and surface brightness profile.
Further investigations on correlations among observables 
	(i.e., the spatial extent of \Lyaa halos and \Lyaa spectral features) and model parameters,
	and spatially resolved spectra
	are presented as well.
We demonstrate that the combination of spectrum and surface brightness profile 
	provides strong constraints on model parameters and thus spatial/kinematic distributions of medium.
\end{abstract}


%%%%%%%%%%%%%%%%%%%%%%%%%%%%%%%%%%%%%%%%%%%%%%%%%%%%%%%%%%%%%%%%%%%%%%%%%%%%%%%%%%%%%%%%%%%%%%%%%%%%
\section{INTRODUCTION}\label{sec-intro}
%{{{
\Lyaa emission is one of the most prominent emission features in the universe.
It is usually generated by the interplay between atomic hydrogen (\ion{H}{1}, the most common element in the universe)
	and ionizing photons from young stars.
It was expected that 
	one would detect strong \Lyaa emission from star-forming galaxies
	and can use it to probe even the first generation of galaxies at high redshifts \citep{PP1967,STM1978}.
However, it had been unsuccessful to detect \Lyaa emission 
	from either nearby galaxies or high-redshift galaxies.
	\citet{MT1981} did the first attempt to search for \Lyaa emission in nearby \ion{H}{2}-selected galaxies,
	and found \Lyaa emission in only one galaxy;
	the first detection of high-redshift \Lyaa emitters that are not physically associated with quasars 
	was made in 1998 by \citet{CH1998}.

Such difficulties in detecting \Lyaa emission are 
	because of particular technical requirements;
	space telescopes for nearby sources
	and high detection sensitivities for high-redshift sources
	are necessary for good detection.
However, it is also due to the resonance scattering nature of \Lya;
	\Lyaa photons are endlessly scattered off by neutral hydrogen atoms
	until they happen to be scattered into wing frequencies, which occurs quite rarely.
\Lyaa photons are forced to travel long distances in the \ion{H}{1} medium
	and have a high chance to be destructed by dust,
	which eventually suppresses the observed intensity.
In addition, this whole process is sensitive to 
	the spatial and kinematic distributions of the \ion{H}{1} medium,
	so it is not easy to understand the observed correlations between physical parameters
	(e.g., between \Lyaa intensity and metallicity, \Lyaa intensity and UV continuum flux, etc.)
	nor to make observational predictions
	\citep{MT1981,Hartmann_etal_1988,Neufeld1990,Charlot_Fall_1993,Valls-Gabaud1993,
		Kunth_etal_1998,Tenorio-Tagle_etal_1999,Mas-Hesse_etal_2003}.

Nevertheless, there have been continuous efforts to search for \Lyaa emission
	from the observation side.
There are targeted observations for individual objects
	as well as large photometric and spectroscopic surveys.
Such \Lyaa surveys include the Large Area Lyman Alpha survey \citep[LALA,][]{Rhoads_etal_2000},
	the \Lyaa Reference Sample survey \citep[LARS,][]{Ostlin_etal_2014},
	the Subaru Deep Field survey \citep[e.g., SILVERRUSH,][]{Ouchi_etal_2018},
	and the survey with Multi-Unit Spectroscopic Explorer \citep[MUSE,][]{Bacon_etal_2015}.
The MUSE at the ESO-VLT is the latest technological advances
	with 24 integral field units and unprecedented sensitivity.
There are several recent observations with MUSE
	that reveal ubiquitous \Lyaa emission in the universe 
	\citep{Bacon_etal_2015,Wisotzki_etal_2016,Drake_etal_2017,Leclercq_etal_2017}.

Observations have shown that the shape of \Lyaa line is diverse,
	which includes broad damped absorption profiles, P-Cygni profiles,
	double-peak profiles, pure symmetric emission profiles,
	and combinations thereof \citep{Kunth_etal_1998,Mas-Hesse_etal_2003,Shapley_etal_2003,
		Moller_etal_2004,Venemans_etal_2005,Wilman_etal_2005,Noll_etal_2004,Tapken_etal_2004}.
Such a variety can be understood through a detailed radiative transfer calculation,
	which is analytically solvable only for simple cases
	(e.g., a static, plane-parallel slab by \citealt{Harrington1973} and \citealt{Neufeld1990},
	a static uniform sphere by \citealt{Dijkstra_etal_2006}).
Later, numerical algorithms based on Monte Carlo techniques 
	had been developed to solve radiative transfer for more general cases.
Now theoretical studies mostly rely on them
	\citep[e.g.,][]{Spaans1996,Loeb_Rybicki_1999,
		Ahn_etal_2000,Ahn_etal_2002, %Ahn_etal_2001,Ahn_etal_2003,Ahn2004,   % choose one of them
		Zheng_Miralda-Escude_2002,Richling2003,
		Cantalupo_etal_2005,
		Dijkstra_etal_2006a, % choose one of them
		Hansen_Oh_2006,Tasitsiomi2006,Verhamme_etal_2006,
		Laursen_etal_2013,Behrens_etal_2014,Duval_etal_2014,
		Gronke_etal_2015,Verhamme_etal_2015,MichelDansac_etal_2020}.  % Verhamme_etal_2015?
Meanwhile, a galaxy model needs to be constructed
	to perform such a radiative transfer calculation. 
One can adopt a realistic galaxy model from hydrodynamical simulations.
Galaxies from such simulations can be useful 
	for performing a statistical study of \Lyaa properties,
	but they can not be directly used to model individual galaxies in observations.
Therefore, it would be better to adopt a simple but manageable toy model
	for the purpose of reproducing observations.
One example for such models is a shell model, 
	in which a central \Lyaa source is surrounded by 
	an constantly-expanding, homogeneous, spherical shell of \ion{H}{1} medium with dust.
Although this shell model has surprisingly well reproduced many observed \Lyaa line profiles
	\citep[e.g.,][]{Ahn2004,Schaerer_Verhamme_2008,Verhamme_etal_2008,Schaerer_etal_2011,
		Gronke_etal_2015,Yang_etal_2016,Karman_etal_2017,Gronke2017},
	because of its extreme simplicity and contrivance,
	there are still rooms to improve \citep[e.g., see Section 7.2 in][]{Yang_etal_2016,Orlitova_etal_2018}.

On the other hand, there are other observables than spectrum,
	which are not yet fully understood.
\citet{Dijkstra_Kramer_2012} is one of few studies that focused on 
	reproducing observed \Lyaa absorption features in the spectra of background galaxies
	and \Lyaa halos around star-forming galaxies
	by considering a galaxy model with outflowing clumpy medium.
Here, \Lyaa halos denote the spatial distribution of \Lyaa emission 
	that is much more extended than that of stellar UV continuum or \Haa
	\citep[e.g.,][]{Fynbo_etal_1999,Steidel_etal_2011,
		Matsuda_etal_2012,Hayes_etal_2013,Yang_etal_2017}.
This indicates rich gas content in the circumgalactic medium (CGM).
Thus, \Lyaa halos might provide us the information
	on the spatial distribution and kinematics of the CGM,
	which is important for understanding galaxy formation and evolution.
However, \Lyaa halos were detectable usually in a stacked image
	because of their low surface brightness \citep{Steidel_etal_2011,Momose_etal_2014,Momose_etal_2016}.
Thanks to the MUSE observations
	that revealed ubiquitous \Lyaa halos around star-forming galaxies at high redshifts
	\citep{Wisotzki_etal_2016,Leclercq_etal_2017},
	we can now study \Lyaa halos for individual galaxies,
	and even the spatial variations of their spectral properties
	 \citep{Erb_etal_2018,Claeyssens_etal_2019,Leclercq_etal_2020}

Therefore we are in a good position to fit 
	both \Lyaa spectrum and \Lyaa surface brightness profile
	for individual high-redshift star-forming galaxies,
	and to test a galaxy model with \Lyaa radiative transfer.
In this study, we construct a galaxy toy model that is improved from the shell model
	to better represent real galaxies:
	a halo in which the density distributions of \Lyaa sources and \ion{H}{1} plus dust medium, 
	as well as medium bulk motion, are modeled with free parameters.
We perform \Lyaa Monte Carlo radiative transfer calculations with this galaxy model
	for a number of model parameter sets.
We then find a parameter set that best reproduces both
	the observed \Lyaa spectrum and surface brightness profile
	of high-redshift star-forming galaxies
	reported in \citet{Leclercq_etal_2017} (hereafter L17).
We discuss the advantage of simultaneous use of these two observables
	in constraining models, 
	and examine various correlations among model parameters and observables.
We also explore spatially resolved \Lyaa spectra,
	and the impact of input \Lyaa spectrum 
	on emerging \Lyaa spectrum and surface brightness profile.
	
This paper is constructed as follow.
In Section \ref{sec-method}, 
	we describe our \Lyaa Monte Carlo radiative transfer simulation, the galaxy model,
	the observational data to fit, and the fitting process.
We present the best-fit results for the galaxy sample in Section \ref{sec-fit-result}.
Further discussions and summaries are presented
	in Sections \ref{sec-discuss} and Section \ref{sec-summary}, respectively.
%}}}

%%%%%%%%%%%%%%%%%%%%%%%%%%%%%%%%%%%%%%%%%%%%%%%%%%%%%%%%%%%%%%%%%%%%%%%%%%%%%%%%%%%%%%%%%%%%%%%%%%%%
\section{METHOD}\label{sec-method}

\subsection{\Lyaa Monte Carlo Radiative Transfer}\label{ssec-model}
%{{{
We describe \Lyaa Monte Carlo radiative transfer calculations we perform in this section.
We use a code called \texttt{LaRT}\footnote{\href{https://seoncafe.github.io/LaRT.html}{https://seoncafe.github.io/LaRT.html}}
	 \citep[Lyman-$\alpha$ Radiative Transfer,][]{Seon_Kim_2020}.
It is written in modern Fortran with the Message Passing Interface,
	and is enabled to consider arbitrary three-dimensional distributions for density, temperature, and kinematics 
	of sources and medium on a regular Cartesian grid.
We set a 128$^3$ grid, which gives a physical resolution of $\sim0.2$kpc.
In the code, we generate and track photons (or photon packets) in real and frequency spaces
	for a given simulation parameter set.
The calculation is performed by distributing photons over processors
	and later is summed together.
The master-slave algorithm is used to implement dynamic load balancing.
The code was extensively tested to reproduce the well-known benchmark cases for
	static slab and spherical geometries as well as
	for Hubble-like expanding spherical media.
The test was performed not only in hydrogen-only media but also in dusty media.
The procedures performed in the code for each photon are as follows.

\begin{enumerate}
\item 
	We generate a photon 
		with a position vector $\mathbf{r}$, a direction vector $\mathbf{k}$, and a frequency $\nu$
		drawn from a \Lyaa source spatial distribution function under consideration, an isotropic distribution function,
		and a \Lyaa input frequency distribution function of interest, respectively.
\item 
	We randomly choose a $\tau$ value following a probability distribution of $\exp(-\tau)$
		to determine the traveling distance $\ell$ for the photon through the relation
		$\tau=\int_0^{\ell}(\sigma_\textrm{\tiny \ion{H}{1}}n_\textrm{\tiny \ion{H}{1}}(\mathbf{x}) +\sigma_\textrm{\cs d}n_\textrm{\cs d}(\mathbf{x}))d|\mathbf{x}|$
		where $\sigma$ and $n$ represent the cross section 
		and the number density of atomic hydrogen (\ion{H}{1}) or dust (d), respectively.
The optical depth due to hydrogen atoms is calculated 
		using the Voigt routine in the VPFIT package \citep{Carswell_Webb_2014}.

\item 
	The position of the photon is updated to $\mathbf{r}+\ell\mathbf{k}$.
	At the new position, we decide which of the two (hydrogen and dust) 
		the photon will interact with
		by considering the probabilities given by their optical depths to the total optical depth.
	If the photon interacts with hydrogen (i.e., the photon being scattered),
		a new direction vector is drawn from the Rayleigh phase function.
	Then, the velocity components of the scattering atom
		are drawn from a function given by Eqn. (4) of \citet{Zheng_Miralda-Escude_2002}
		and the Maxwell-Boltzmann distribution \citep[see][for more details]{Seon_Kim_2020}.
	This velocity of the scattering atom and the old and new direction vectors of the photon
		determine a new frequency following the energy-momentum conservation law.
	If it interacts with dust,
		the photon will be either scattered with the probability of $a$ (dust albedo) 
		or absorbed with the probability of $1-a$.
	When the photon is scattered by dust, the frequency is not changed, 
		and a new direction is drawn from a proper angular redistribution function 
		\citep[the Henyey-Greenstein phase function by default,][]{Henyey_Greenstein_1941,Witt1977}.
\item 
	We repeat Steps 2 and 3 until either the photon escapes the system considered (i.e., a galaxy) 
		or is absorbed by dust. 
	Then, we go to Step 1 for a new photon.
\end{enumerate}

Although the absorption and scattering by dust can be explicitly simulated 
	using our code with various choices of albedo and scattering phase function
	as described in Step 3,
	we post-process the effect of dust to reduce the computational time.
Whenever each photon escapes the galaxy, 
	we record its initial position ($r_i$), initial frequency ($\nu_i$), 
	cumulative (the photon has gone through) hydrogen optical depth ($\hat{\tau}_\textrm{\tiny \ion{H}{1}}$), 
	cumulative hydrogen column density ($\hat{N}_\textrm{\tiny \ion{H}{1}}$),
	final (projected) position in the observed image ($R_f$), final frequency ($\nu_f$),
	and distance to an observer from the final position ($d_\textrm{\cs obs}$). 
The initial position/frequency and the cumulative hydrogen column density 
	are particularly recorded for post-processing 
	the spatial/frequency distributions of \Lyaa source and dust effect.
Such post-processing is adopted to reduce the number of simulation runs
	that are required to study the effect of the three.
Following \citet{Gronke_etal_2015}, 
	we post-process the spatial/frequency distributions of \Lyaa and dust effect
	by adjusting the weight ($w$) of each photon to
	\begin{equation}
		w =	\frac{\mathcal{R}_f(r_i)}{\mathcal{R}_i(r_i)} 
			\frac{\mathcal{S}_f(\nu_i)}{\mathcal{S}_i(\nu_i)} 
			\,\exp( -(1-a) \tilde\sigma_\textrm{\tiny d,MW} 
			\,\hat{N}_\textrm{\tiny \ion{H}{1}} 
			\,\textrm{DGR}/\textrm{DGR}_\textrm{\tiny MW}).
		\label{eqn-weight}
	\end{equation}
$\mathcal{R}$ and $\mathcal{S}$ are 
	the input spatial and spectral distributions of \Lyaa sources, respectively.
Here the subscripts $i$ and $f$ indicate the initial (that is chosen as an input for the simulation run)
	 and final choices, respectively.
$\textrm{DGR}/\textrm{DGR}_\textrm{\tiny MW}$ is the amount of dust relative to gas 
	(dust-to-gas ratio, DGR) normalized by the value of the Milky Way,
	and $\tilde\sigma_\textrm{\tiny d,MW}$ is the dust cross section {\it per neutral hydrogen} of the Milky Way,
	which is the product of $\sigma_\textrm{\tiny d,MW}$ and DGR$_\textrm{\tiny MW}$.
Our choices of $\mathcal{R}_i$ and $\mathcal{S}_i$ are 
	uniform distribution functions of $r$ and $\nu$, respectively, within given ranges,
	which can be easily modified for other distributions through post-processing.
The exponential term can be simply rewritten as $\exp(-(1-a)\hat{\tau}_\textrm{\cs d})$, 
	which depicts how much fraction of each photon would not be destructed by dust,
	 but be survived after its travel in the galaxy.
In the post-processing, we assume that 
	dust scatters photons to the direction of its original propagation (perfect forward scattering),
	of which validation is presented in Appendix \ref{sec-validate-pp}.
At far ultraviolet wavelengths, scattering is indeed strongly forward-directed \citep{Seon_Draine_2016}.
The dust extinction cross-section and albedo of the Milky Way are
	$\tilde\sigma_\textrm{\tiny d,MW}=1.61\times10^{-21}\,\textrm{cm}^2/H$
	and $a_\textrm{\tiny MW}=0.325$, respectively \citep{Weingartner_Draine_2001,Draine2003}.

To make observables, we use the information of {\it all} photons 
	at the moment when each of them escapes the galaxy.
This method works
	because we set the system to be spherically symmetric (Section \ref{ssec-galmodel}),
	and thus observation from all directions will be identical.
%}}}

\subsection{Galaxy Model}\label{ssec-galmodel}
%{{{
We construct a galaxy model by improving the widely used shell model,
	which is more realistic but still simple.
We adopt a spherically symmetric halo model for the distribution of hydrogen atom (no deuterium) plus dust medium,
	of which density follows an exponential function with a scale radius (\rscale) that is a free parameter.
The overall density level is another free parameter, 
	that is the optical depth from the center to the edge of the halo 
	for a photon at the \Lyaa central frequency ($\tau_0$\footnote{
		The subscript, 0, indicates the \Lyaa central frequency.}).
We fix the medium temperature at $10^4\textrm{K}$.

We assume that all the \Lyaa photons are generated through recombinations of electrons and hydrogen ions
	that are ionized by UV photons from young stars (i.e., \ion{H}{2} regions around young stars).
We thus model the spatial distribution of \Lyaa sources
	using the observed UV continuum surface brightness profiles.
The UV surface brightness profile of the target galaxies in this study 
	is well described by an exponential function 
	of the projected distance from the galaxy center
	(see the rightmost panels of Figures 2 and 3 in L17),
	which is reconstructed as a modified Bessel function of the second kind 
	with $n=0$ ($K_0(x)$) when the distance is measured in three-dimensional space.
We therefore assume that the spatial distribution of \Lyaa sources follows $K_0(r/rs_\textrm{\cs cont})$
	where \rscontt is the scale radius of the UV continuum surface brightness profile.
\Lyaa photons are assumed to follow a Voigt profile with temperature $10^4\textrm{K}$ 
	(a typical value for \ion{H}{2} regions) in frequency space.
As mentioned in Section \ref{ssec-model},
	the spatial and frequency distributions of \Lyaa photons are post-processed with Eqn. (\ref{eqn-weight})
	by inserting the Bessel function and the Voigt function for $\mathcal{R}_f$ and $\mathcal{S}_f$, respectively.

We consider the bulk motion of the medium as most preexisting models do.
An observed \Lyaa emission line is typically singly red-peaked or red-peak dominated,
	which are expected to emerge from outflowing medium 
	\citep[e.g.,][]{Verhamme_etal_2006,Verhamme_etal_2008,Vanzella_etal_2010}.
There are other observational evidences such as the offsets between redshifts of 
	interstellar absorption lines, \Lyaa lines, and nebular emission lines
	\citep[][and references therein]{Steidel_etal_2010},
	and blueshifted absorption and redshifted emission for resonance lines 
	\citep[][and references therein]{Prochaska_etal_2011}.
In particular, we employ an outflow model in which velocity increases and then decreases 
	as radial distance from the center increases.
This is motivated by \citet[see their Section 6.1 and references therein]{Dijkstra_Kramer_2012}
	to describe a momentum-driven wind from the center 
	that is decelerating in the galactic gravitational potential well.
However, instead of adopting a functional form 
	by solving a momentum equation in a gravitational potential well as in \citet{Dijkstra_Kramer_2012},
	we rather consider a linearly increasing and then decreasing function for simplicity.
It is expressed as
	\begin{equation}
		V(r) = 
		\begin{cases}
			V_\textrm{\cs peak}\,\, r/r_\textrm{\cs peak} \quad\textrm{if } r\le r_\textrm{\cs peak} \\
			V_\textrm{\cs peak} + \Delta V\, (r-r_\textrm{\cs peak})/(r_\textrm{\cs max}-r_\textrm{\cs peak})
			\quad\textrm{if } r>r_\textrm{\cs peak}
		\end{cases}
	\end{equation}
	where \Vpeak, \rpeak, and \delVV
	are the peak velocity, the radius where velocity reaches the peak value, 
	and the velocity difference between those at \rmaxx and \rpeakk
	(\delVV is allowed in the range between $-$\Vpeakk and 0 
	to represent decelerated outflowing motion but prevent inflowing motion),
	which are free parameters in our galaxy model.
It should be noted that in our model 
	the initial generation of \Lyaa photons is not from this expanding medium (i.e., non-comoving sources).
Therefore, the frequency of \Lyaa photons observed by a hydrogen atom in the medium will be seen shifted
	by the amount that is proportional to the medium velocity.

The abovementioned five free parameters 
	(\rscalee and $\tau_0$ for medium density structure,
	\Vpeak, \rpeak, and \delVV for medium kinematic structure)
	are simulation parameters,
	each of which is assigned a value from a range of interest for a simulation run.
In addition, the relative amount of dust with respect to gas (DGR) 
	and redshift ($z$) of a model galaxy are also considered as free parameters in our model.
Their effects are implemented through post-processing, so called post-processing parameters.
In post-processing,
	the weight of each photon in a simulation output
	is adjusted following Eqn. (\ref{eqn-weight}) for a chosen DGR value,
	and their final wavelengths are redshifted by a factor of $1+z$ in the observer's frame due to the cosmic expansion.
Simulated observables are constructed in the observer's frame.

Table \ref{tab-pargrid} shows the parameter grid of simulation.
The simulation is run on the grid of the five simulation parameters,
	and outputs are additionally post-processed on the grid of the two post-processing parameters.
In total, we run 13,230 simulations with $10^6$ photons per simulation,
	and each simulation is post-processed over additional 110 post-processing parameter grid points.

%%%%%%%%%%%%%%%%%%%%%%
% Table
\begin{deluxetable}{cc}
	\tablecaption{Parameter Grid of Simulation\label{tab-pargrid}}
	\tablehead{Parameter & Values}
	\startdata
		\rscale\tablenotemark{a}		& $[0.1,0.2,0.3,0.4,0.5,0.6,0.7,0.8,0.9]$ \\
		\rpeak\tablenotemark{b}			& $[0.0,0.1,0.2,0.3,0.4,0.5,0.6]$ \\
		\Vpeak\tablenotemark{c}			& $[100,200,300,400,500]$ \\
		\delV\tablenotemark{d}			& $[-500,-450,-400,\ldots,-100,-50,0]$ \\
		$\log \tau_0$\tablenotemark{e}	& $[5.7,6.0,6.3,6.6,6.9,7.2]$ \\
		DGR\tablenotemark{g} 			& $[0.0,0.2,0.4,\ldots,1.6,1.8,2.0]$ \\
		$z$\tablenotemark{g} 			& $z_0+[0.001,0.002,0.003,\ldots,0.008,0.009,0.010]$ \\
	\enddata
	\tablecomments{If it is not mentioned, each parameter is in the unit stated here for the rest of the paper.}
	\tablenotetext{a}{The scale radius of the medium density distribution that is described by an exponential function.
					It is in the normalized unit by the maximum scale of the system (\rmax) we set in the simulation.
					All the length scales in this study are in the normalized unit by \rmax.}
	\tablenotetext{b}{The radius when the expanding velocity of the medium reaches its peak,
						in the normalized unit by \rmax.}
	\tablenotetext{c}{The peak velocity of the expanding velocity profile of the medium
						in the unit of $\textrm{km s}^{-1}$.}
	\tablenotetext{d}{The difference between the velocities at \rmaxx and \Vpeakk
						in the unit of $\textrm{km s}^{-1}$.
					\delVV is allowed in the range between $-$\Vpeakk and 0.}
	\tablenotetext{e}{The optical depth measured at the \Lyaa central frequency,
						which is given by the product of the hydrogen column density of the system
						and the cross section of a \Lyaa photon at the central frequency with a neutral hydrogen atom.}
	\tablenotetext{f}{Dust-to-Gas Ratio, which is the amount of dust relative to gas (in terms of mass).
					It is in the normalized unit by the value of the Milky Way (DGR$_\textrm{\tiny MW}$).}
	\tablenotetext{g}{Redshift. $z_0$ is an approximate estimate for the redshift of the system
						derived from observation.}
\end{deluxetable}
%%%%%%%%%%%%%%%%%%%%%%
%}}}

\subsection{Observational Data and Fitting Procedure}\label{ssec-muse}
%{{{
The targets that we aim to model with our simulation are taken from L17.
They investigated 250 Lyman Alpha Emitters (LAEs) in the {\it Hubble} Ultra Deep Field,
	which were observed with the MUSE to study the extended \Lyaa halo around individual galaxies.
Because the MUSE covers a wavelength range between 4750\AA$\,$ and 9350\AA,
	the targeted LAEs are mainly at a redshift range $2.75\lesssim z\lesssim6.5$.
L17 summarized all the measurements from their analysis in Table B.1,
	which are good enough to fully reconstruct observed surface brightness profiles, but not spectra
	(only equivalent width (EW) and full width at half maximum (FWHM) of \Lyaa lines are available in their table).
Therefore, we use only 14 LAEs for which 
	\Lyaa spectrum and surface brightness profile are fully presented 
	in their Figures 2 and 3, as a pilot study.
Because we set an outflow model to reproduce red peak-dominated spectra 
	that are typical \Lyaa features in observations,
	we exclude three LAEs with double-peak spectrum (MUSE \#1087, 106, and 6297).
We also exclude three LAEs with few (less than 15) data points in their surface brightness profiles;
	two of which have no counterparts in the UV continuum (MUSE \#6498, 6534, and 218).
In the result, we have eight LAEs, 
	which are MUSE \#1185, 82, 6905, 1343, 53, 171, 547, and 364.
We read data points of their spectrum and surface brightness profile 
	using \texttt{Engauge Digitizer}\footnote{\href{http://markummitchell.github.io/engauge-digitizer/}{http://markummitchell.github.io/engauge-digitizer/}},
	a free software that extracts data points from images with graphs.

On the simulation side, 
	we construct a spectrum and a surface brightness profile
	by counting photons (with their weight) 
	in a given wavelength bin and a radial bin, respectively.
To better match the observation,
	it is necessary to follow the actual observational setups
	including the aperture for spectrum, imaging band width, 
	spectral resolution, and point spread function (PSF).
Following the processes described in L17,
	we apply their aperture size 
	($r_\textrm{\cs lim}$, the white contour in the second column of their Figures 2 and 3)
	and imaging band width (the purple shaded range in the third column)
	when we count photons to construct spectrum and surface brightness profile, respectively.
We also convolve the spectrum 
	with a Gaussian kernel with the MUSE spectral resolution of $R\sim3000$,
	and convolve the surface brightness profile
	with a Moffat kernel with a fixed power index ($\beta$) of 2.8 
	and the MUSE PSF FWHM $\sim0.7\arcsec$.
The MUSE spectral resolution and PSF FWHM are dependent on wavelength, 
	but the dependence is weak \citep{Bacon_etal_2017}.
Therefore we ignore their wavelength dependence for simplicity.
The aperture size ($r_\textrm{\cs lim}$) applied to each MUSE galaxy 
	is summarized in Table \ref{tab-MUSEinfo}
	together with the scale radius of the UV surface brightness profile (\rscont).
Both are in the unit normalized by \rmaxx
	that is the maximum angular extent of $5\arcsec$ as displayed in Figures 2 and 3 in L17,
	and that corresponds to the maximum angular size of the galaxy in the simulation.

%%%%%%%%%%%%%%%%%%%%%%
% Table
\begin{deluxetable*}{ccccccc}
	\tablecaption{Observed Properties of Our Target Galaxies\label{tab-MUSEinfo}}
	\tablehead{MUSE \# 
			& \rscont\tablenotemark{a} 
			& $r_\textrm{\cs lim}$\tablenotemark{b} 
			& $[\lambda_1,\lambda_2]$\tablenotemark{c}
			& $\langle S/N \rangle_\textrm{\cs sp}$\tablenotemark{d}
			& $\langle S/N \rangle_\textrm{\tiny SBP}$\tablenotemark{e}
			& $\langle S/N \rangle_\textrm{\cs tot}$\tablenotemark{f}}
	\startdata
		1185& 0.041 & 0.24 & [6681.179, 6692.522] &  7.747 & 7.051 & 7.482 \\
	 	82	& 0.017 & 0.20 & [5598.684, 5611.14]  &  3.344 & 6.025 & 4.327 \\
		6905& 0.029 & 0.20 & [4982.524, 4990.045] &  3.835 & 4.096 & 3.943 \\
		1343& 0.016 & 0.20 & [6038.929, 6050.097] &  1.427 & 2.070 & 1.748 \\
		53	& 0.030 & 0.24 & [7021.233, 7029.938] & 10.158 & 7.192 & 9.205 \\
		171 & 0.025 & 0.24 & [5934.927, 5941.174] &  2.471 & 4.171 & 3.168 \\
		547 & 0.011 & 0.24 & [8477.564, 8486.28]  &  3.891 & 4.398 & 4.103 \\
		364 & 0.014 & 0.24 & [6004.97, 6009.919]  &  1.213 & 3.008 & 1.801 \\
	\enddata
	\tablenotetext{a}{Scale radius of UV continuum radial profile in the unit of \rmax. 
					Borrowed from \rscontt in Figure 4 (also see Table B.1) in L17.}
	\tablenotetext{b}{Aperture size for spectrum obtained by roughly measuring the radius of 
					the HST segmentation mask that is convolved with the MUSE PSF
					(white contour in each panel of the second column of Figures 2 and 3 in L17).}
	\tablenotetext{c}{Wavelength range for surface brightness profile (i.e., image bandwidth)
					that corresponds to the purple area in each panel of the third column of Figures 2 and 3 in L17.}
	\tablenotetext{d}{Mean signal-to-noise of observed spectrum data}
	\tablenotetext{e}{Mean signal-to-noise of observed surface brightness profile data}
	\tablenotetext{f}{Mean signal-to-noise of observed spectrum and surface brightness profile data}
\end{deluxetable*}
%%%%%%%%%%%%%%%%%%%%%%

To quantify how well the spectrum and the surface brightness profile of simulation 
	describe the observational data, we define a likelihood as 
	\begin{equation}
		\ln \mathcal{L} \propto -\frac{1}{2} \sum_{i} \left(\frac{O_i-M_i}{\sigma(O_i)}\right)^2.
	\end{equation}
Here, $i$ denotes $i$-th bin (a wavelength bin for spectrum and a radial bin for surface brightness profile),
	$O$ and $M$ denote values from observation and model, respectively,
	and $\sigma(O)$ is the error in observational data.
Assuming that spectrum and surface brightness profile are independent from each other,
	the total likelihood ($\mathcal{L}_\textrm{\cs tot}$) of a model with a given parameter set
	can be defined by a product of the likelihoods of spectrum and surface brightness profile 
	($\mathcal{L}_\textrm{\cs sp}$ and $\mathcal{L}_\textrm{\tiny SBP}$, respectively,
	and the subscripts sp and SBP for spectrum and surface brightness profile, respectively).
By calculating $\mathcal{L}_\textrm{\cs tot}$ for all permitted parameter sets,
	we construct a surface of $\mathcal{L}_\textrm{\cs tot}$
	in the seven (five simulation parameters plus two post-processing parameters) dimensional parameter space.
Then, we derive a one-dimensional (1D) (marginal) posterior distribution of a parameter 
	and a two-dimensional (2D) (marginal) posterior map of two parameters
	by marginalizing over other parameters assuming uniform priors.
A best-fit parameter is found as the value 
	where the maximum of its 1D posterior distribution appears,
	and the degeneracy between parameters is inferred from 2D posterior maps.
We note that the normalizations of the simulated spectrum and the simulated surface brightness profile 
	are adjusted arbitrarily to minimize the difference between model and observation.
{\it Because we compare only their shapes,
	a parameter that is more sensitive to the absolute levels (e.g., DGR)
	could be poorly constrained by our modeling.}

The error of each best-fit value is roughly measured as follows.
We linearly connect adjacent $\mathcal{L}$ values 
	and adjust the normalization to make the area below the $\mathcal{L}$ profile 
	within the allowed parameter range be 1.
We then make a 1$\sigma$ range that includes a best-fit value 
	and covers 68\% of the total area.
%}}}

%%%%%%%%%%%%%%%%%%%%%%%%%%%%%%%%%%%%%%%%%%%%%%%%%%%%%%%%%%%%%%%%%%%%%%%%%%%%%%%%%%%%%%%%%%%%%%%%%%%%
\section{RESULTS: FIT TO OBSERVED \Lyaa SPECTRA AND SURFACE BRIGHTNESS PROFILES}\label{sec-fit-result}
In this section, we show 1D and 2D posterior distributions of the seven parameters,
	constraints on their best-fit value,
	and the corresponding best model spectrum and surface brightness profile
	for each of our eight target galaxies.
We first fit spectrum and surface brightness profile separately (Section \ref{ssec-fit-sep}),
	and later fit both simultaneously (Section \ref{ssec-fit-sim}).

\subsection{To Fit Spectra and Surface brightness profiles Separately}\label{ssec-fit-sep}
\begin{figure*}
	\center
	\includegraphics[width=0.9\textwidth]{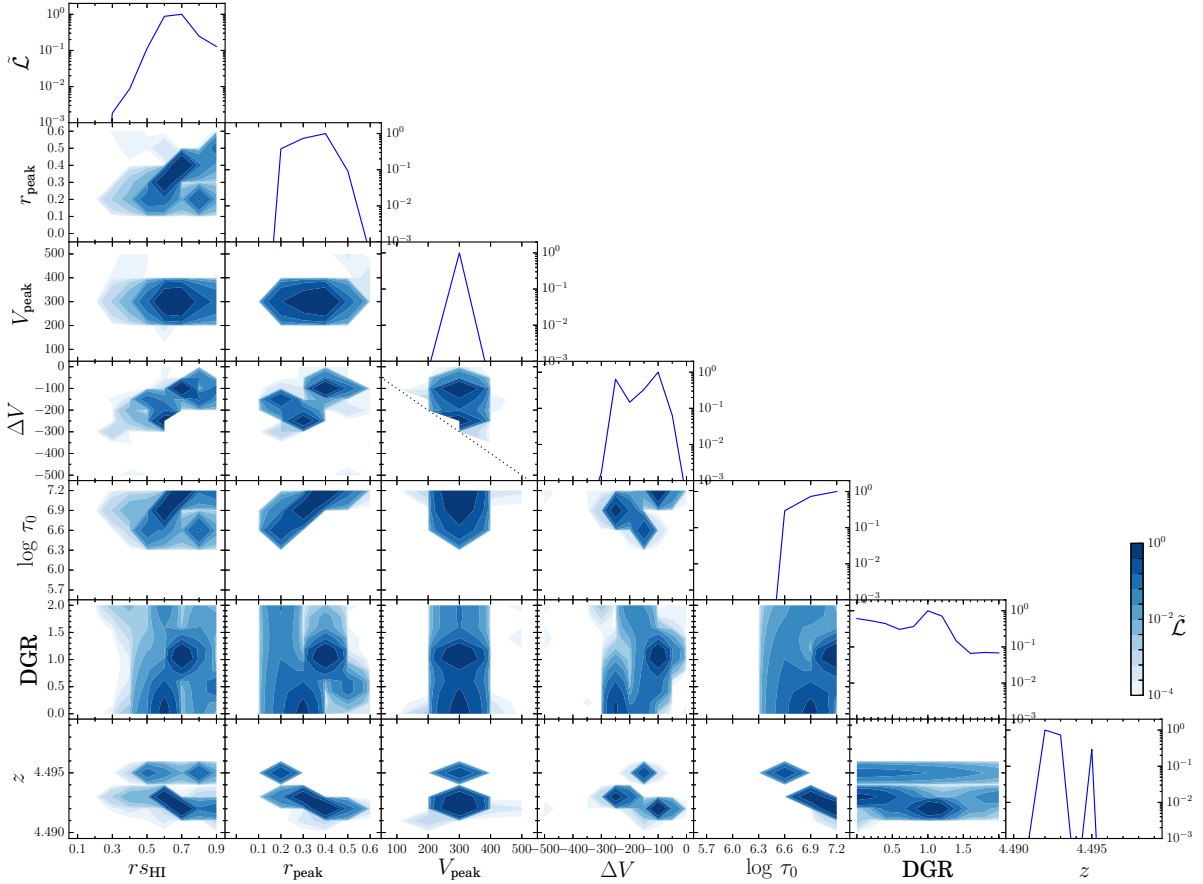}
	\caption{
		1D (the top panel of each column) and 2D (the rest) marginal posterior distributions
			for the spectrum ($\tilde{\mathcal{L}}_\textrm{\cs sp}$) of MUSE \#1185.
			The tilde indicates that posterior values are normalized by their maximum.
		In the panel of \delV--\Vpeak, the dotted line indicates the allowed \delVV range
			(i.e., $-V_\textrm{\cs peak}<\Delta V<0$) for a given \Vpeakk value
			to represent decelerated outflowing motion but prevent inflowing motion.
	}\label{fig-lnliksp-1185}
\end{figure*}

\begin{figure*}
	\center
	\includegraphics[width=0.9\textwidth]{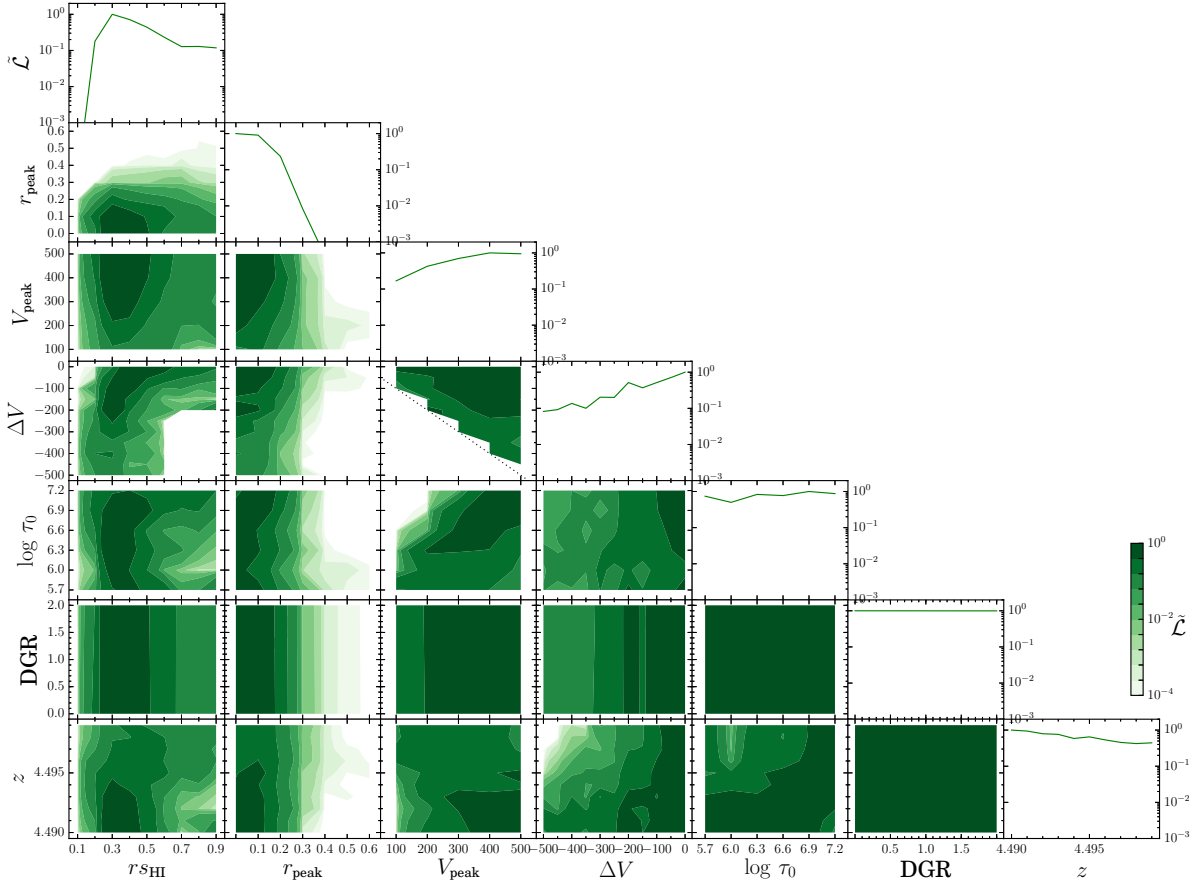}
	\caption{
		Similar to Figure \ref{fig-lnliksp-1185}, 
			but (normalized) posterior distributions for the surface brightness profile 
			($\tilde{\mathcal{L}}_\textrm{\tiny SBP}$) of MUSE \#1185.
	}\label{fig-lnliksbp-1185}
\end{figure*}

%{{{
Figures \ref{fig-lnliksp-1185} and \ref{fig-lnliksbp-1185} show
	the posterior distributions of the parameters 
	for the spectrum ($\mathcal{L}_\textrm{\cs sp}$) 
	and surface brightness profile ($\mathcal{L}_\textrm{\tiny SBP}$), respectively, 
	of MUSE \#1185.
Note that posterior values are normalized by their maximum
	(denoted by $\tilde{\mathcal{L}}$).
Overall, spectrum provides tighter constraints on the parameters 
	than surface brightness profile does.
Among the seven parameters,
	redshift is most tightly constrained by spectrum,
	but neither spectrum nor surface brightness profile constrains DGR.
It should be noted that 
	surface brightness profile also provides a constraint on redshift.
The results are similar for other targets.
%%%%%%%%%%%%%%%%%%%%%
% Figure 
%%%%%%%%%%%%%%%%%%%%%
% Figure
%%%%%%%%%%%%%%%%%%%%%

The best-fit parameter sets for spectrum and surface brightness profile, respectively,
	are found based on their 1D posterior distribution
	as the maximum posterior location for each parameter.
We draw two sets of model spectrum and surface brightness profile (solid line) 
	with observational data (filled circles with error bar) in the left and middle panels:
	one with the best-fit parameter set for spectrum (Figure \ref{fig-bestfit-sp-1185})
	and the other with that for surface brightness profile (Figure \ref{fig-bestfit-sbp-1185}).
Each best-fit parameter set reproduces well one, not the other.

The right panels show model profiles of medium density and velocity for each given best-fit parameter set.
It is clearly seen that 
	the models for spectrum and surface brightness profile do not necessarily agree with each other;
the discrepancy is removed by fitting the spectrum and surface brightness profile simultaneously.
%}}}

\subsection{To Fit Spectra and Surface brightness profiles Simultaneously}\label{ssec-fit-sim}
\begin{figure*}
	\center
	\includegraphics[width=0.9\textwidth]{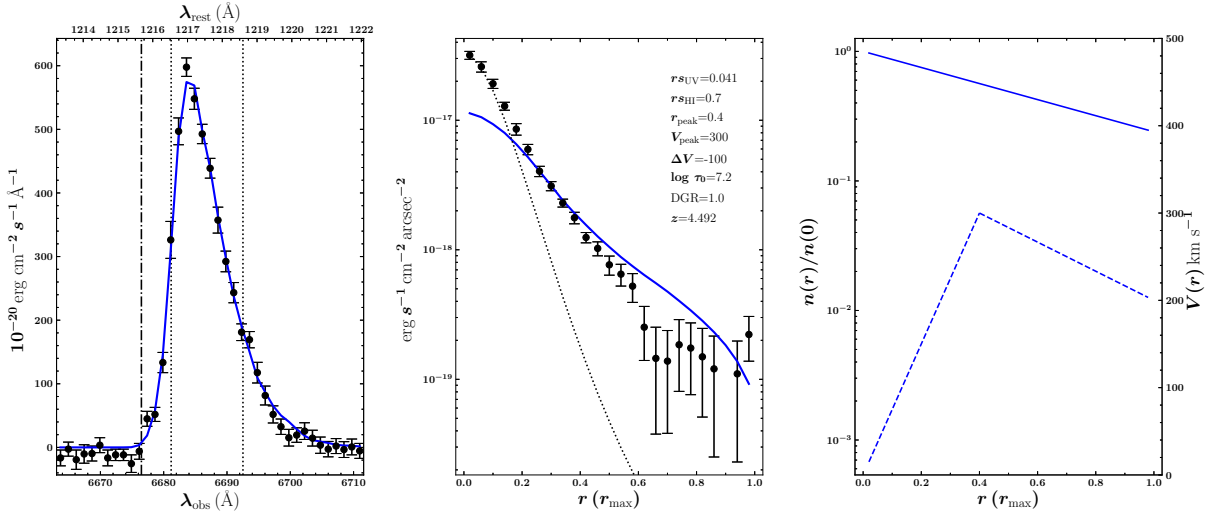}
	\caption{
		The spectrum (left) and surface brightness profile (middle) are presented:
			filled circles with error bar are observational data
			and solid line is the model with the best-fit parameter set
			for the spectrum of MUSE \#1185 
			(denoted in the upper right of the right panel).
		In the left panel,
			two dotted vertical lines denote the wavelength range 
			used for surface brightness profile.
		The dash-dotted line denotes the \Lyaa central wavelength
			in the observer's frame of the best-fit redshift (lower x-axis)
			and in the rest frame (upper x-axis, at 1215.67$\AA$).
		In the middle panel,
			the dotted line represents the surface brightness profile of the observed UV continuum
			that is used to reconstruct the spatial distribution of \Lyaa source,
			which is an input in our model.
		In the right panel,
			the models for density (solid line, y-axis on the left) 
			and velocity (dashed line, y-axis on the right) profiles of medium 
			for the given parameter set are presented.
	}\label{fig-bestfit-sp-1185}
\end{figure*}

\begin{figure*}
	\center
	\includegraphics[width=0.9\textwidth]{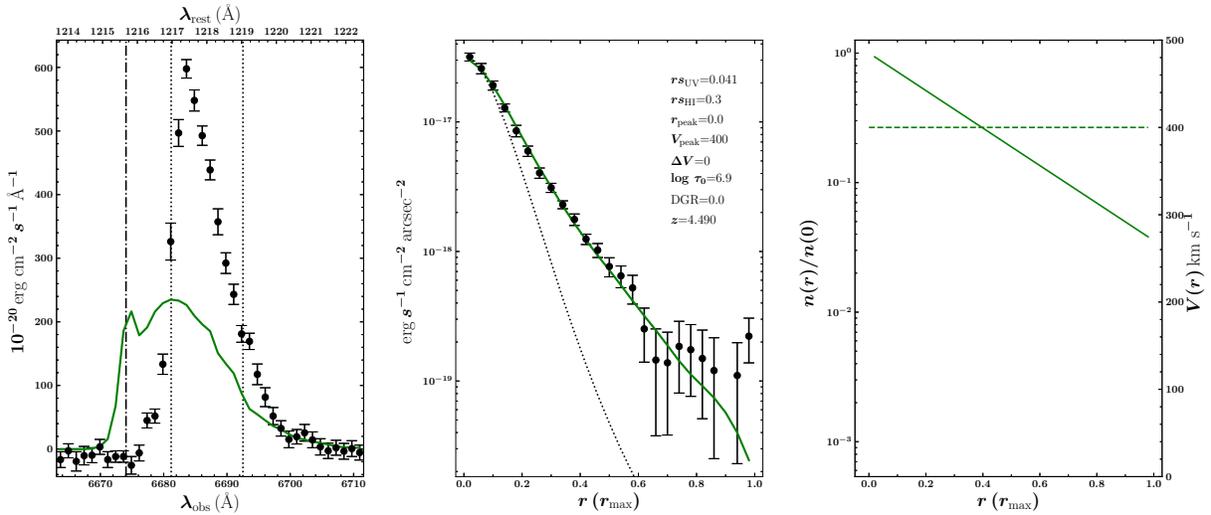}
	\caption{
		Similar to Figure \ref{fig-bestfit-sp-1185},
			but with the best-fit model for the surface brightness profile of MUSE \# 1185.
	}\label{fig-bestfit-sbp-1185}
\end{figure*}

%{{{
To find a best-fit parameter set that explains well both spectrum and surface brightness profile simultaneously,
	we now explore the total posterior distributions, $\mathcal{L}_\textrm{\cs tot}$,
	which are given by the product of $\mathcal{L}_\textrm{\cs sp}$ and $\mathcal{L}_\textrm{\tiny SBP}$.
Figure \ref{fig-lnliktot-1185} shows those for MUSE \#1185
	(refer to similar figures in Appendix \ref{sec-othertargets} for other target galaxies).
The total posterior distributions become much sharper than those of spectrum or surface brightness profile.
The model spectrum (left) and surface brightness profile (middle)
	with the best-fit parameter set obtained from $\mathcal{L}_\textrm{\cs tot}$
	are shown in Figure \ref{fig-bestfit-tot-1185}.
This parameter set reproduces nicely both the spectrum and surface brightness profile.
The model profiles of medium density and velocity obtained from this simultaneous fit
	are shown in the right panel. 
The density profile is the same as that obtained from the surface brightness profile fit,
	while the velocity profile is different from both of the two obtained from the separate fits
	(see the right panels of Figures \ref{fig-bestfit-sp-1185} and \ref{fig-bestfit-sbp-1185}).
The	medium density profile tends to follow that determined by surface brightness profile,
	as indicated in Table \ref{tab-bestfit}
	that summarizes fitting results.
%%%%%%%%%%%%%%%%%%%%%
% Figure
%%%%%%%%%%%%%%%%%%%%%
% Figure

%%%%%%%%%%%%%%%%%%%%%
% Figure
\begin{figure*}
	\center
	\includegraphics[width=0.9\textwidth]{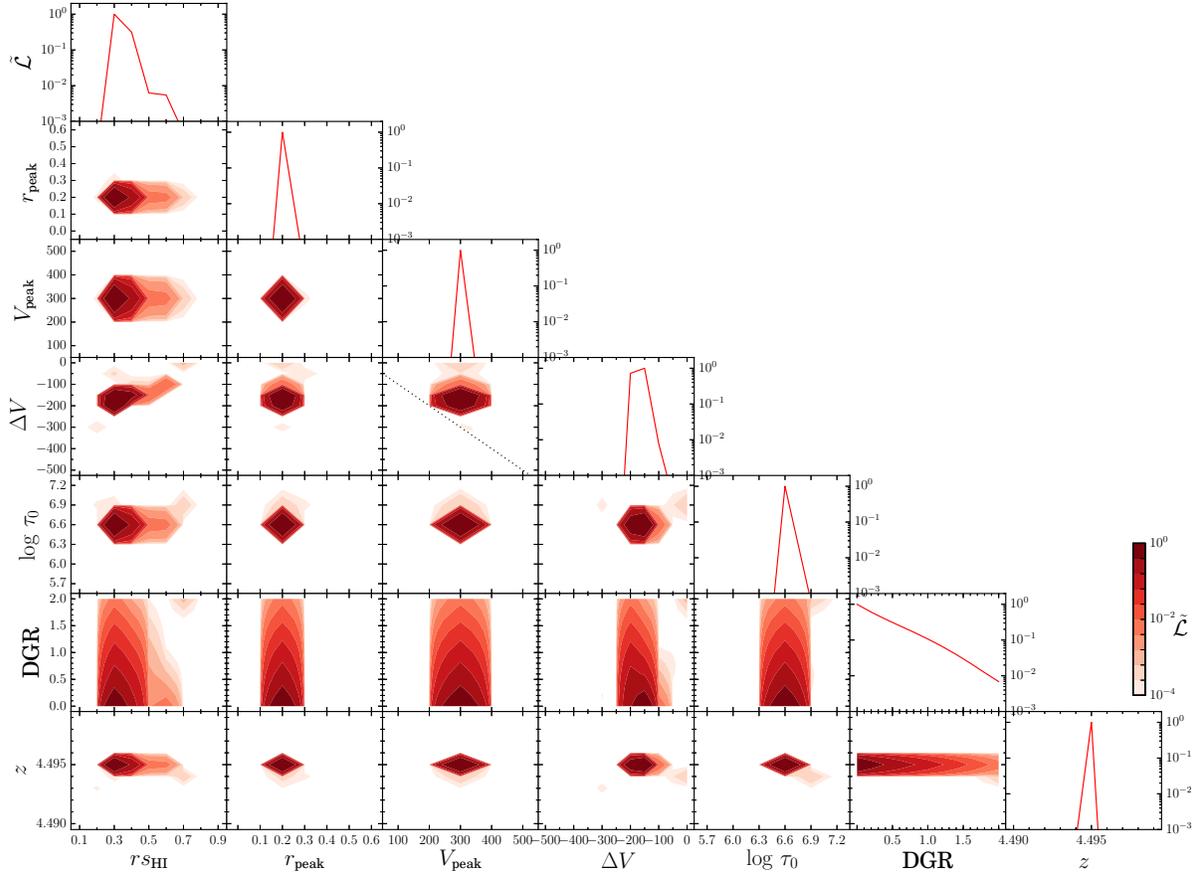}
	\caption{
		Similar to Figure \ref{fig-lnliksp-1185}, 
			but for total posterior distributions ($\tilde{\mathcal{L}}_\textrm{\cs tot}$)
			of MUSE \#1185.
	}\label{fig-lnliktot-1185}
\end{figure*}
%%%%%%%%%%%%%%%%%%%%%
% Figure
\begin{figure*}
	\center
	\includegraphics[width=0.9\textwidth]{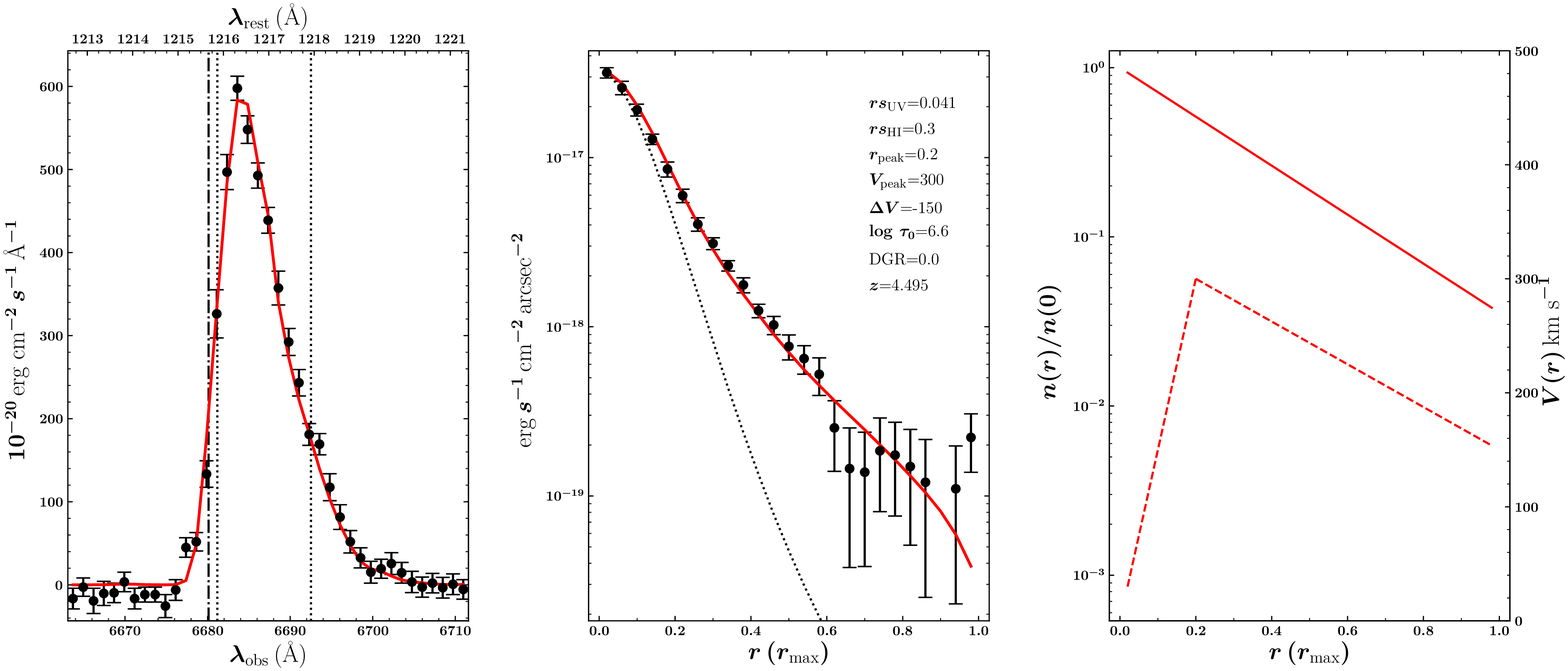}
	\caption{
		Similar to Figure \ref{fig-bestfit-sp-1185},
			but with the best-fit model 
			obtained by considering both spectrum and surface brightness profile
			of MUSE \#1185.
	}\label{fig-bestfit-tot-1185}
\end{figure*}
%%%%%%%%%%%%%%%%%%%%%

We summarize best-fit parameter sets determined by
	$\mathcal{L}_\textrm{\cs sp}$, $\mathcal{L}_\textrm{\tiny SBP}$, and $\mathcal{L}_\textrm{\cs tot}$,
	respectively, in Table \ref{tab-bestfit} for all the target galaxies. 
In most cases, the best-fit parameter set determined by $\mathcal{L}_\textrm{\cs tot}$
	is different from that by either $\mathcal{L}_\textrm{\cs sp}$ or $\mathcal{L}_\textrm{\tiny SBP}$.
In the case of MUSE \#53,
	the parameters are constrained entirely by spectrum,
	which can be expected from their posterior distributions given by spectrum (Figure \ref{fig-lnliksp-53})
	that are much sharper than those by surface brightness profile (Figure \ref{fig-lnliksbp-53}).
Although the surface brightness profile does not alter the best-fit parameter values determined by spectrum,
	it still makes the parameter constraints tighter.
This is clearly seen in the total posterior distributions (Figure \ref{fig-lnliktot-53})
	in comparison with the spectrum posterior distributions (Figure \ref{fig-lnliksp-53}).

The constraints of parameters are, of course, largely affected by data quality.
The peak values of spectrum and surface brightness profile
	of MUSE \#1185, 82, 6905, and 53 are 
	in the orders of $10^{-18}$erg cm$^{-2}$s$^{-1}$\AA$^{-1}$
	and $10^{-17}$erg cm$^{-2}$s$^{-1}$arcsec$^{-2}$, respectively,
	(refer to filled circles with error bars in Figures 
	\ref{fig-bestfit-tot-1185}, \ref{fig-bestfit-tot-82}, \ref{fig-bestfit-tot-6905}, and \ref{fig-bestfit-tot-53})
	with the mean signal-to-noise ratio ($\langle S/N\rangle$) $\gtrsim 4$
	(see the last three columns of Table \ref{tab-MUSEinfo}).
On the other hand,
	those of MUSE \#1343, 171, 547, and 364 
	are lower by an order of magnitude with $\langle S/N\rangle \lesssim 4$
	(refer to Figures \ref{fig-bestfit-tot-1343}, \ref{fig-bestfit-tot-171}, \ref{fig-bestfit-tot-547}, and \ref{fig-bestfit-tot-364}
	and Table \ref{tab-MUSEinfo}).
The posterior distributions for the target galaxies with stronger signals
	are sharper in general,
	and therefore parameter constraints are tighter
	(e.g., Figures \ref{fig-lnliktot-1185}, \ref{fig-lnliktot-82}, \ref{fig-lnliktot-6905}, \ref{fig-lnliktot-53}
	versus Figures \ref{fig-lnliktot-1343}, \ref{fig-lnliktot-171}, \ref{fig-lnliktot-547}, \ref{fig-lnliktot-364}).

Table \ref{tab-bestfit} also shows
	the reduced chi-square value of the model spectrum ($\chi_{\nu,\textrm{\cs sp}}^2$)
	for a given best-fit parameter set.
Similarly, the 11th column is that for a model surface brightness profile ($\chi_{\nu,\textrm{\tiny SBP}}^2$).
The best-fit parameter set based on $\mathcal{L}_\textrm{\cs tot}$ (the last row of each object)
	gives fairly good reduced chi-square values ($\lesssim1$) for both spectrum and surface brightness profile,
	which quantitatively shows that the model successfully fits the observation.

%%%%%%%%%%%%%%%%%%%%%%
% Table
\begin{deluxetable*}{ccccccccccc}
	\tablecaption{Best-fit Parameter Values\label{tab-bestfit}}
	\tablehead{MUSE \#
			& $\mathcal{L}$\tablenotemark{a} 
			& \rscale
			& \rpeak
			& \Vpeak
			& $\Delta V$
			& $\log \tau_0$
			& DGR
			& $z$ 
			& $\chi_{\nu,\textrm{\cs sp}}^2$\tablenotemark{b}
			& $\chi_{\nu,\textrm{\tiny SBP}}^2$\tablenotemark{c}}
	\startdata
		\multirow{3}{*}{1185}	& sp	& $0.7^{+0.2}_{-0.1}$	& $0.4^{+0.2}_{-0.1}$	& $300^{+43}_{-43}$		& $-100^{+100}_{-104}$	& $7.2^{}_{-0.6}$		& $1.0^{+0.8}_{-0.6}$	& $4.492^{+0.001}_{-0.002}$		& $  1.21$ & $16.18$ \\
								& SBP	& $0.3^{+0.2}_{-0.2}$	& $0.0^{+0.1}_{}$		& $400^{+100}_{-148}$	& $   0^{}_{-238}$		& $6.9^{+0.4}_{-0.7}$	& $0.0^{+1.4}_{}$		& $4.490^{+0.005}_{-0.001}$		& $120.48$ & $ 0.84$ \\ 
								& tot	& $0.3^{+0.1}_{-0.1}$	& $0.2^{+0.0}_{-0.0}$	& $300^{+43}_{-43}$		& $-150^{+150}_{-37}$	& $6.6^{+0.1}_{-0.1}$	& $0.0^{+0.4}_{}$		& $4.495^{+0.000}_{-0.000}$		& $  1.88$ & $ 1.14$ \\ 
		\hline                                                                                                                                                    
		\multirow{3}{*}{82}		& sp	& $0.9^{}_{-0.5}$		& $0.2^{+0.1}_{-0.1}$	& $300^{+200}_{-82}$	& $   0^{}_{-297}$		& $7.2^{}_{-0.6}$		& $0.0^{+1.2}_{}$		& $3.602^{+0.001}_{-0.006}$		& $  3.70$ & $ 5.21$ \\
								& SBP	& $0.7^{+0.2}_{-0.2}$	& $0.0^{+0.1}_{}$		& $500^{}_{-277}$		& $-200^{+97}_{-300}$	& $5.7^{+0.5}_{}$		& $0.0^{+1.4}_{}$		& $3.601^{+0.003}_{-0.003}$		& $ 44.19$ & $10.41$ \\
								& tot	& $0.5^{+0.4}_{-0.1}$	& $0.1^{+0.0}_{-0.0}$	& $300^{+43}_{-43}$		& $-250^{+47}_{-30}$	& $6.6^{+0.1}_{-0.1}$	& $2.0^{}_{-1.3}$		& $3.604^{+0.000}_{-0.000}$		& $  0.94$ & $ 1.16$ \\
		\hline                                                                                                                                                    
		\multirow{3}{*}{6905}	& sp	& $0.5^{+0.3}_{-0.3}$	& $0.2^{+0.2}_{-0.2}$	& $200^{+55}_{-62}$		& $-150^{+108}_{-154}$	& $7.2^{}_{-0.6}$		& $1.0^{+0.7}_{-0.7}$	& $3.096^{+0.001}_{-0.001}$		& $  5.05$ & $20.58$ \\
								& SBP	& $0.3^{+0.3}_{-0.2}$	& $0.0^{+0.2}_{}$		& $500^{}_{-253}$		& $-100^{+100}_{-139}$	& $6.0^{+0.6}_{-0.3}$	& $2.0^{}_{-1.4}$		& $3.098^{+0.002}_{-0.005}$		& $ 48.80$ & $ 0.38$ \\
								& tot	& $0.1^{+0.0}_{}$		& $0.0^{+0.1}_{}$		& $300^{+52}_{-53}$		& $-300^{+44}_{-83}$	& $6.3^{+0.3}_{-0.6}$	& $0.0^{+1.3}_{}$		& $3.098^{+0.001}_{-0.001}$		& $  0.82$ & $ 0.49$ \\
		\hline                                                                                                                                                    
		\multirow{3}{*}{1343}	& sp	& $0.1^{+0.5}_{}$		& $0.4^{+0.2}_{-0.2}$	& $500^{}_{-286}$		& $   0^{}_{-264}$		& $7.2^{}_{-0.7}$		& $1.4^{+0.7}_{-0.8}$	& $3.964^{+0.002}_{-0.004}$		& $  0.17$ & $ 1.51$ \\
								& SBP	& $0.9^{}_{-0.4}$		& $0.3^{+0.2}_{-0.2}$	& $400^{+100}_{-164}$	& $-200^{+141}_{-168}$	& $7.2^{}_{-1.0}$			& $1.0^{+0.7}_{-0.7}$	& $3.963^{+0.003}_{-0.003}$		& $  1.02$ & $ 0.34$ \\
								& tot	& $0.8^{+0.1}_{-0.3}$	& $0.4^{+0.2}_{-0.2}$	& $200^{+168}_{-100}$	& $-200^{+125}_{-167}$	& $6.9^{+0.3}_{-0.4}$	& $2.0^{}_{-1.4}$		& $3.964^{+0.002}_{-0.004}$		& $  0.96$ & $ 1.36$ \\
		\hline                                                                                                                                                    
		\multirow{3}{*}{53}		& sp	& $0.4^{+0.0}_{-0.0}$	& $0.1^{+0.0}_{-0.0}$	& $300^{+43}_{-43}$		& $ -50^{+18}_{-18}$	& $6.3^{+0.1}_{-0.1}$	& $2.0^{}_{-1.0}$		& $4.776^{+0.000}_{-0.000}$		& $  1.65$ & $ 1.50$ \\
								& SBP	& $0.2^{+0.2}_{-0.1}$	& $0.1^{+0.1}_{-0.1}$	& $500^{}_{-253}$		& $   0^{}_{-227}$		& $6.0^{+0.6}_{-0.3}$	& $0.0^{+1.4}_{}$		& $4.775^{+0.002}_{-0.003}$		& $117.90$ & $ 6.84$ \\
								& tot	& $0.4^{+0.1}_{-0.1}$	& $0.1^{+0.0}_{-0.1}$	& $300^{+43}_{-43}$		& $ -50^{+22}_{-23}$	& $6.3^{+0.1}_{-0.1}$	& $2.0^{}_{-1.0}$		& $4.776^{+0.000}_{-0.000}$		& $  1.65$ & $ 1.50$ \\
		\hline                                                                                                                                                    
		\multirow{3}{*}{171}	& sp	& $0.1^{+0.4}_{}$		& $0.6^{}_{-0.4}$		& $200^{+191}_{-100}$	& $-100^{+100}_{-108}$	& $7.2^{}_{-0.7}$		& $2.0^{}_{-1.4}$		& $3.880^{+0.002}_{-0.001}$		& $  2.24$ & $ 6.86$ \\
								& SBP	& $0.5^{+0.3}_{-0.2}$	& $0.0^{+0.1}_{}$		& $500^{}_{-259}$		& $-200^{+117}_{-205}$	& $6.6^{+0.5}_{-0.6}$	& $2.0^{}_{-1.4}$		& $3.881^{+0.004}_{-0.002}$		& $ 10.22$ & $ 0.79$ \\
								& tot	& $0.8^{+0.1}_{-0.4}$	& $0.0^{+0.1}_{}$		& $200^{+48}_{-51}$		& $   0^{}_{-175}$		& $6.3^{+0.2}_{-0.2}$	& $2.0^{}_{-1.4}$		& $3.882^{+0.003}_{-0.001}$		& $  1.20$ & $ 0.81$ \\
		\hline                                                                                                                                                    
		\multirow{3}{*}{547}	& sp	& $0.9^{}_{-0.5}$		& $0.2^{+0.2}_{-0.2}$	& $200^{+219}_{-100}$	& $-150^{+84}_{-254}$	& $6.6^{+0.7}_{-0.2}$	& $0.0^{+1.4}_{}$		& $5.973^{+0.002}_{-0.001}$		& $ 13.46$ & $12.09$ \\
								& SBP	& $0.5^{+0.2}_{-0.1}$	& $0.1^{+0.1}_{-0.1}$	& $500^{}_{-225}$		& $-500^{+137}_{}$		& $5.7^{+0.5}_{}$		& $0.0^{+1.4}_{}$		& $5.978^{+0.001}_{-0.005}$		& $ 39.42$ & $ 1.30$ \\
								& tot	& $0.7^{+0.2}_{-0.2}$	& $0.1^{+0.1}_{-0.1}$	& $300^{+200}_{-78}$	& $-200^{+88}_{-188}$	& $6.3^{+1.0}_{-0.1}$	& $2.0^{}_{-1.4}$		& $5.974^{+0.001}_{-0.001}$		& $  1.25$ & $ 1.07$ \\
		\hline                                                                                                                                        
		\multirow{3}{*}{364}	& sp	& $0.5^{+0.3}_{-0.2}$	& $0.1^{+0.4}_{-0.1}$	& $500^{}_{-236}$		& $   0^{}_{-231}$		& $6.3^{+0.3}_{-0.5}$	& $2.0^{}_{-1.4}$		& $3.939^{+0.001}_{-0.001}$		& $  1.81$ & $ 3.42$ \\ 
								& SBP	& $0.5^{+0.2}_{-0.2}$	& $0.0^{+0.2}_{}$		& $400^{+100}_{-167}$	& $-200^{+108}_{-174}$	& $6.6^{+0.5}_{-0.6}$	& $0.0^{+1.4}_{}$		& $3.935^{+0.004}_{-0.001}$		& $  3.55$ & $ 0.64$ \\ 
								& tot	& $0.9^{}_{-0.4}$		& $0.3^{+0.4}_{-0.1}$	& $200^{+77}_{-70}$		& $-200^{+67}_{-300}$	& $6.0^{+0.3}_{-0.3}$	& $2.0^{}_{-1.4}$		& $3.939^{+0.004}_{-0.001}$		& $  0.68$ & $ 0.36$ \\ 
	\enddata
	\tablecomments{Note that \rscalee and \rpeakk are in the normalized unit by \rmax, 
				and DGR is in the unit of the value of the Milky Way (DGR$_\textrm{\tiny MW}$).}
	\tablenotetext{a}{The marginal posterior distribution of parameters 
				that is used to determine a given best-fit parameter set
				among those obtained using spectrum ($\mathcal{L_\textrm{\cs sp}}$), 
				surface brightness profile ($\mathcal{L_\textrm{\tiny SBP}}$), and both ($\mathcal{L_\textrm{\cs tot}}$).}
	\tablenotetext{b}{The reduced chi-square value of the model spectrum with a given best-fit parameter set (of a given row in the table).}
	\tablenotetext{c}{The reduced chi-square value of the model surface brightness profile with a given best-fit parameter set.}
\end{deluxetable*}
%%%%%%%%%%%%%%%%%%%%%%
%}}}

%%%%%%%%%%%%%%%%%%%%%%%%%%%%%%%%%%%%%%%%%%%%%%%%%%%%%%%%%%%%%%%%%%%%%%%%%%%%%%%%%%%%%%%%%%%%%%%%%%%%
\section{DISCUSSION}\label{sec-discuss}
\subsection{Interpretations on Best-fit Results}\label{ssec-fit-interpret}
%{{{
In this section,
	we compare the best-fit values of some of our model parameters such as \Vpeakk and $\tau_0$
	with those from other studies.
\citet{Garel_etal_2012} constructed high-redshift LAEs in a dark matter-only cosmological simulation
	by implementing a semi-analytic model with an expanding shell model.
Their model galaxies, which are tuned to give a good match to observed UV and \Lyaa luminosity functions,
	have the expanding velocity of $\sim 150$--200$\,\textrm{km s}^{-1}$
	and \ion{H}{1} column density of $\sim 10^{20}\,\textrm{cm}^{-2}$.
We also compare our results with those of \citet{Verhamme_etal_2008} and \citet{Yang_etal_2016}.
\citet{Verhamme_etal_2008} reproduced the observed \Lyaa spectra of 11 high-redshift ($z\sim3$) Lyman break galaxies
	using an expanding shell model,
	and they found the best models for these galaxies with the expanding velocity of $\sim$150--200$\,\textrm{km s}^{-1}$
	and \ion{H}{1} column density of $\sim 2\times10^{19}$--$7\times10^{20}\,\textrm{cm}^{-2}$.
\citet{Yang_etal_2016} modeled 12 low-redshift ($z\sim0.2$) green pea galaxies\footnote{ 
		Green pea galaxies are nearby compact starburst galaxies
			and are thought to be analogous to high-redshift star-forming galaxies.}
	as an expanding shell, too.
Their best models for these galaxies give the expanding velocity of $\sim$30--350$\,\textrm{km s}^{-1}$
	and \ion{H}{1} column density of $\sim10^{19}$--$10^{20}\,\textrm{cm}^{-2}$.
All these results are similar to our results.

There are other estimates on the kinematics of medium in galaxies
	using interstellar/circumgalactic absorption lines.
\citet{Steidel_etal_2010} measured absorption lines in QSO spectra
	made by circumgalactic gas of star-forming galaxies at redshift $2\lesssim z\lesssim3$
	that are on the sightlines to the QSOs.
They then examined properties of absorption lines as a function of the galactocentric impact parameter up to $\gtrsim100$kpc,
	and modeled them using outflowing halos with constant/decelerating velocity profiles and a covering factor profile.
A constant velocity profile gives an outflowing velocity of 650--820$\,\textrm{km s}^{-1}$,
	and a decelerating velocity profile gives a mean outflowing velocity of $\sim$200--300$\,\textrm{km s}^{-1}$.
We could not directly compare because of different assumptions,
	but the estimate with a constant velocity profile is larger than our best-fit values
	and the estimate with a decelerating velocity profile is more or less similar to ours.
Because the estimate of the outflowing velocity varies largely depending on which velocity profile is used,
	the disagreement between these estimates could be not an issue.
It should be also noted that the best-fit velocity varies depending on the choice of a covering factor profile,
	and possibly other factors such as optical depth.
Moreover, different emission/absorption lines trace different kinds of gas, thus different kinematics.
Therefore, these suggest that 
	our constraints on the outflowing velocity are consistent with previous studies in general.

The last thing we would like to address is the constraint on redshift.
We set redshift as a free parameter, and find its best-fit values to have shifts of 0.001--0.008 to lower redshifts
	than the estimates when the peak of \Lyaa spectrum is assumed to be at the rest-frame \Lyaa central wavelength.
This means that \Lyaa line is more redshifted than that by the systemic redshift of galaxies,
	which is due to the resonance scattering of \Lya.
Compared to the typical redshift error of Sloan Digital Sky Survey
	(i.e., 30$\,\textrm{km s}^{-1}$, \citealt{Strauss_etal_2002}),
	the shifts of 0.001--0.008 are non-negligible.
More importantly, the accuracy of redshift estimate is transferred to that of the estimates on $\tau_0$
	and subsequently other parameters.
Reminding that \Lya-based redshift estimates
	are not always correct \citep[e.g.,][]{Orlitova_etal_2018},
	it is necessary to secure other reliable estimates of redshift if available.
%}}}

\subsection{Advantages of Using Both Spectrum and Surface Brightness Profile}\label{ssec-synergy}
%{{{
We examine the degeneracy between parameters based on the 2D posterior distributions
	(e.g., see Figures \ref{fig-lnliksp-1185} and \ref{fig-lnliksbp-1185}).
In the case of spectrum,
	degeneracies of $\tau_0$--\rpeakk and $z$--$\tau_0$
	appear prominently and commonly for all the target galaxies.
A weak degeneracy between $z$ and \rpeakk
	also appears in the majority of the target galaxies.
The posterior distributions of surface brightness profile also show
	degeneracies of \delV--\rscale, $\tau_0$--\Vpeak, 
	$z$--\Vpeak, and $z$--$\tau_0$ for most target galaxies,
	which are overall less prominent than those of spectrum.
Interestingly, the degeneracy between $z$ and $\tau_0$
	appears to be opposite for spectrum (negative correlation) 
	and surface brightness profile (positive).
Although parameter degeneracy itself hinders precise parameter constraints,
	such opposite behaviors of parameter degeneracy for spectrum and surface brightness profile
	together can provide tighter constraints on
	$z$, $\tau_0$, and consequently on other parameters as well.

The opposite behaviors of parameter degeneracy is partly attributed to
	the fact that the posterior distributions of spectrum and surface brightness profile
	prefer different parts of parameter spaces.
This results in quite different best-fit parameter sets for spectrum and surface brightness profile
	as summarized in Table \ref{tab-bestfit}.
The difference is non-negligible 
	as seen in Figures \ref{fig-bestfit-sp-1185} and \ref{fig-bestfit-sbp-1185};
	the best-fit parameter for one of spectrum and surface brightness profile
	completely fails at reproducing the other.
The comparison between the $\chi_{\nu,\textrm{\cs sp}}^2$ and the $\chi_{\nu,\textrm{\tiny SBP}}^2$
	with the best-fit parameter set for either spectrum or surface brightness profile
	shows this failure in a quantitative way.
The difference (i.e., ratio) between these two chi-squares can be as large as by a factor of a hundred.
This indicates that spectrum and surface brightness profile are complementary and orthogonal 
	in constraining our model parameters.
Therefore, these two observables together can a lot better constrain parameters,
	which is shown in the previous section with the total posterior distributions
	that become much sharper than those of spectrum or surface brightness profile.
Even the redshift constraint, which is expected to be done primarily by spectrum,
	is affected when surface brightness profile is taken into account.
This constraining power of surface brightness profile on redshift
	comes from the restriction on wavelength range for surface brightness profile
	(i.e., image bandwidth).
In the cases of MUSE \#1343, 53, and 364,
	surface brightness profile does not alter the redshift constraint by spectrum.
However, surface brightness profile still contributes to the parameter constraint
	by making it tighter, as seen in the total posterior distribution for redshift
	compared to its spectrum posterior distribution
	(e.g., Figure \ref{fig-lnliktot-53} versus Figure \ref{fig-lnliksp-53}).
%}}}

\subsection{\Lyaa Spectrum and Surface Brightness Profile Variation in Model Parameter Space}\label{ssec-parvar}
%{{{
We explore the variations of \Lyaa spectrum and of surface brightness profile in the model parameter space
	to better understand the fitting results and the parameter degeneracy in the previous sections.
Here, we have convolved raw model spectrum and surface brightness profile 
	with a Gaussian kernel and a Moffat kernel, respectively,
	as described in Section \ref{ssec-muse}.
However, we have applied neither the size of aperture (to spectrum)
	nor the image bandwidth (to surface brightness profile)
	to first understand pure impacts of the physical parameters.
The effect of the aperture size will be indirectly discussed later in Section \ref{ssec-SpResolvedSp}
	with spatially resolved spectra.
We examine the impacts of six parameters except redshift.
The results for $\tau_0$ and \rscalee are described in detail here,
	but those for other parameters are presented in Appendix \ref{sec-parvar-rest}.
It should be noted that
	we explore the impacts of parameters on the {\it shapes} of
	spectrum and surface brightness profile.

Although \Lyaa radiative transfer process
	is a bunch of random scattering events,
	we can roughly guess the shapes of emerging \Lyaa spectrum and surface brightness profile
	by inferring the frequency change and the last scattering position of photons, respectively.
The rule of thumb is that
	the larger the {\it effective} optical-depth\footnote{ 
		Conventionally optical depth refers to
			the cumulative optical depth that is measured at a fixed frequency.
		The {\it effective} optical-depth refers to an optical depth
			that takes into account the frequency change (thus the change of cross section) of photons,
			due to the relative motion between the photons and the medium,
			in the rest frame of the medium.}
	is, the larger the frequency change is.
A larger frequency change results in a broader spectrum 
	with its peak farther away from the \Lyaa central wavelength,
	from which the correlation between peak shift and FWHM is naturally expected,
	as discussed in Section \ref{ssec-peakshift-fwhm}.
The last scattering position can be guessed by inferring 
	the location where the cumulative effective optical-depth reaches a certain value.
We try to understand the impact of each parameter on \Lyaa spectrum and surface brightness profile
	based on this rule of thumb.
However, it should be noted that 
	the details of the impact of each parameter can manifest differently
	depending on the values of other parameters.

We show changes in the \Lyaa spectrum and surface brightness profile with each parameter.
	%for some of our target galaxies.
We choose two galaxies for which the impact of a parameter on the spectrum and surface brightness profile
	is apparent most significantly.
Our choice can differ depending on a parameter of interest.
For illustration, we fix parameters other than the one under consideration at their best-fit values.
However, DGR is fixed at zero unless it is the parameter of interest (i.e., Figure \ref{fig-DGR})
	to make problems simpler.
Spectrum is normalized by its total intensity,
	and surface brightness profile is normalized by its maximum value (i.e., the value at $r=0$)
	for convenience.

%%%%%%%%%%%%%%%%%%%%%
% Figure
\begin{figure*}
	\center
	\includegraphics[width=\textwidth]{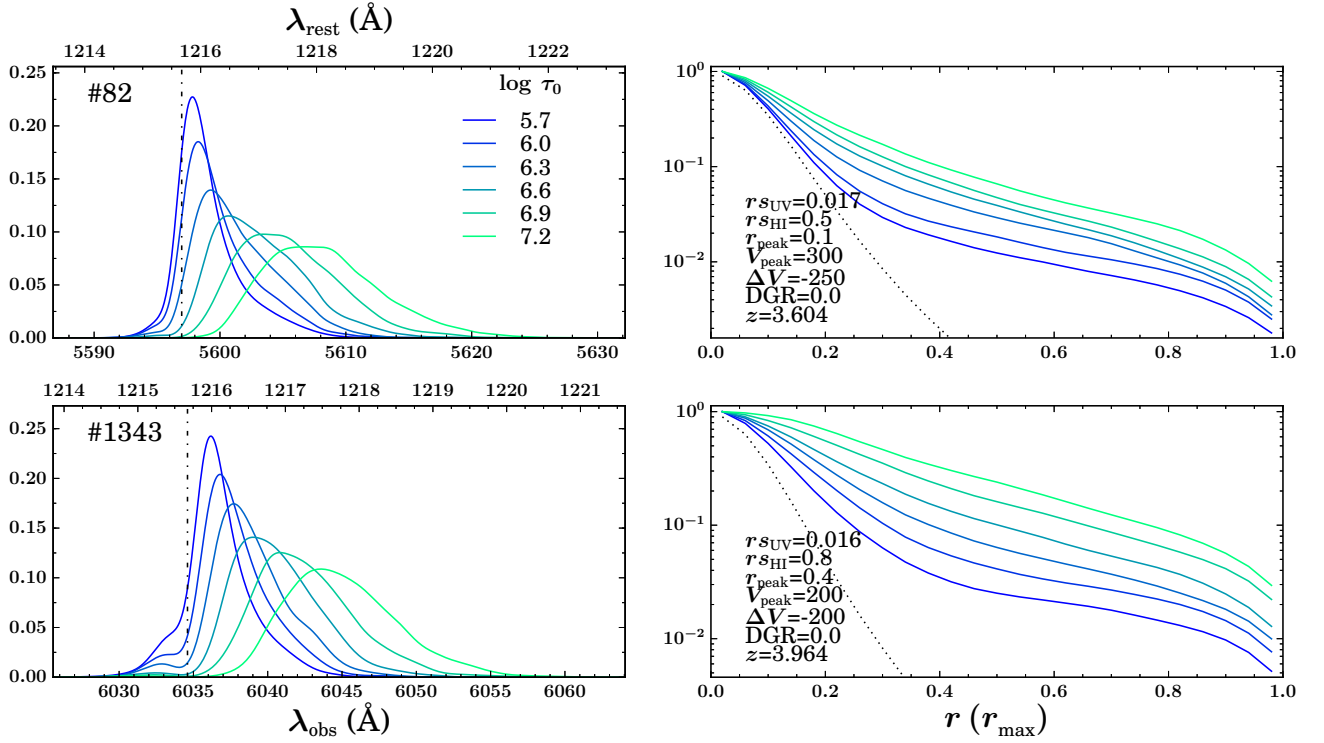}
	\caption{
		Spectra (left panel) and surface brightness profiles (right panel) with varying $\tau_0$ 
			for two of our target galaxies.
		The MUSE id of the chosen galaxies is written in top left of each spectrum panel.
		The parameters other than $\tau_0$ are fixed at their best-fit values
			except DGR that is fixed at zero.
		The dot-dashed line in the spectrum panels represents the \Lyaa central wavelength, 1215.67$\AA$,
			and the dotted line in the surface brightness profile panels 
			represents the surface brightness profile of UV continuum
			(i.e., \Lyaa source distribution).
	}\label{fig-tau0}
\end{figure*}
%%%%%%%%%%%%%%%%%%%%%
We start from the changes of spectrum and surface brightness profile with $\tau_0$,
	which is easy to understand.
Figure \ref{fig-tau0} shows the cases of MUSE \#82 and 1343.
The peak of spectrum moves farther away from the central wavelength
	and the width becomes broader as $\tau_0$ increases.
A surface brightness profile becomes flatter with increasing $\tau_0$.
These trends are because the number of scatterings in general increases accordingly.

%%%%%%%%%%%%%%%%%%%%%
% Figure
\begin{figure*}
	\center
	\includegraphics[width=\textwidth]{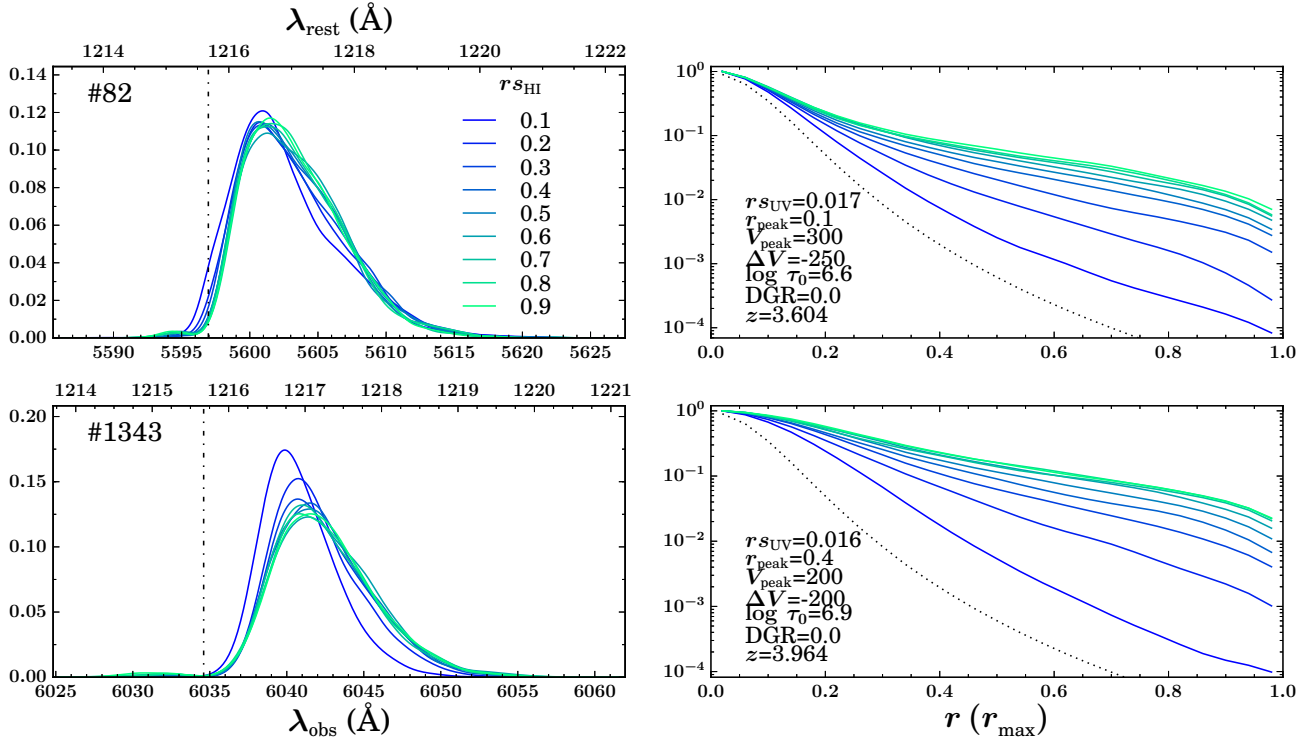}
	\caption{
		Similar to Figure \ref{fig-tau0}, but with varying \rscalee.
	}\label{fig-rscale}
\end{figure*}
%%%%%%%%%%%%%%%%%%%%%
Figure \ref{fig-rscale} shows the changes of \Lyaa spectrum and surface brightness profile with \rscalee
	for the same galaxies.
Spectrum does not change significantly with \rscale,
	but surface brightness profile becomes clearly flatter as \rscalee increases.
In a less centrally-concentrated medium (larger \rscale),
	\Lyaa photons can diffuse out of the central region more easily.
In turn, last scatterings tend to happen at larger radii, 
	which results in a flatter surface brightness profile.
Meanwhile, it is not obvious to understand the behavior of spectral shape with \rscale.
Although the column density remains the same,
	redistributing matter can either effectively increase or decrease optical depth
	depending on the velocity structure of the medium.
It is not only effective optical-depth but also last scattering positions that change with \rscale,
	which means that the velocities at which \Lyaa photons are scattered off also change.
All of these factors together make the prediction on the change of spectral shape with \rscalee more complicated.

There are two competing factors playing roles in forming the spectral shape in an expanding medium:
	(1) the Doppler frequency shift and (2) the effective optical-depth.
In an expanding medium, the frequency of a photon being scattered would be Doppler-shifted
	when transformed from the fixed frame to the medium frame;
	as a result, the final photon frequency after many scatterings 
	will be shifted by an amount proportional to the expanding velocity.
Therefore, the frequency changes due to the Doppler effect
	(if photons undergo enough number of scatterings)
	will be more significant in a faster medium than in a slower medium.
However, as the medium's velocity increases,
	the effective optical-depth would decrease,
	and thus the number of scatterings the photons undergo
	before escaping the medium will decrease as well;
	consequently, we expect that a faster medium yields a smaller frequency change.
This latter effect will be particularly significant in a fast-moving medium.
We found, as shown in Appendix \ref{sec-peak-Vmax},
	that the frequency shift due to the Doppler effect in an expanding medium
	is essential in a relatively slow-moving medium;
	however, the effective optical-depth becomes more critical in a fast-moving medium.

For example, if the medium velocity is largely decelerated at large radii,
	putting more matter at larger radii (i.e., larger \rscale) 
	will result in a larger effective optical-depth.
Therefore, as seen for surface brightness profiles in Figure \ref{fig-rscale},
	more scatterings tend to happen at larger radii, thus by medium at smaller velocities.
However, this does not make spectrum necessarily broader and its peak farther away from the central wavelength.
The two competing effects (the Doppler frequency shift and the effective optical-depth)
	seem to be canceled out in the case of MUSE \#82,
	while the latter effect (i.e., enhanced effective optical-depth by increasing \rscale)
	overwhelms the former effect (i.e., smaller frequency shift due to smaller medium velocity)
	in the case of MUSE \#1343.
%\textbf{When the medium velocity is smaller,
%	the frequency is shifted less in the medium frame,
%	and so the final frequency after scatterings is
%	despite increased effective optical depth
%	(we demonstrate this in Appendix \ref{sec-peak-Vmax}).}
%These two effects seem to be canceled out in the case of MUSE \#82,
%	while the former effect (i.e., enhanced effective optical depth by increasing \rscale)
%	overwhelms the latter effect (i.e., \textbf{smaller frequency shift due to smaller medium velocity})
%	in the case of MUSE \# 1343.
The impact of \rscalee on spectrum could be more significant
	when there is non-negligible amount of \Lyaa photons emitted at large radii (i.e., larger \rscont).
Indeed, all of our target galaxies are modeled to have a very compact \Lyaa source distribution
	as inferred from their UV distribution (see Table \ref{tab-MUSEinfo}),
	which could limit the impact of \rscalee on spectrum.

As \rpeakk decreases, 
	spectrum becomes broader with higher intensities at the central wavelength and at the red wing,
	and surface brightness profile tends to be steeper at inner radii (Figure \ref{fig-rpeak}).
Increasing \Vpeakk has an effect similar to but stronger than that of decreasing \rpeak;
	spectrum shows a much clearer peak shift toward the central wavelength 
	and surface brightness profile becomes steeper over the entire range (Figure \ref{fig-Vpeak}).
Here, we need to note the \Vpeakk in Figure \ref{fig-Vpeak}
	is similar to the high velocities considered in Appendix \ref{sec-peak-Vmax},
	for which no considerable scatterings happen 
	and thus the Doppler frequency shift effect is insignificant.
As \delVV decreases (i.e., the edge velocity decreases),
	surface brightness profile becomes flatter,
	and spectrum has lower intensity at the central wavelength
	with its peak shift toward the red side (Figure \ref{fig-delV}).
When the edge velocity becomes zero, 
	the emergent spectrum shows a small blue bump.
The spectrum becomes slightly sharper and its peak comes closer to the central wavelength as DGR increases;
	the impact of DGR appears more apparent when optical depth is larger,
	as shown in the bottom row of Figure \ref{fig-DGR}.
The impact of DGR on surface brightness profile is in general minor.
Please refer to Appendix \ref{sec-parvar-rest}
	for the interpretations on these behaviors.

Multiple parameters show impacts on spectrum or surface brightness profile.
Therefore, a degeneracy between parameters is naturally expected,
	and we mentioned some of them in the 2D posterior distributions of Section \ref{ssec-synergy}.
We now can understand these degeneracies in a qualitative way
	from the spectral and surface brightness profile variations in the parameter space examined above.
The most obvious degeneracy is found to be the one for $\tau_0$ and $z$ in spectrum.
It is because their impacts on spectrum are strong enough 
	to manifest themselves consistently for any combinations of other parameters.
A redshift $z$ stretches a wavelength range by a factor of $(1+z)$.
Thus, both $\tau_0$ and $z$ cause a larger shift of the peak 
	and a broader width when their value increases.
This leads to the negative degeneracy between $\tau_0$ and $z$ as noted in Section \ref{ssec-synergy}.
Meanwhile, \rpeakk and \Vpeakk also have impacts on the spectral shape,
	but their impacts are not as significant as those of $\tau_0$ and $z$.
Moreover, the changes of the peak shift and the width with \rpeakk or \Vpeakk 
	do not appear consistently for all cases;
	sometimes a larger shift comes with a broader width and sometimes not.
Therefore, a degeneracy involved with \rpeakk or \Vpeakk could be either strong or weak.
The parameter degeneracy in the surface brightness profile
	is weak as in Figure \ref{fig-lnliksbp-1185}.
This is not because parameters play unique roles in shaping surface brightness profile,
	but because their roles are too simple.
Although most parameters have a clear impact on the surface brightness profile,
	the variation of surface brightness profile made by each parameter 
	is mostly about the change in the steepness.
Such a simple variation is not enough to discriminate the effects of different parameters,
	which leads to poor parameter constraints from surface brightness profile.
Still, we can better constrain parameters using both spectrum and surface brightness profile
	as seen in the posterior distributions of $\mathcal{L}_\textrm{\cs tot}$
	sharper than those of $\mathcal{L}_\textrm{\cs sp}$ or $\mathcal{L}_\textrm{\tiny SBP}$
	(e.g., Figure \ref{fig-lnliktot-1185} versus Figures \ref{fig-lnliksp-1185} or \ref{fig-lnliksbp-1185}).
This can be understood from the fact that 
	the variations of spectrum and surface brightness profile with parameters are quite diverse.
Some parameters change only one of the two, while others change both (e.g., \rscalee versus \rpeak);
	some parameters change the spectrum in a way similar to other parameters,
	but not the surface brightness profile (e.g., $\tau_0$ versus $z$).
Such different behaviors make total posterior distributions tighter and break parameter degeneracy.
%}}}

\subsection{Spatial Extent of \Lyaa Halos and Its Dependence on Model Parameters}\label{ssec-lyahalo}
%{{{
One notable feature of \Lyaa halos around high-$z$ star-forming galaxies is
	their size that is typically much larger than the size observed in the UV continuum.
The ratio of the scale radii of the brightness distributions in \Lyaa and UV continuum
	(\rshalorscont) of our target galaxies covers a wide range from 3.5 to 21, 
	with a mean value of 10.2.
Indeed, our target galaxies have a size much larger in \Lyaa than in UV continuum.
We perform a simple analysis to examine correlations 
	between each model parameter and \rshalorscontt
	for a hint on which physical parameter is responsible for
	the property of the extended \Lyaa halos.

%%%%%%%%%%%%%%%%%%%%%
% Figure
\begin{figure*}
	\center
	\includegraphics[width=0.85\textwidth]{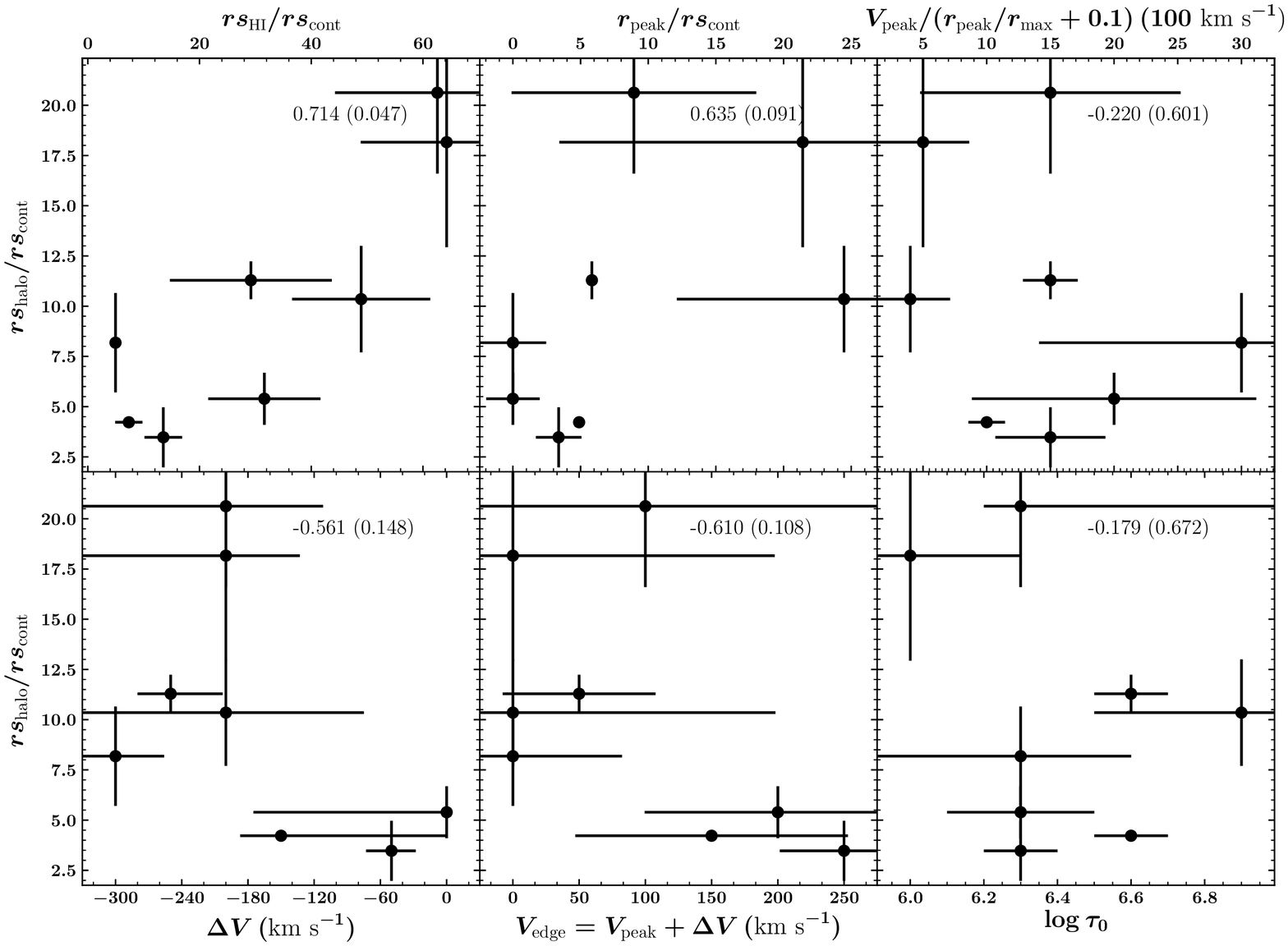}
	\caption{
		Correlation between \rshalorscontt and each of
			six parameter values of \rscale, \rpeak, $V_\textrm{\cs peak}/r_\textrm{\cs peak}$,
			 \delV, $V_\textrm{\cs edge}\,(=V_\textrm{\cs peak}+\Delta V)$, and $\tau_0$
			of our eight target galaxies.
		The Spearman rank-order correlation coefficient and $p$-value of each correlation 
			are given in the top right of each panel
			(one in the parenthesis is $p$-value).
		In the top left panel for \rscont, all galaxies in L17 are also presented 
			with open gray circles and their Spearman rank-order correlation coefficient and $p$-value
			are also written in gray.
	}\label{fig-parcorr-rshalo}
\end{figure*}
%%%%%%%%%%%%%%%%%%%%%

Figure \ref{fig-parcorr-rshalo} shows the correlations between
	\rshalorscontt and each of six parameters that are 
	\rscalerscont, \rpeak$/$\rscont, \Vpeak$/$\rpeak, \delV, $V_\textrm{\cs edge}$, and $\tau_0$
	for our target galaxies (see filled circles with error bars in each panel).
We exclude DGR and $z$ because they are certainly not responsible for the extent of \Lyaa emission.
Errors are calculated from the likelihood distributions of model parameters,
	as summarized in Table \ref{tab-bestfit}.
Although the errors are asymmetric, we assume symmetric errors 
	by taking the mean of lower and upper errors
	when we need to apply the error propagation for the final error estimation
	(except \delVV and $\tau_0$).
In any case, the error estimation is rough, and thus it should not be taken at face value.
Among the model parameters, 
	\rscalerscontt and \rpeakrscontt show non-negligible correlations
	with \rshalorscontt ($p$-value smaller than 0.1),
	and their Spearman rank-order correlation coefficients are 0.74 and 0.64, respectively.
These correlations suggest that
	the large extent of \Lyaa emission results from
	wide distribution of neutral hydrogen medium  
	and from slow increase of outflowing velocity in the inner region.
Another correlation that might be worth paying attention to is the one 
	between \rshalorscontt and $V_\textrm{\cs edge}$
	(Spearman rank-order correlation coefficient is $-0.61$ and $p$-value is 0.11),
	which indicates that \Lyaa emission is more extended
	when the velocity of outer medium is smaller
	(i.e., the velocity decrease in the outer region is larger).
This is because more \Lyaa photons, especially those that freely escape from the inner region,
	can be scattered and escape at large radii, which results in more \Lyaa emission at large radii
	and thus a larger extent of \Lyaa halos.

To summarize,
	large spatial extents of \Lyaa halos seem relevant to
	large spatial extents of UV emission (a result of L17) and neutral hydrogen medium,
	and small bulk velocities of the medium at small and large radii.
This needs to be further examined using a larger galaxy sample in future studies.
%}}}

\subsection{Correlation between Peak Shift and FWHM of \Lyaa Spectrum}\label{ssec-peakshift-fwhm}
%{{{
The correlation between peak shift and FWHM of \Lyaa spectrum
	is naturally expected from the resonant scattering process of \Lyaa photons. 
This correlation has been recognized as a way to derive a correct systemic redshift of a galaxy
	when only the \Lyaa emission line is available \citep[e.g.][]{Zheng_Wallace_2014}. 
Recently, \citet{Verhamme_etal_2018} found an empirical relation between red peak shift and FWHM
	using 45 galaxies for which lines other than \Lyaa line can be used for the systemic redshift estimate.
They showed that this relation recovers the systemic redshift estimate 
	with an accuracy of $\leq100\,\textrm{km s}^{-1}$.

\begin{figure*}
	\center
	\includegraphics[width=0.6\textwidth]{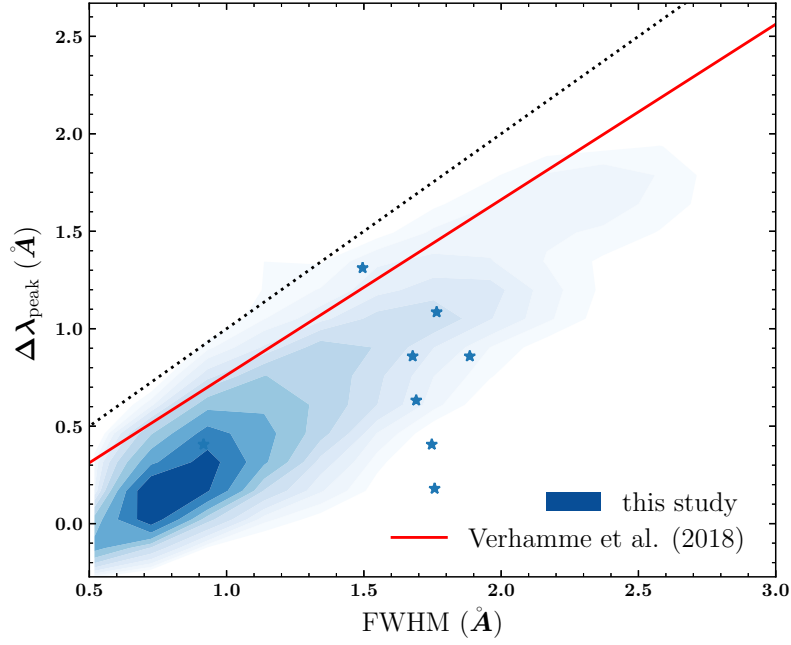}
	\caption{
			Distribution of peak shift and FWHM of simulated \Lyaa spectra
				of our 13,230 simulation runs (blue contours).
			Contour levels are in log scale.
			Blue stars are for the best-fit spectra of the MUSE galaxies examined in this study.
			Red solid line represents the empirical relation 
				found by \citet{Verhamme_etal_2018} in observation,
				and black dotted line represents $y=x$ line.
	}\label{fig-peakshift-FWHM}
\end{figure*}

We also confirm this correlation using all of our 13,230 simulations
	(blue density contours in Figure \ref{fig-peakshift-FWHM}).
However, the distribution of peak shifts and FWHMs of our simulated spectra 
	has an offset from the empirical relation found by \citet[][red line]{Verhamme_etal_2018}.
%Such a discrepancy could be mainly due to 
%	the difference in the models used to analyze the observational data.
%For example, we use all parameter combinations,
%	and obviously not all of them are realistic.
\citet{Verhamme_etal_2018} compared in their Figure 4 
	the correlation from observations with that from models 
	such as expanding shells, spheres or bi-conical outflows \citep{Schaerer_etal_2011,Zheng_Wallace_2014};
	there also exists an offset between the two,
	which is however in the opposite direction of ours.
This implies that the correlation between peak shift and FWHM 
	can be a model discriminator.
%%%%%%%%%%%%%%%%%%%%%
% Figure
%%%%%%%%%%%%%%%%%%%%%

%%%%%%%%%%%%%%%%%%%%%
% Figure
\begin{figure*}
	\center
	\includegraphics[width=\textwidth]{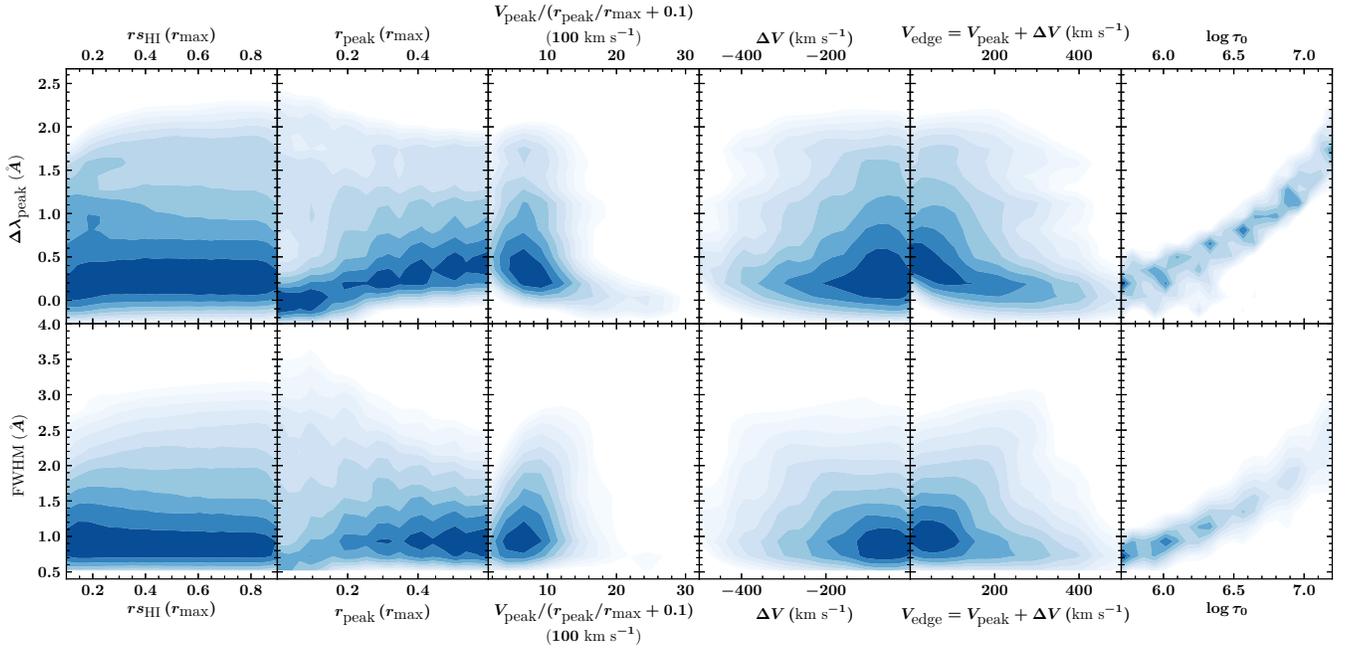}
	\caption{
		Correlation between peak shift ($\Delta\lambda_\textrm{peak}$) and other parameters (upper row),
			and FWHM and other parameters (lower row).
	}\label{fig-parcorr-peakshift-FWHM}
\end{figure*}
%%%%%%%%%%%%%%%%%%%%%
We explore correlations between model parameters 
	and peak shift (top) or FWHM (bottom) in Figure \ref{fig-parcorr-peakshift-FWHM}.
Both peak shift and FWHM show strong correlations with optical depth (rightmost panel in each row),
	which is the main driver of the correlation between peak shift and FWHM in Figure \ref{fig-peakshift-FWHM}.
While their correlations with other parameters are relatively weak,
	the correlation with velocity profile parameters appears interesting (middle panels);
	for example, a peak shift is negatively correlated with medium velocity, and
	FWHM, on the contrary, is positively correlated.
The negative correlation between peak shift and medium velocity is because 
	\Lyaa photons tend to experience less scatterings when medium velocity is larger
	(the medium velocity in our model is relatively large, and thus
	the effect of effective optical-depth is larger than that of the Doppler frequency shift).
This may also cause a negative correlation between FWHM and medium velocity,
	but this does not seem to be the case.
It is because our velocity profile spans a range of velocity 
	(unlike a constantly-expanding velocity profile)
	that becomes broader as peak velocity increases.
Therefore, when peak velocity is larger,
	\Lyaa photons are Doppler-shifted 
	to a larger range of wavelength in the rest frame of medium,
	which ends up with a larger FWHM.
This again indicates that the peak shift--FWHM correlation could play a role in 
	discriminating models, especially regarding velocity profile.
%}}}

\subsection{Spatially Resolved \Lyaa Spectra}\label{ssec-SpResolvedSp}
\begin{figure*}
	\center
	\includegraphics[width=\textwidth]{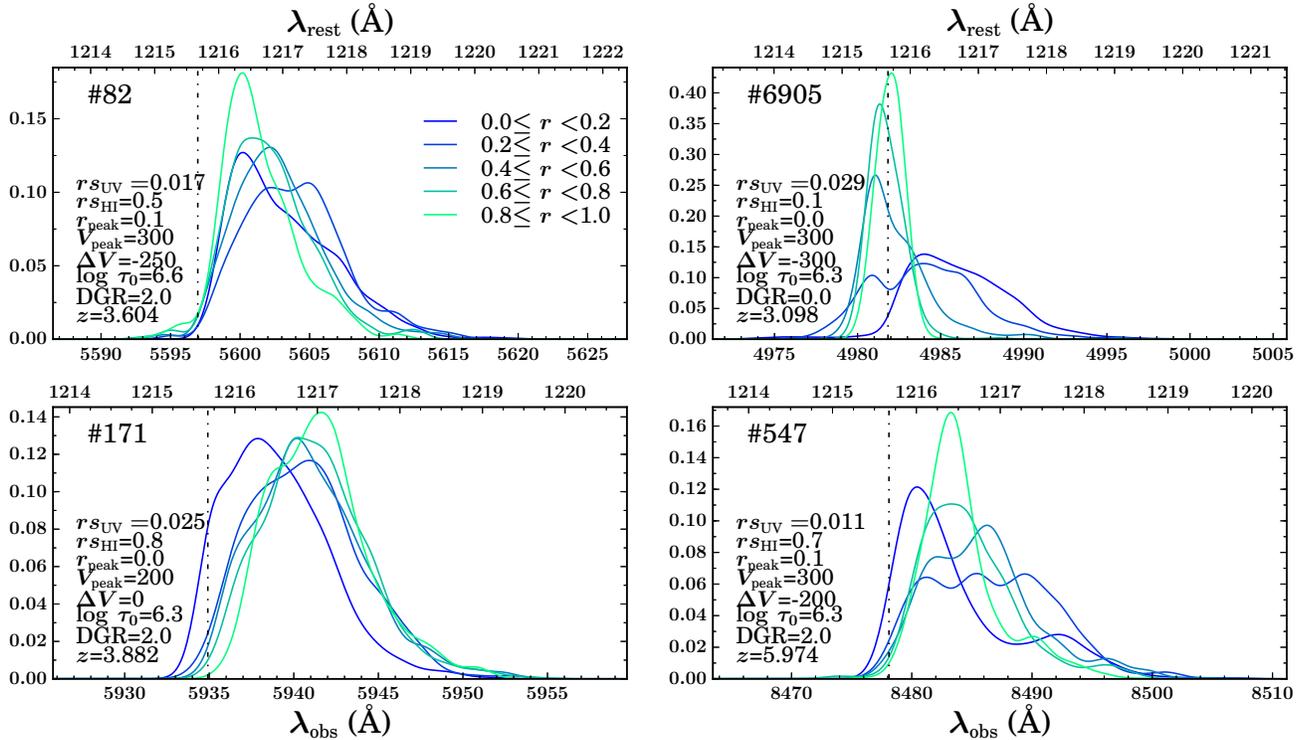}
	\caption{
		Simulated spectra at 5 annuli of different radii from the galaxy center
			for four target galaxies with their best-fit parameter set.
		Each MUSE id is written in top left of each panel
			and dot-dashed line represents the \Lyaa central wavelength.
	}\label{fig-resolvedSp}
\end{figure*}

%{{{
One advantage of using the MUSE observational data is that
	we can obtain spatially resolved spectra.
Unfortunately only (spatially) integrated spectra are publicly available for now,
	but it is still worth examining simulated spectra
	as a function of (projected) distance from the galaxy center. %for future studies.
Figure \ref{fig-resolvedSp} shows the simulated \Lyaa spectra 
	measured at annuli of different radii
	with the best-fit parameter set for four target galaxies.
Each spectrum is normalized by the total number of \Lyaa photons at each annulus.
Because the true intensity level of spectra is imprinted in the surface brightness profile,
	here we examine only their shape.
The results show that the shape of these spectra varies significantly with the radius,
	indicating that such spatially resolved spectra can be additionally used to infer
	spatial and kinematic distributions of medium in more detail.
%%%%%%%%%%%%%%%%%%%%%
% Figure
%%%%%%%%%%%%%%%%%%%%%

In the case of MUSE \#171, 
	which is a relatively simple case with a constant expanding velocity,
	the different spectra at different locations
	are because of difference in the number of scatterings
	that photons have undergone until their escape.
The peak of the spectrum tends to appear at a longer wavelength
	when the spectrum is measured at a larger distance from the galaxy center.
Therefore, a larger shift of the spectral peak at a larger distance
	indicates that these photons escape after more scatterings.
This makes sense because these photons come across longer distances from the central part,
	being scattered more to reach that the larger radii
	and then being able to escape there.\footnote{
		Although we assume a spatially extended \Lyaa source distribution,
			emission is mostly from the central region
			(i.e., the scale radius for UV continuum emission of our target galaxies is
			smaller than 5\% of the maximum extent of the system).
		Therefore, on-site emission at large radii accounts for only a small fraction,
			and the majority are those scattered from the central region,
			which had been scattered many times to reach large radii.}

The frequency shift (with respect to the initial frequency) of escaping photons
	could be larger not only when optical depth is larger
	but also when the medium moves faster,
	if the medium velocity is not too high (see Appendix \ref{sec-peak-Vmax}).
For other target galaxies,
	the peak shift of spectrum tends to become larger and then smaller as the radius increases.
Such a behavior of the peak shift could be attributed to 
	the velocity profile that increases and then decreases
	as the distance from the galaxy center increases.
However, it should be noted that
	if the medium velocity is too large, photons can escape the system without many scatterings,
	so the behavior of the peak shift as a function of radius
	could not be easily understood. 
It should be also noted that 
	the annuli where we obtain the spectra are in the projected space.
Therefore, the spectrum at a small radius
	consists of photons that are emerging at a large range of three-dimensional distance,
	which possibly makes multiple peaks in the spectrum
	as seen in most of our target galaxies.
Such a feature will be useful for extracting kinematic information of the medium, 
	especially its spatial variation, from observational data.

Although the parameters of the medium velocity field
	help us to understand the overall trend of spectra from different annuli, 
	we need to consider other factors as well,
	especially \rscalee and $\tau_0$ to understand the phenomena in more detail.
For example, 
	the combination of small \rscalee and small $\tau_0$ for the case of MUSE \#6905
	leads to so few scatterings at large radii
	that the spectrum at the outermost annuli shows almost no shift
	even though the medium is completely static at the edge.
The peak of spectra from the outer part of the galaxy
	shows a slight shift to the blue side
	because the expanding velocity profile of medium always decreases (i.e., $r_\textrm{\cs peak}=0$).

In summary, \Lyaa spectra from different positions in and around a galaxy are quite distinguishable,
	of which degree is determined by the interplay of multiple parameters
	regarding the spatial and kinematic distributions of medium.
Therefore, spatially resolved \Lyaa spectra could give 
	much tighter constraints on the parameters than an integrated spectrum.
We will continue our work in this direction.
%}}}

\subsection{Impacts of \Lyaa Input Spectrum on Emerging Spectrum and Surface Brightness Profile}\label{ssec-intsp}
%{{{
To perform \Lyaa Monte Carlo radiative transfer calculations,
	we set the input \Lyaa spectrum as a Voigt profile
	with a temperature of $10^4\textrm{K}$,
	the typical temperature of \ion{H}{2} regions.
%A Voigt profile is a convolution of a Lorentzian profile and a Gaussian profile,
%	and represents a typical frequency distribution of photons 
%	that are emitted from atoms in thermal motion (no bulk motion).
%Our calculations are based on several assumptions;
%	(1) \ion{H}{2} regions are the only source of \Lyaa photons,
%	(2) \ion{H}{2} regions have a temperature of $10^4\textrm{K}$ and have no turbulent or bulk motion,
%	and (3) the optical depth of \ion{H}{2} regions are low enough
%%	for generated \Lyaa photons to avoid scatterings
%	(so the emerging spectrum from \ion{H}{2} region is not a double-peak profile).
All the results we show in this paper are based on this input spectrum.
In other words, we performed the \Lyaa radiative transfer  calculation
	only in the diffuse interstellar medium and the circumgalactic medium,
	ignoring the radiative transfer in \ion{H}{2} regions.
However, in reality, for instance, 
	some bulk outflowing motion induced by stellar winds in \ion{H}{2} regions
	could give rise to asymmetric double-peak spectra \citep{TKimm_etal_2019}.

There are many possible options for the input spectrum,
	which include a Gaussian profile, a symmetric double-peak profile, 
	and one including continuum with different line strengths (i.e., different EWs).
However, we consider only an asymmetric double-peak profile of which red peak is stronger than the blue one,
	as suggested by \citet{TKimm_etal_2019}.
We approximate the asymmetric double-peak profiles in \citet{TKimm_etal_2019}
	using a sum of two Voigt profiles;
	each of them is shifted by $\Delta x$ toward red and blue, respectively,
	and the red peak is two times stronger than the blue one.
%We impose a larger weight (e.g., by a factor of two) on the red side than on the blue side.
Although our approximation is not precisely mimicking those in \citet{TKimm_etal_2019},
	we do expect to draw a qualitatively consistent conclusion 
	with the case by adopting the exact form of their spectra.

\begin{figure*}
	\center
	\includegraphics[width=\textwidth]{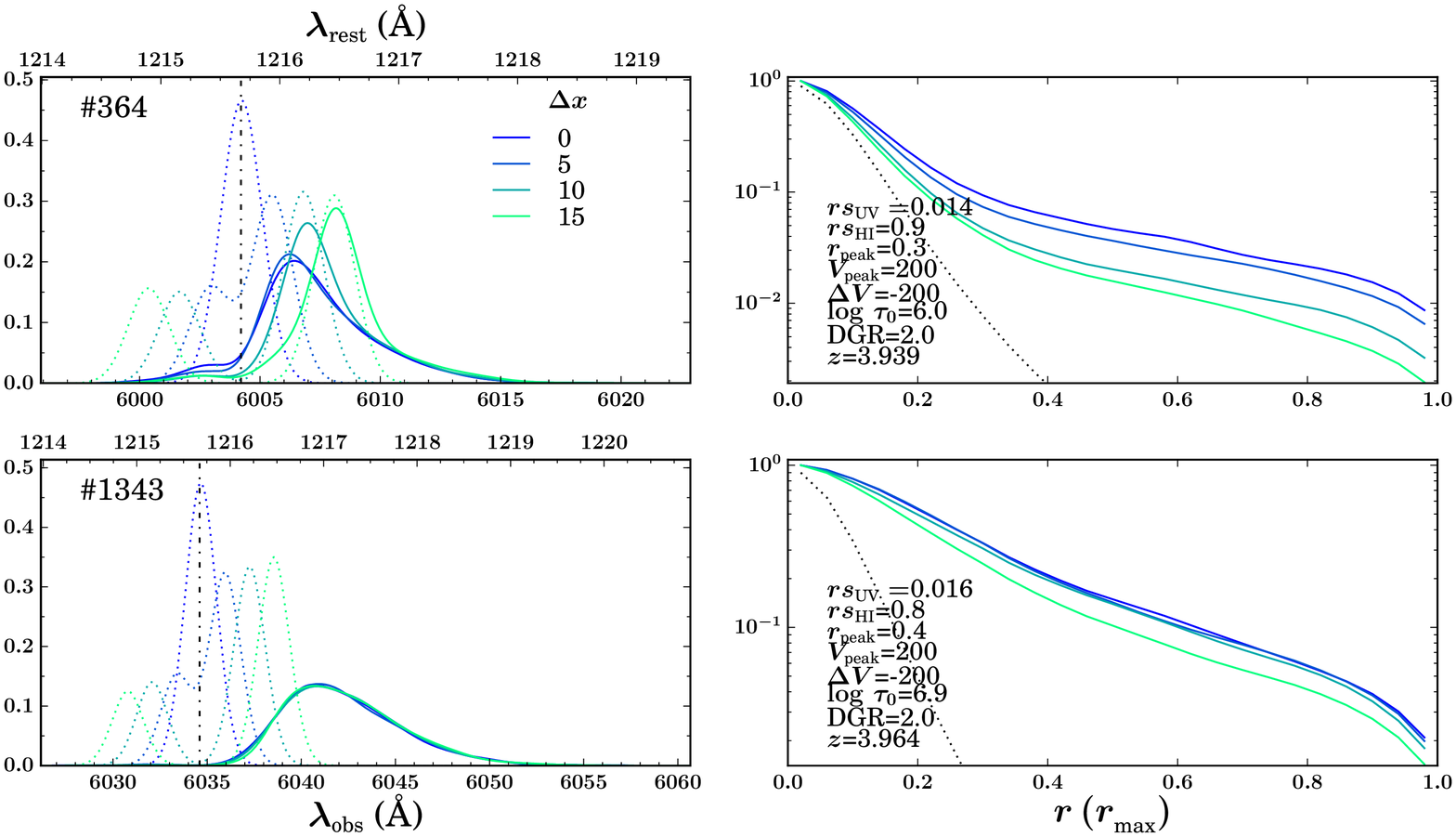}
	\caption{
		Similar to Figure \ref{fig-rscale}, but with varying $\Delta x$ 
			(the offset of each peak from the central wavelength for the input \Lyaa spectrum).
		In the spectrum panels, dotted lines are for different input spectra of different $\Delta x$ values.
	}\label{fig-inSp}
\end{figure*}

We construct \Lyaa spectrum and surface brightness profile 
	using the asymmetric double-peak profile as an input spectrum,
	which is easily done by adjusting the weight of each photon,
	following the procedure described in Section \ref{ssec-model}.
We consider different cases with four peak separations 
	($2\Delta x$ where $\Delta x=0, 5, 10, 15$, and $\Delta x=0$ corresponds to a pure Voigt profile),
	and implement each case by putting $V(x+\Delta x) +0.5V(x-\Delta x)$ 
	for $\mathcal{S}_f$ in Eqn. (\ref{eqn-weight})
	where $V(x)$ represents a Voigt profile. 
Figure \ref{fig-inSp} shows the \Lyaa spectra and surface brightness profiles
	of the four different peak separations for two galaxies with their best-fit parameter set.
As the two peaks of the input spectrum are farther away from the central wavelength 
	(i.e., as $\Delta x$ increases),
	the emerging spectrum becomes sharper and the surface brightness profile becomes steeper.
The photons on the red side in the input spectrum 
	become even redder (i.e., farther away from the central wavelength) 
	in the medium frame because of the expanding velocity;
	they are rarely scattered and emerging at their initial frequencies.
On the other hand, 
	the photons on the blue side become less blue (i.e., closer to the central wavelength);
	they escape after a lot of scatterings, forming a broad component.
In the result,
	the emerging spectrum tends to have a sharper peak with a larger shift
	as the peak separation increases.
In the case of surface brightness profile, 
	as the two peaks of the input spectrum are farther away from the central wavelength,	
	the majority (red photons) are less scattered, which results in a steeper profile.
It should be noted that the details will be determined by effective optical-depth.
When the effective optical-depth is large, the change in the emerging spectrum 
	due to the variation of the input spectrum will be small 
	because a large number of scatterings tend to erase 
	the information of the initial frequency distribution.
	(e.g., the case of MUSE \#1343).
%%%%%%%%%%%%%%%%%%%%%
% Figure
%%%%%%%%%%%%%%%%%%%%%

The best fit for each target galaxy will be quite different 
	from that in Table \ref{tab-bestfit}
	if we adopt such an asymmetric (red-dominant) double-peak spectrum
	as an input spectrum.
We expect that a parameter set that gives a larger effective optical-depth
	(i.e., larger optical depth and smaller expanding velocities)
	will be preferred to remove a sharp peak with a large shift.
It is important to have a better idea on the \Lyaa input spectrum
	for more accurate modeling in small spatial scales.
One more thing we should notice is that previous studies varied 
	the width of the input \Lyaa spectrum, incorporating turbulent motion.
However, we could successfully model the data without considering this effect.
%}}}

%%%%%%%%%%%%%%%%
\section{SUMMARY}\label{sec-summary}
%{{{
We perform \Lyaa radiative transfer calculations with an outflowing halo model
	that is improved from a simple, constantly-expanding shell model\footnote{
		The numerical data of the present models will be made available at 
		\href{https://data.kasi.re.kr/vo/LaRT\_models/}{https://data.kasi.re.kr/vo/LaRT\_models/}.}.
We reproduce successfully \Lyaa properties (i.e., spectrum and surface brightness profile) 
	of eight star-forming galaxies at $z=$3--6 observed with MUSE,
	of which results are consistent with the results from other studies.
We summarize our results as follows.

\begin{itemize}
\item
The constraints on the model parameters are largely improved
	by using spectrum and surface brightness profile simultaneously.
This is because
	spectrum and surface brightness profile change diversely and independently
	with each model parameter,
	which helps breaking degeneracies between the model parameters.

\item 
We examine spatially resolved \Lyaa spectra emerging at different distances from the galactic center.
The spectral shape (e.g., the position and the number of spectral peaks and width) changes with distance.
The changes can be understood with a given model parameter set,
	which provides a basis of using spatially resolved spectra
	to better infer density and velocity fields of circumgalactic medium from observations.

\item
The individual model parameters change \Lyaa spectrum and surface brightness profile in a various way.
The changes in \Lyaa spectrum and surface brightness profile by each model parameter
	could be significant or not depending on other parameter values.
However, the impacts of optical depth and redshift on \Lyaa spectrum and surface brightness profile
	are always apparent.

\item
There is degeneracy between model parameters,
	which can be understood from the changes of \Lyaa spectrum and surface brightness profile with each model parameter.
The most prominent ones are 
	optical depth--the radius of peak expanding velocity 
	and optical depth--redshift in determining the spectrum.
The parameter degeneracy in determining the surface brightness profile
	appears weak in general, which is mainly because of the poor parameter constraints of surface brightness profile.
	
\item 
Dust does have impacts on the shape of \Lyaa spectrum
	due to differential intensity reductions depending on wavelengths.
This is because \Lyaa spectrum consists of \Lyaa photons
	escaping after different numbers of scatterings 
	that systematically change with emerging wavelengths.
However, such impacts of dust become significant 
	when the effective optical-depth is large.

\item
We examine correlations between model parameters 
	and the observed properties of high-redshift \Lyaa galaxies.
The spatial extent of \Lyaa halos shows strong correlations
	with the spatial extents of UV emission and neutral hydrogen medium,
	and bulk velocities of medium.
The positive correlation between peak shift and FWHM of \Lyaa spectrum
	is well reproduced by our simulations.
The peak shift--FWHM correlation could provide an additional constraint 
	on the velocity profile of medium.

\item
\Lyaa input spectrum is an important factor
	that determines the shapes of emerging spectrum and surface brightness profile
	(in particular for \ion{H}{2} regions with relatively low optical depth).
We test the case of a red-dominant double-peak input spectrum for various peak separations,
	which results in systematic changes in the emerging spectrum and the surface brightness profile.
This suggests that model parameter constraints 
	could vary largely depending on the choice of \Lyaa input spectrum.
\end{itemize}

This study is the first attempt 
	to model the observed \Lyaa spectrum and surface brightness profile simultaneously.	
We will extend our analyses to a larger sample
	to characterize physical parameters of high-redshift \Lya-emitting galaxies
	with better statistics.
%}}}

\acknowledgments
We thank to the referee for his/her careful reading and the constructive comments.
This work was supported by the National Research Foundation of Korea (NRF) grant
	funded by the Korean government (MSIP) (No. 2017R1A2B4008291).
This work was also supported by the National Institute of supercomputing 
	and Network/Korea Institute of Science and Technology Information
	with supercomputing resources including technical supporting (KSC-2018-S1-0005).
HSong would like to thank Minsu Shin, Yujin Yang, Max Gronke, and Clotilde Laigle
	for useful discussions and advices.

\bibliography{ms}{}

\appendix
\section{Validation of Post-processing Dust Effect}\label{sec-validate-pp}
\begin{figure*}
	\center
	\includegraphics[width=0.60\textwidth]{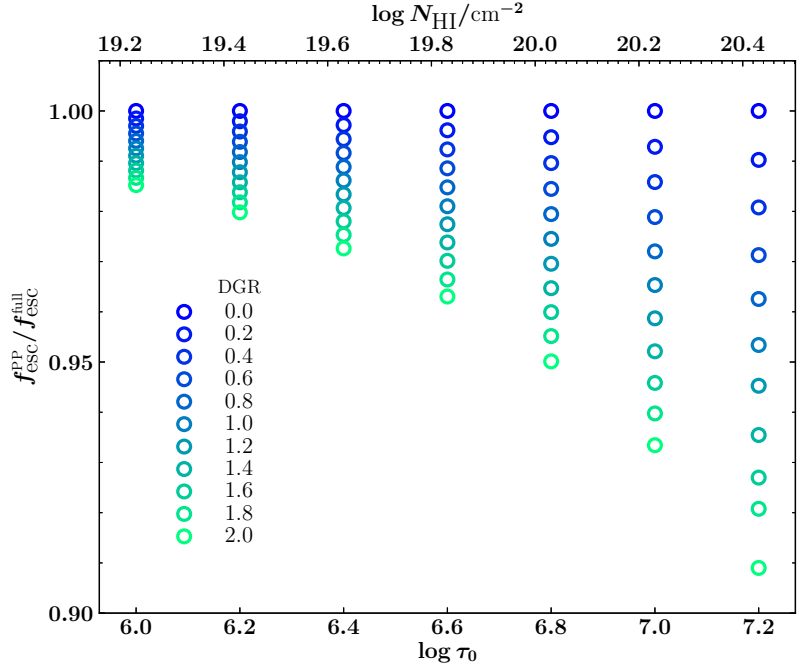}
	\caption{
		The ratio of two estimates on the escape fraction of \Lyaa photons
			that are emitted from a point, monochromatic source
			and traveling through an infinity slab of temperature $10^4\textrm{K}$
			as functions of $\tau_0$ (x-axis) 
			and DGR (denoted by colors):
			$f_\textrm{esc}^\textrm{full}$ is obtained when the effect of dust is fully simulated
			and $f_\textrm{esc}^\textrm{\tiny PP}$ is when the effect of dust is post-processed.
	}\label{fig-fe-full-pp}
\end{figure*}

%{{{
We assume that dust scattering is perfectly forward-directed
	for the post-processing of the dust effect.
We examine whether this assumption is valid or not
	by comparing the escape fractions of \Lyaa photons
	when the effect of dust is fully taken into account in simulation from the beginning ($f_\textrm{esc}^\textrm{full}$)
	and when it is implemented through post-processing ($f_\textrm{esc}^\textrm{\tiny PP}$).
The \Lyaa radiative transfer simulation is performed in 
	an infinite, static slab of medium at a temperature $10^4\textrm{K}$
	for the ranges of $\tau_0$ and DGR of interest of this study.
We consider a point, monochromatic source of \Lya.
Figure \ref{fig-fe-full-pp} shows the ratio of $f_\textrm{esc}^\textrm{\tiny PP}/f_\textrm{esc}^\textrm{full}$
	as functions of $\tau_0$ and DGR.
The post-processing approach tends to slightly underestimate the escape fraction,
	and the degree of underestimate increases with $\tau_0$ and DGR.
However, the amount of underestimation is less than $\sim$9\% 
	even for the largest $\tau_0$ and DGR (i.e., $\log \tau_0=7.2$ and DGR$=2\textrm{DGR}_\textrm{\tiny MW}$),
	and these two treatments of dust show a good agreement in general.
Therefore, the post-processing approach can be claimed 
	as a good approximation to the full simulation approach.
This is expected because
	the angular redistribution function for dust scattering in simulation
	is set to be the Henyey-Greenstein phase function, which is strongly forward-throwing.
In summary, the assumption of perfect forward scattering by dust
	is valid at least for the ranges of $\tau_0$ and DGR considered here, 
	and so the application of post-processing the effect of dust is.
%%%%%%%%%%%%%%%%%%%%%
% Figure
%%%%%%%%%%%%%
%}}}

\section{Posterior Distributions of Model Parameters (Continued)}\label{sec-othertargets}
\begin{figure*}
	\center
	\includegraphics[width=0.86\textwidth]{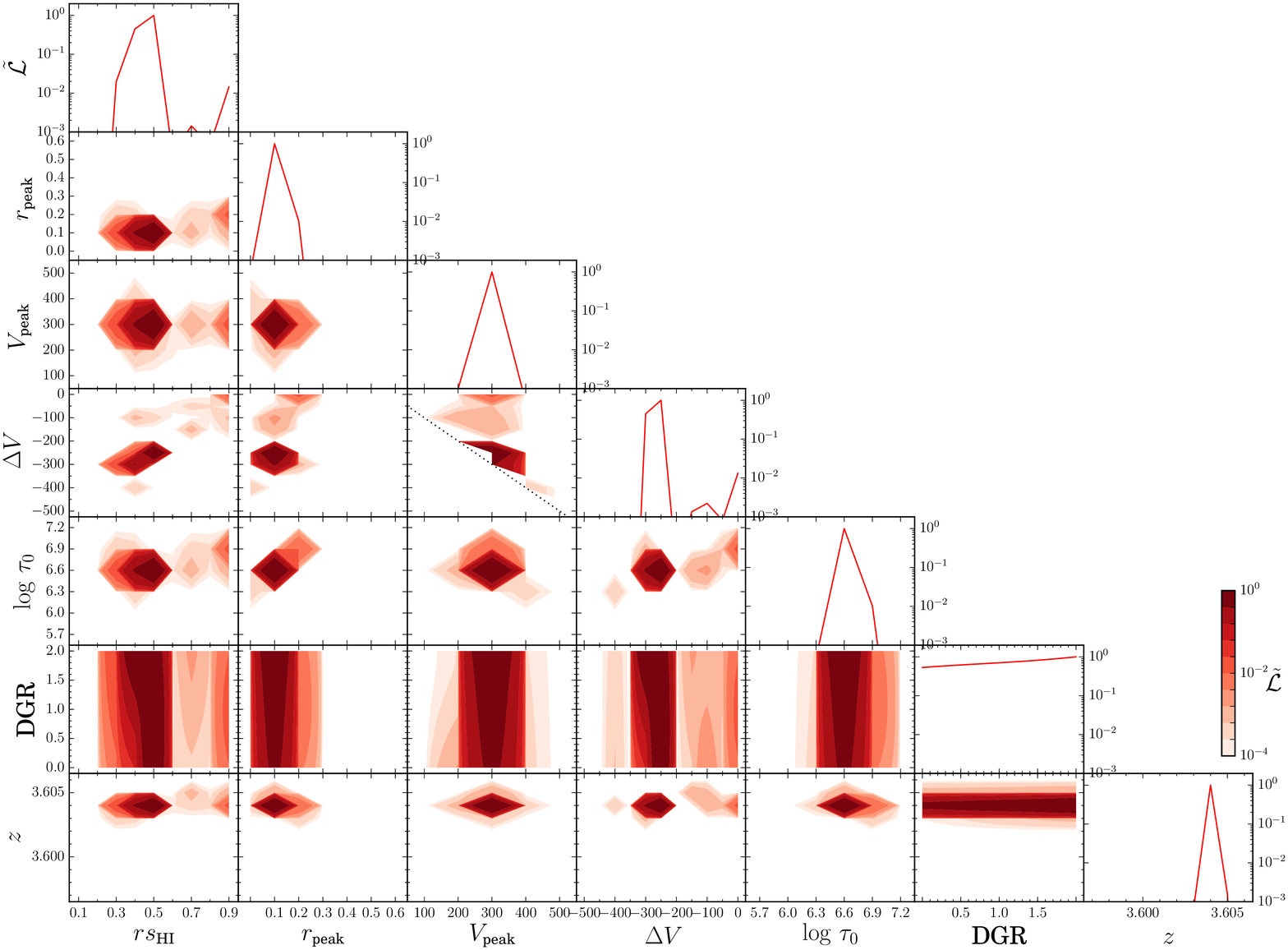}
	\caption{
		Similar to Figure \ref{fig-lnliktot-1185}, 
			but for MUSE \#82.
	}\label{fig-lnliktot-82}
\end{figure*}

\begin{figure*}
	\center
	\includegraphics[width=0.86\textwidth]{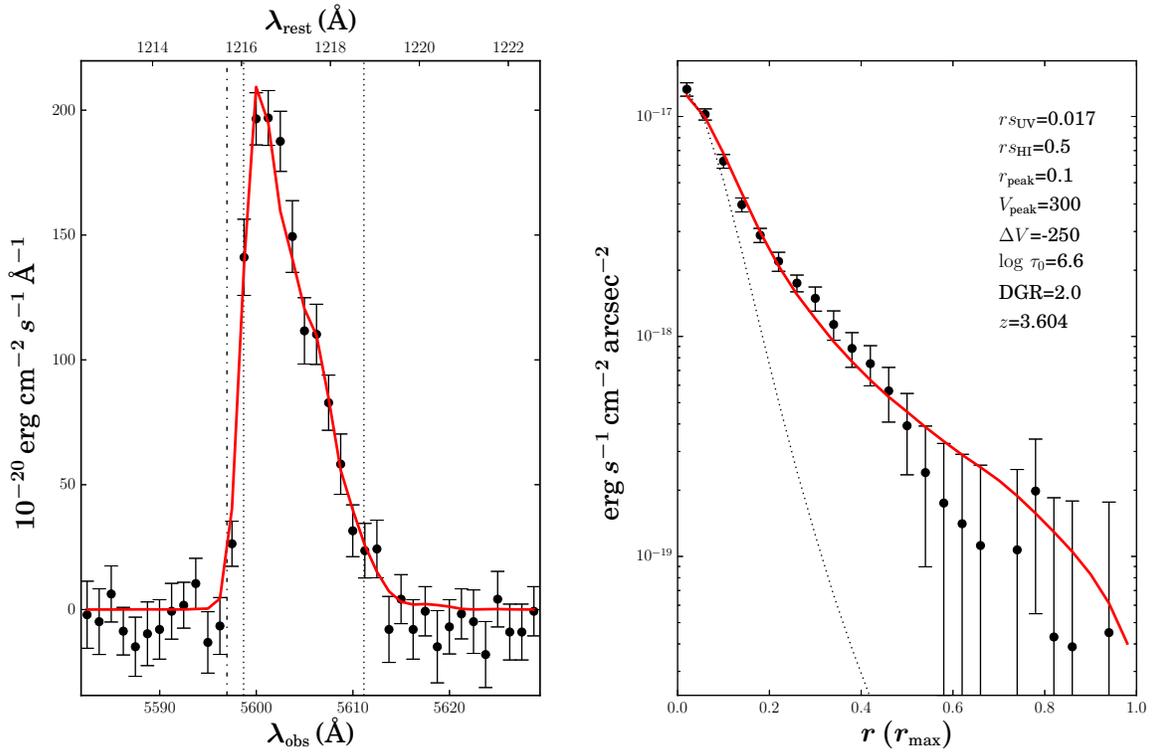}
	\caption{
		Similar to Figure \ref{fig-bestfit-tot-1185}, 
			but for MUSE \#82.
	}\label{fig-bestfit-tot-82}
\end{figure*}

\begin{figure*}
	\center
	\includegraphics[width=0.86\textwidth]{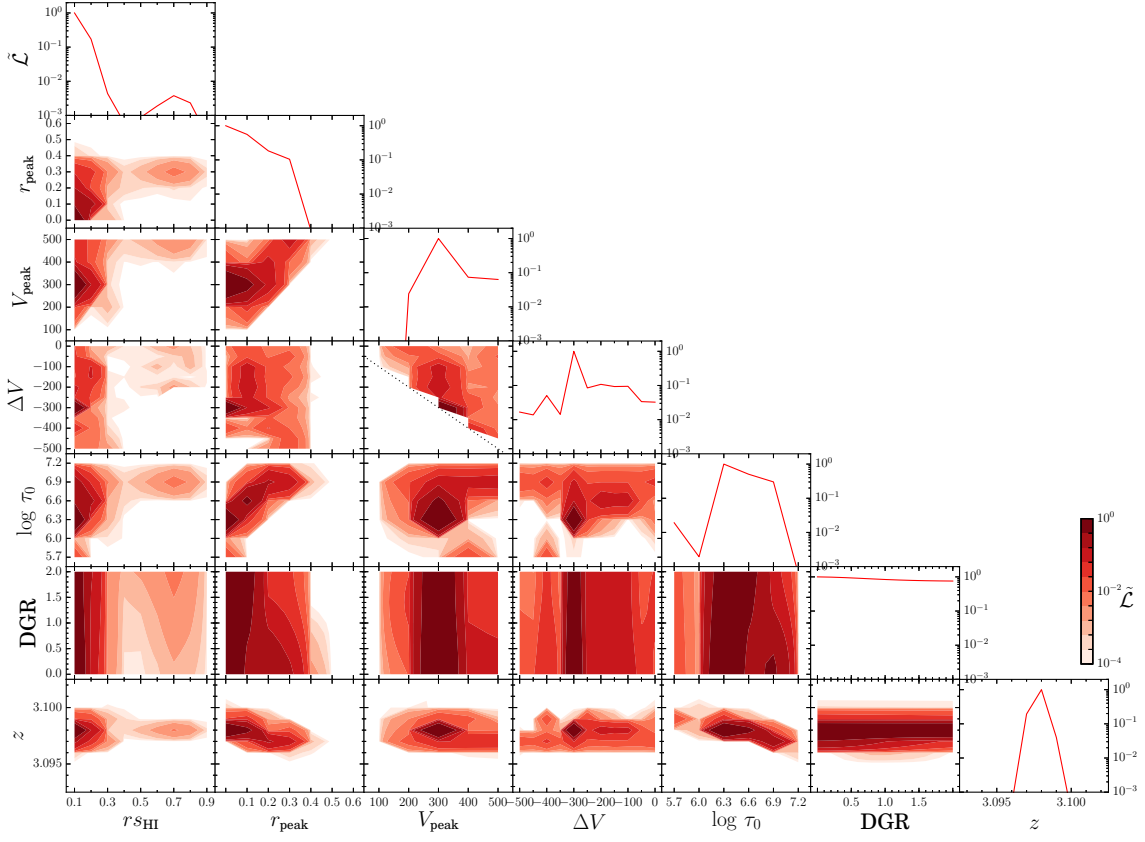}
	\caption{
		Similar to Figure \ref{fig-lnliktot-1185}, 
			but for MUSE \#6905.
	}\label{fig-lnliktot-6905}
\end{figure*}

\begin{figure*}
	\center
	\includegraphics[width=0.86\textwidth]{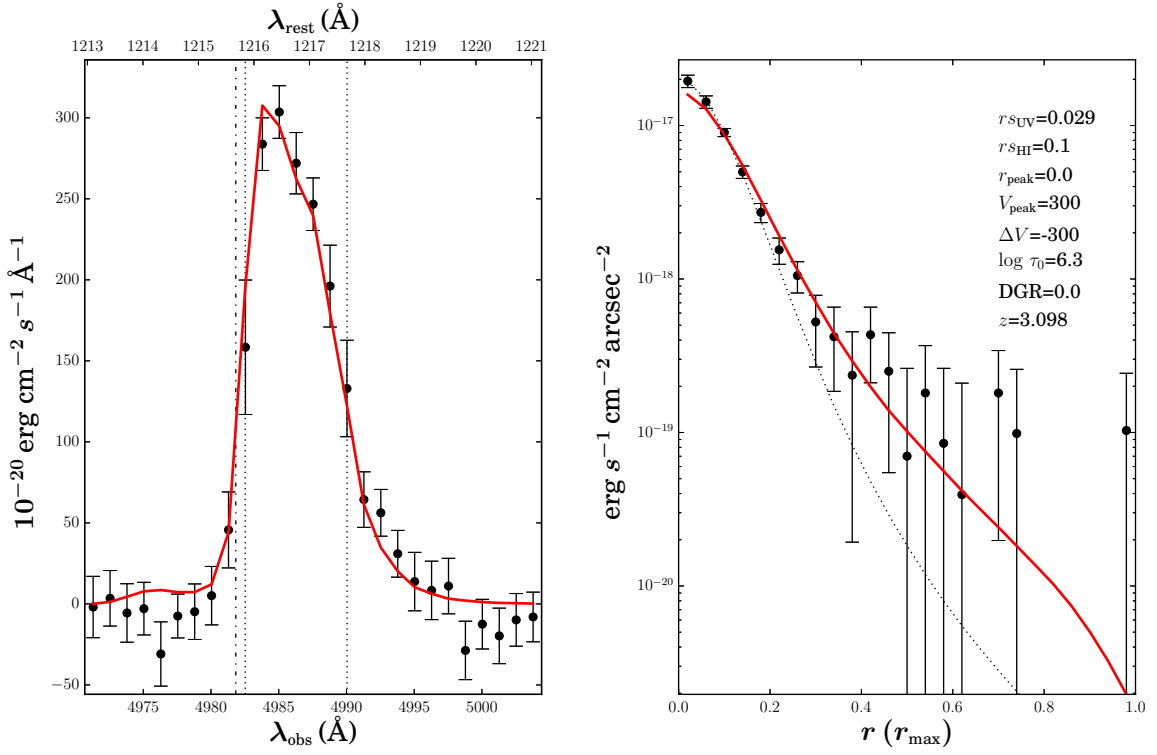}
	\caption{
		Similar to Figure \ref{fig-bestfit-tot-1185}, 
			but for MUSE \#6905.
	}\label{fig-bestfit-tot-6905}
\end{figure*}

\begin{figure*}
	\center
	\includegraphics[width=0.86\textwidth]{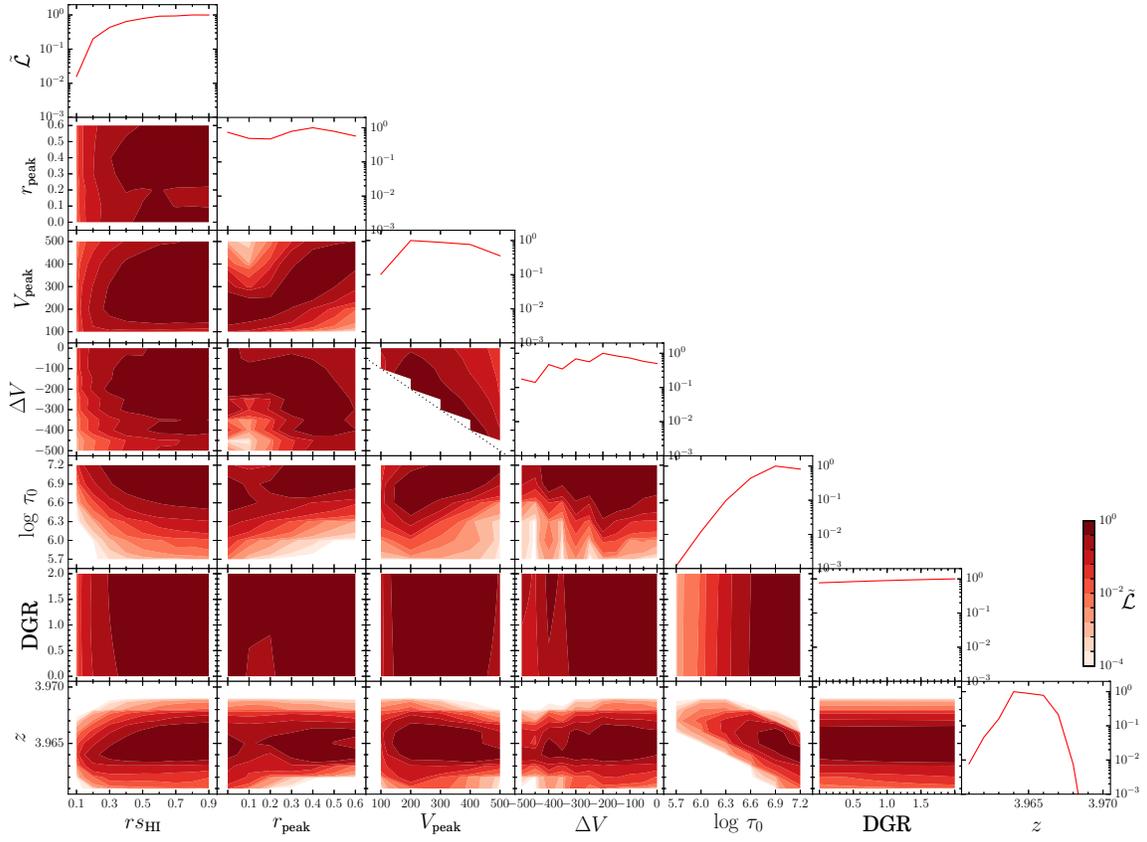}
	\caption{
		Similar to Figure \ref{fig-lnliktot-1185}, 
			but for MUSE \#1343.
	}\label{fig-lnliktot-1343}
\end{figure*}

\begin{figure*}
	\center
	\includegraphics[width=0.86\textwidth]{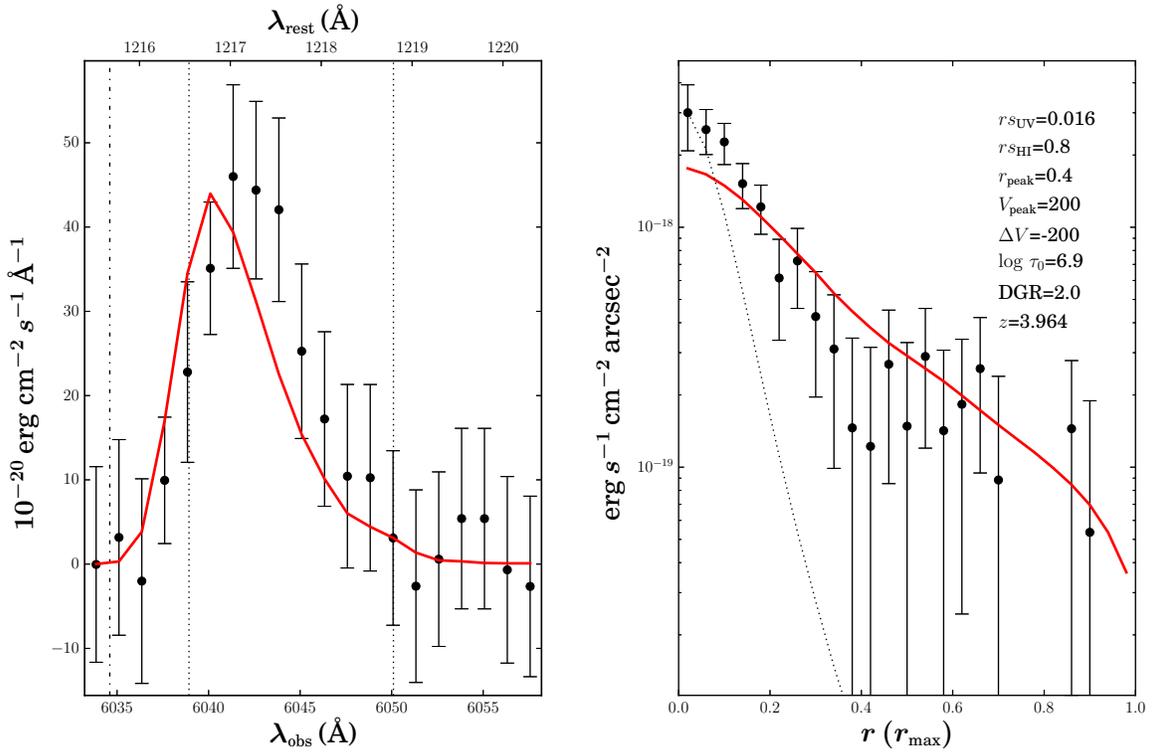}
	\caption{
		Similar to Figure \ref{fig-bestfit-tot-1185}, 
			but for MUSE \#1343.
	}\label{fig-bestfit-tot-1343}
\end{figure*}

\begin{figure*}
	\center
	\includegraphics[width=0.82\textwidth]{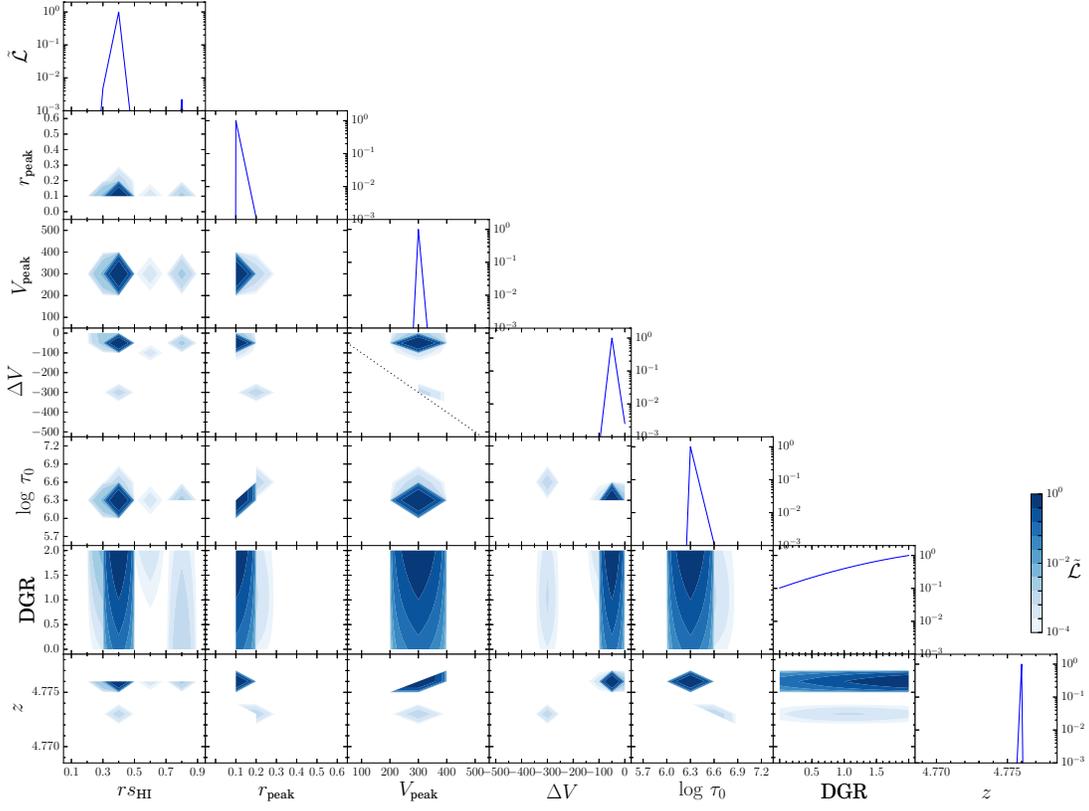}
	\caption{
		Similar to Figure \ref{fig-lnliksp-1185}, 
			but for MUSE \#53.
	}
	\label{fig-lnliksp-53}
\end{figure*}

%{{{
We present only total posterior distributions ($\mathcal{L}_\textrm{\cs tot}$)
	for the rest seven target galaxies 
	(Figures \ref{fig-lnliktot-82}, \ref{fig-lnliktot-6905}, \ref{fig-lnliktot-1343}
	\ref{fig-lnliktot-53}, \ref{fig-lnliktot-171}, \ref{fig-lnliktot-547}, and \ref{fig-lnliktot-364})
 	for the sake of brevity.
However, we present posterior distributions for spectrum ($\mathcal{L}_\textrm{\cs sp}$)
	and surface brightness profile ($\mathcal{L}_\textrm{\tiny SBP}$) for MUSE \#53
	because it is a special case of its best-fit parameter set being determined solely by spectrum.
Nevertheless, parameter constraints become tighter by including surface brightness profile,
	which is revealed by comparing Figures \ref{fig-lnliktot-53} and \ref{fig-lnliksp-53}.
The best-fit model spectrum and surface brightness profile determined with ($\mathcal{L}_\textrm{\cs tot}$)
	for the seven galaxies are also shown in Figures 
	\ref{fig-bestfit-tot-82}, \ref{fig-bestfit-tot-6905}, \ref{fig-bestfit-tot-1343}, \ref{fig-bestfit-tot-53}, 
	\ref{fig-bestfit-tot-171}, \ref{fig-bestfit-tot-547}, and \ref{fig-bestfit-tot-364}. 

%%%%%%%%%%%%%%%%%%%%%
% Figures
%%%%%%%%%%%%%%%%%%%%%
% 82
%%%%%%%%%%%%%
%%%%%%%%%%%%%%%%%%%%%
% 6905
%%%%%%%%%%%%%%
%%%%%%%%%%%%%%%%%%%%%
% 1343
%%%%%%%%%%%%%%
%%%%%%%%%%%%%%%%%%%%%
% 53
%%%%%%%%%%%%%
\begin{figure*}
	\center
	\includegraphics[width=0.82\textwidth]{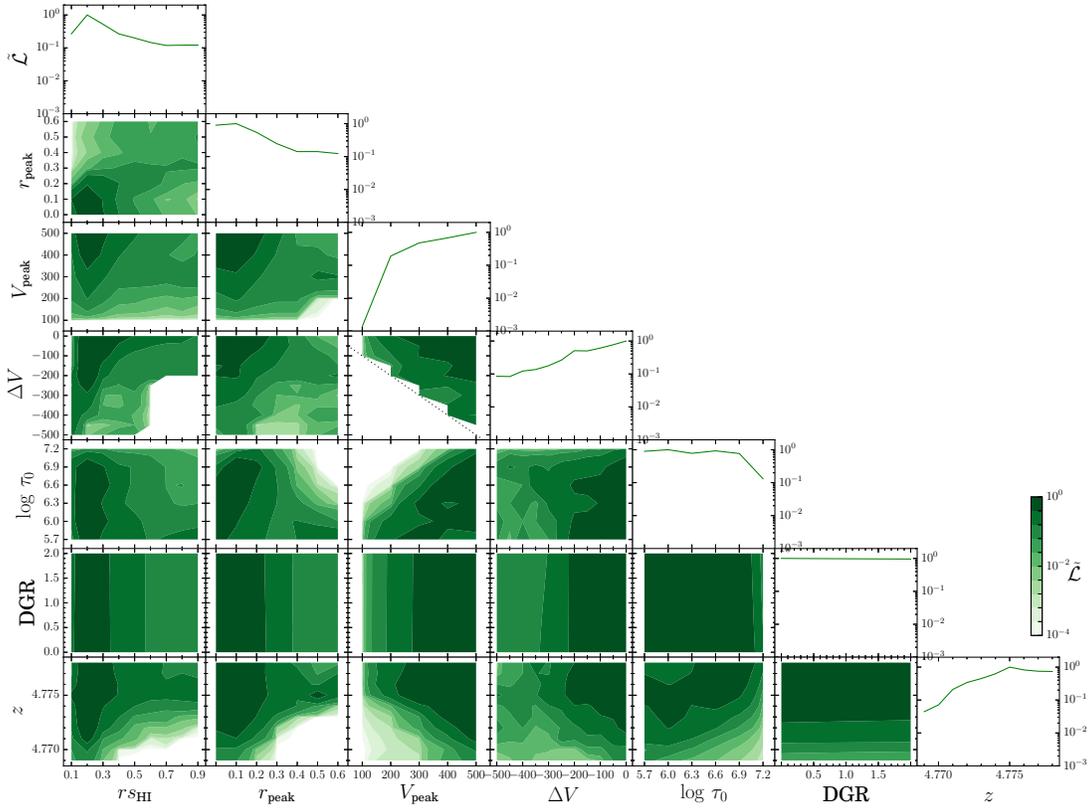}
	\caption{
		Similar to Figure \ref{fig-lnliksbp-1185}, 
			but for MUSE \#53.
	}
	\label{fig-lnliksbp-53}
\end{figure*}
%%%%%%%%%%%%%
\begin{figure*}
	\center
	\includegraphics[width=0.86\textwidth]{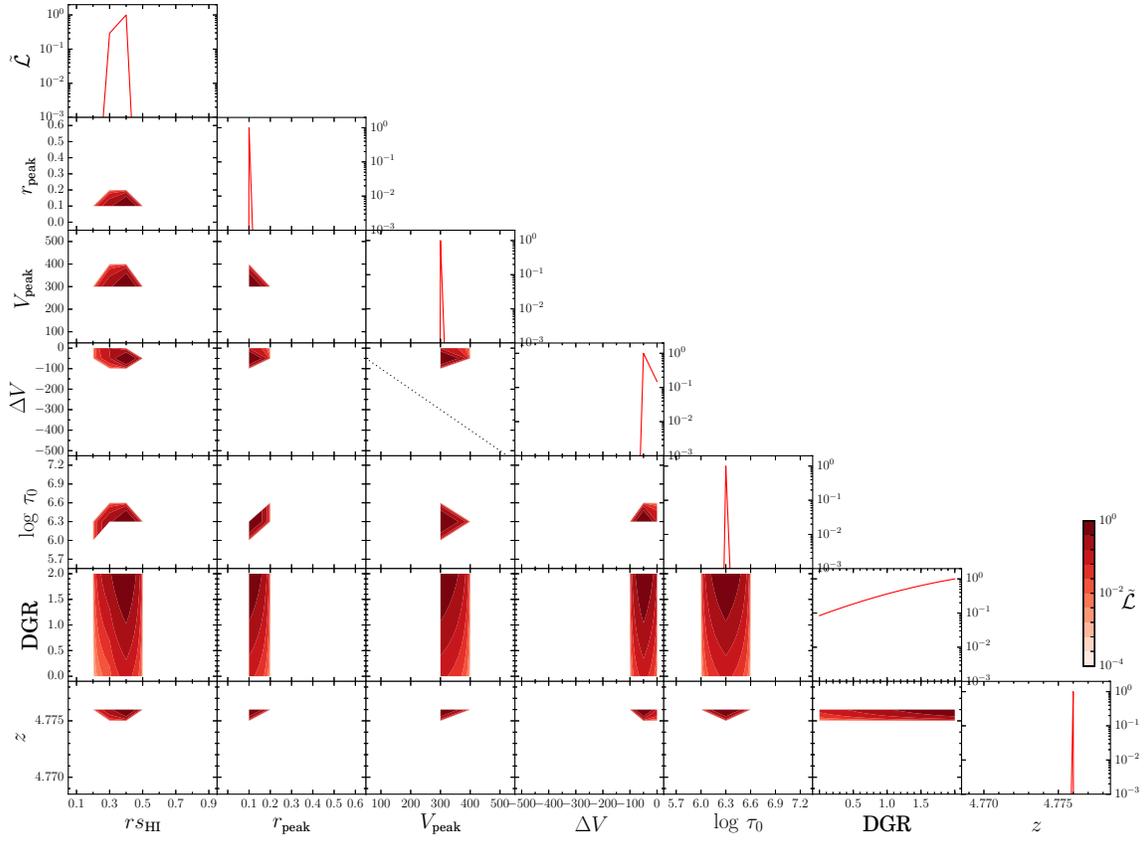}
	\caption{
		Similar to Figure \ref{fig-lnliktot-1185}, 
			but for MUSE \#53.
	}\label{fig-lnliktot-53}
\end{figure*}
%%%%%%%%%%%%%%
\begin{figure*}
	\center
	\includegraphics[width=0.86\textwidth]{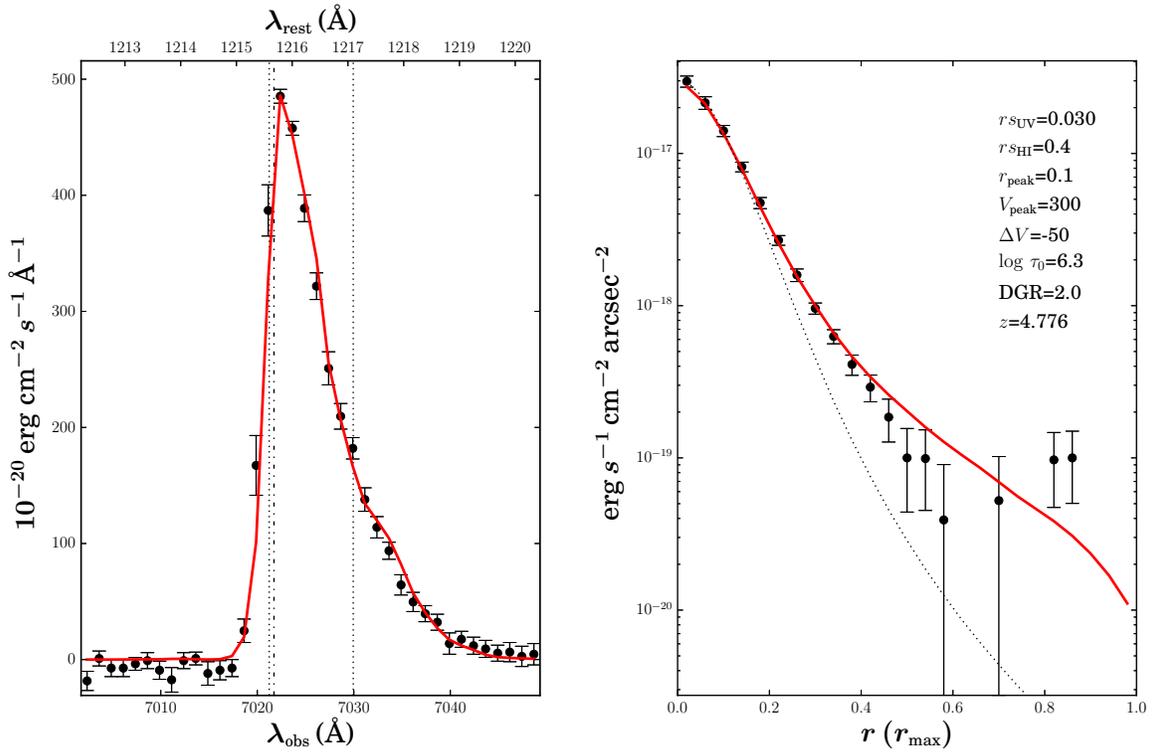}
	\caption{
		Similar to Figure \ref{fig-bestfit-tot-1185}, 
			but for MUSE \#53.
	}\label{fig-bestfit-tot-53}
\end{figure*}
%%%%%%%%%%%%%%%%%%%%%
% 171
\begin{figure*}
	\center
	\includegraphics[width=0.86\textwidth]{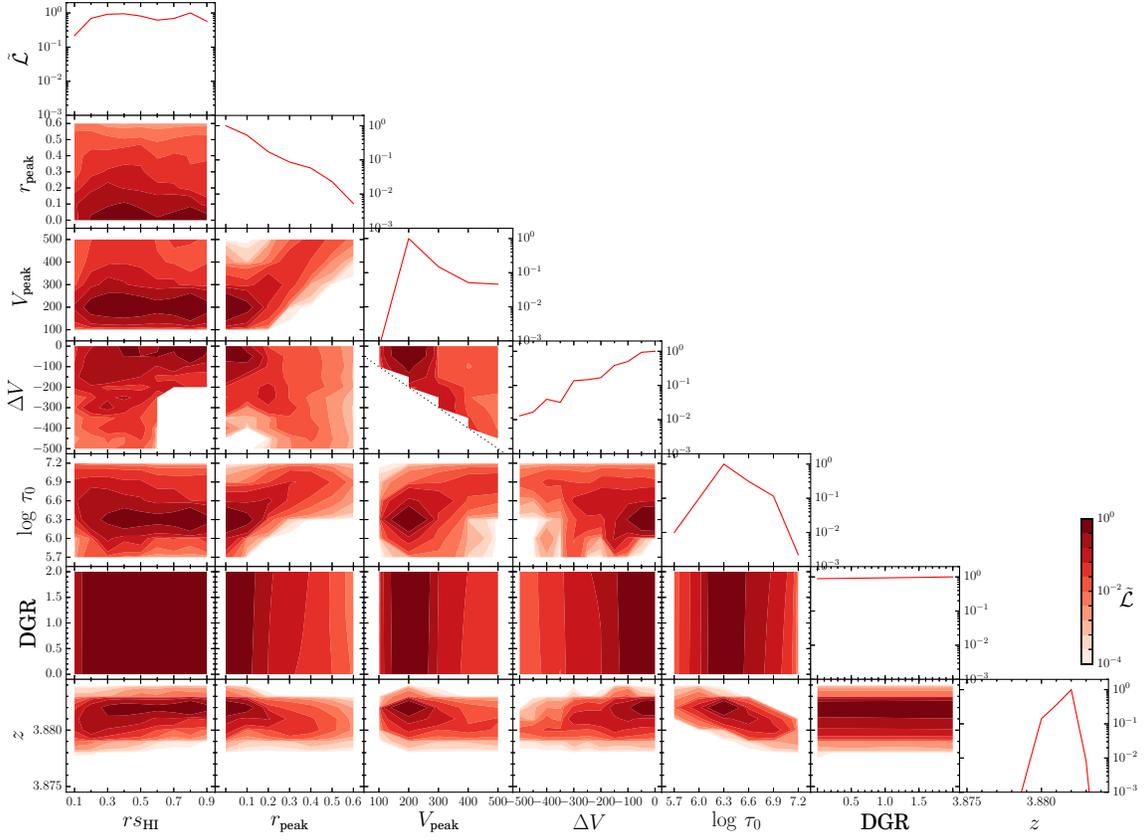}
	\caption{
		Similar to Figure \ref{fig-lnliktot-1185}, 
			but for MUSE \#171.
	}\label{fig-lnliktot-171}
\end{figure*}
%%%%%%%%%%%%%%
\begin{figure*}
	\center
	\includegraphics[width=0.86\textwidth]{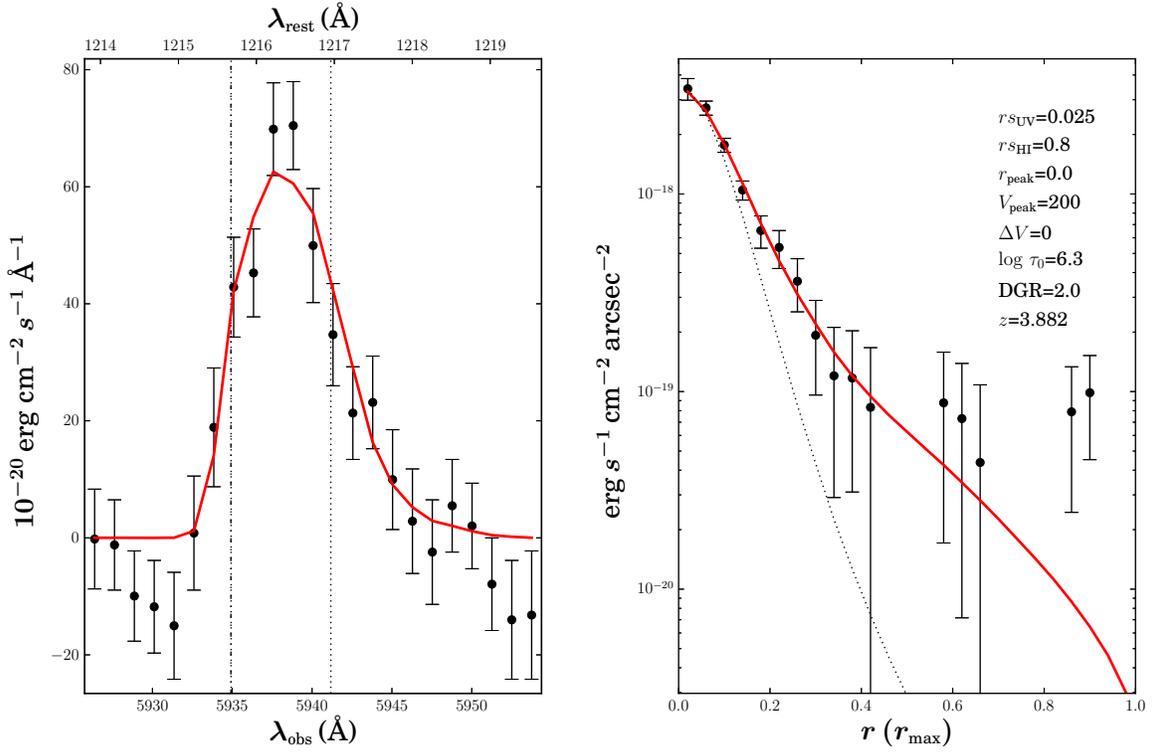}
	\caption{
		Similar to Figure \ref{fig-bestfit-tot-1185}, 
			but for MUSE \#171.
	}\label{fig-bestfit-tot-171}
\end{figure*}
%%%%%%%%%%%%%%%%%%%%%
% 547
\begin{figure*}
	\center
	\includegraphics[width=0.86\textwidth]{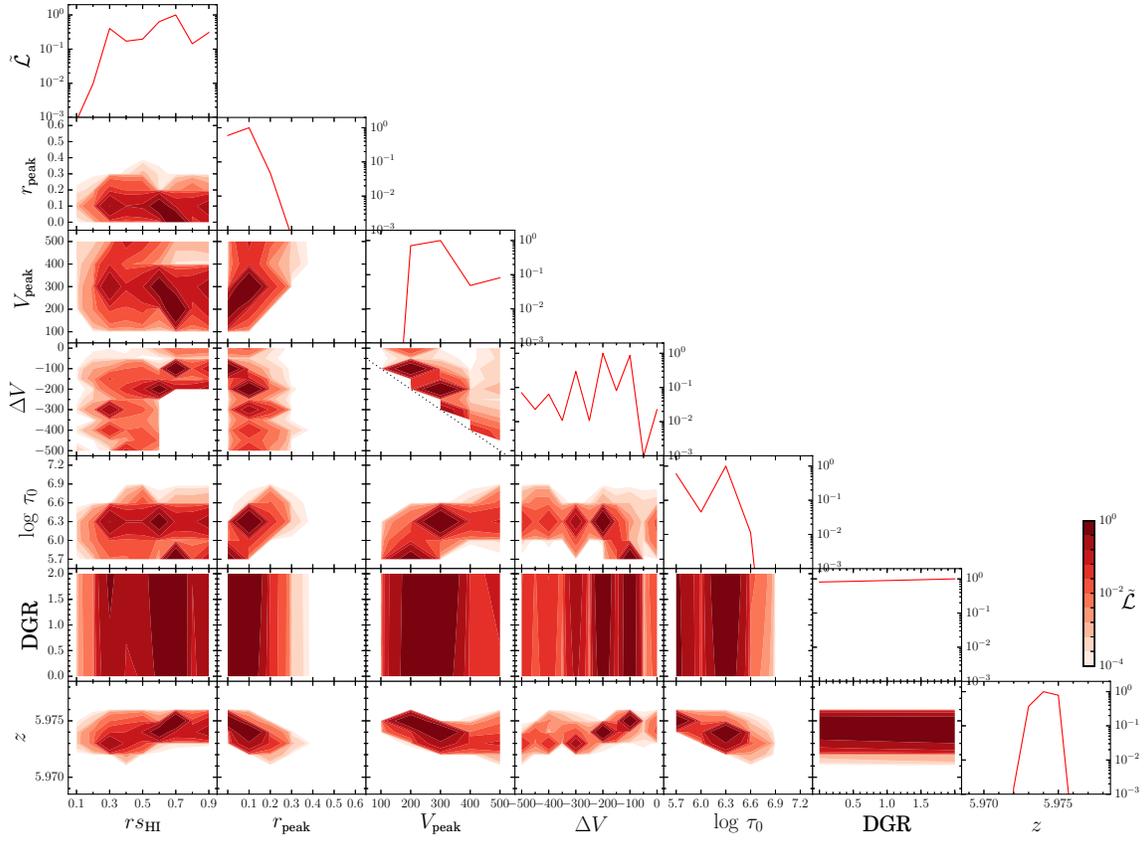}
	\caption{
		Similar to Figure \ref{fig-lnliktot-1185}, 
			but for MUSE \#547.
	}\label{fig-lnliktot-547}
\end{figure*}
%%%%%%%%%%%%%%
\begin{figure*}
	\center
	\includegraphics[width=0.86\textwidth]{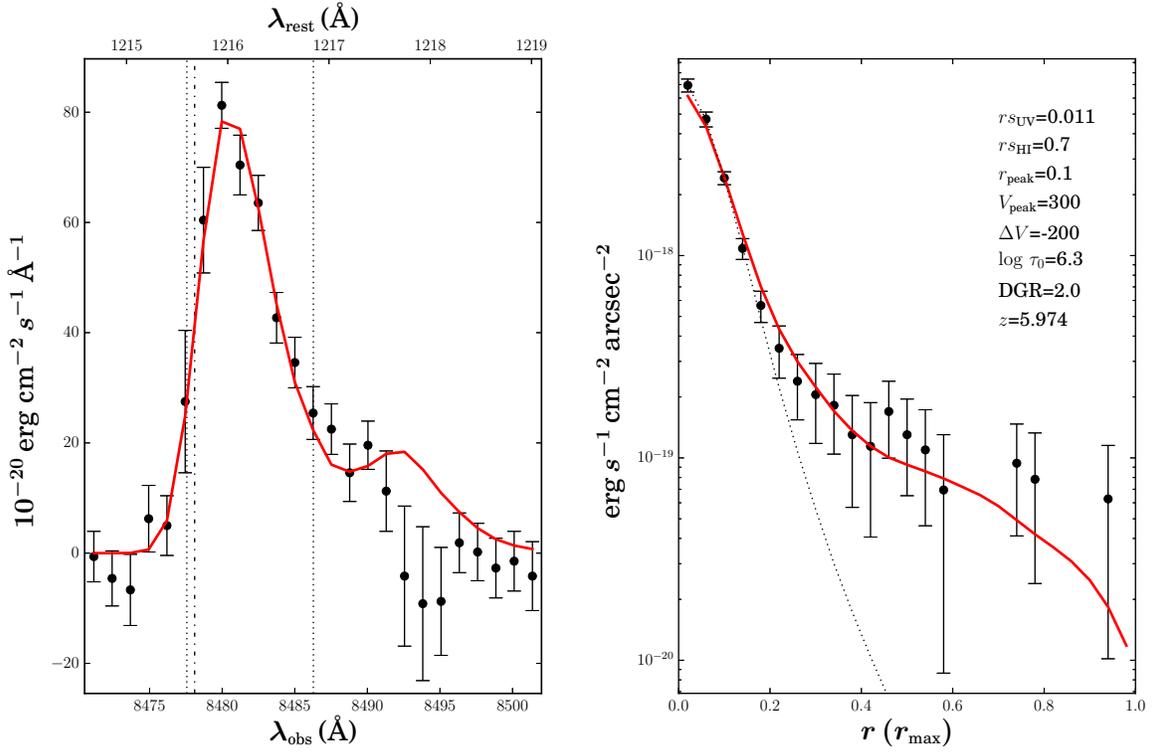}
	\caption{
		Similar to Figure \ref{fig-bestfit-tot-1185}, 
			but for MUSE \#547.
	}\label{fig-bestfit-tot-547}
\end{figure*}
%%%%%%%%%%%%%%%%%%%%%
% 364
\begin{figure*}
	\center
	\includegraphics[width=0.86\textwidth]{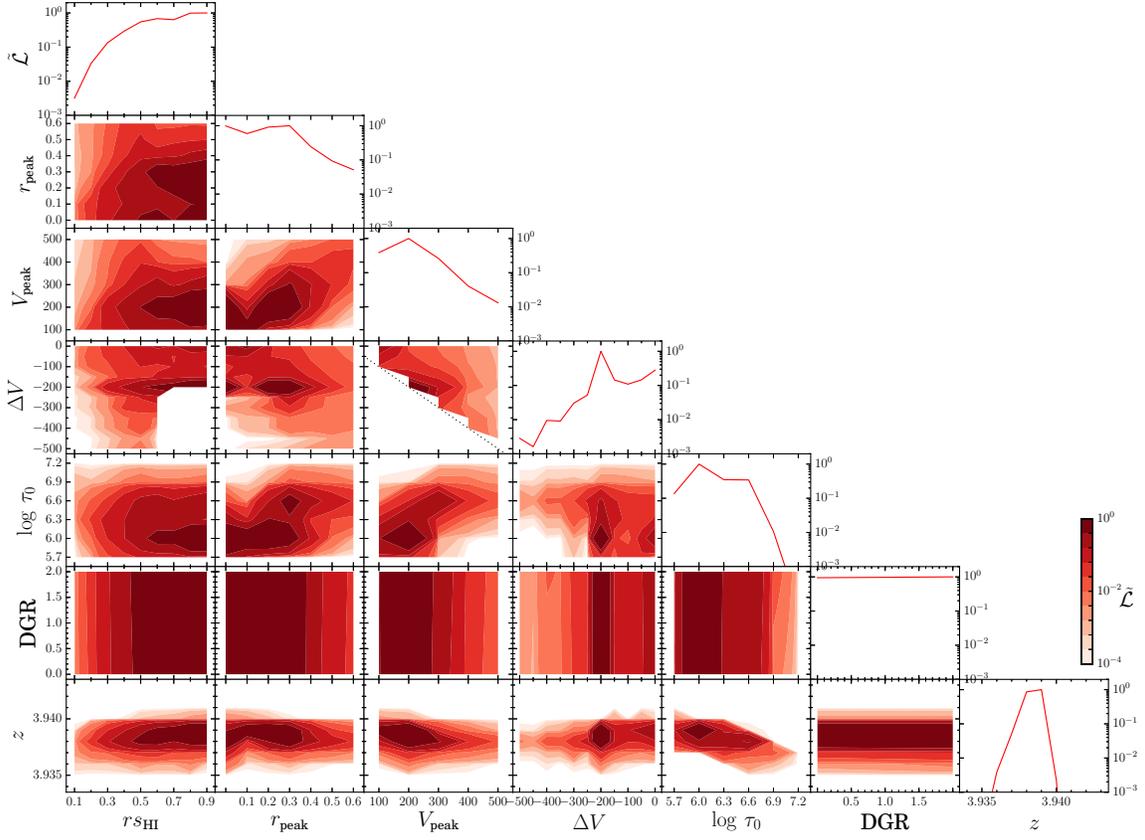}
	\caption{
		Similar to Figure \ref{fig-lnliktot-1185}, 
			but for MUSE \#364.
	}\label{fig-lnliktot-364}
\end{figure*}
%%%%%%%%%%%%%%
\begin{figure*}
	\center
	\includegraphics[width=0.86\textwidth]{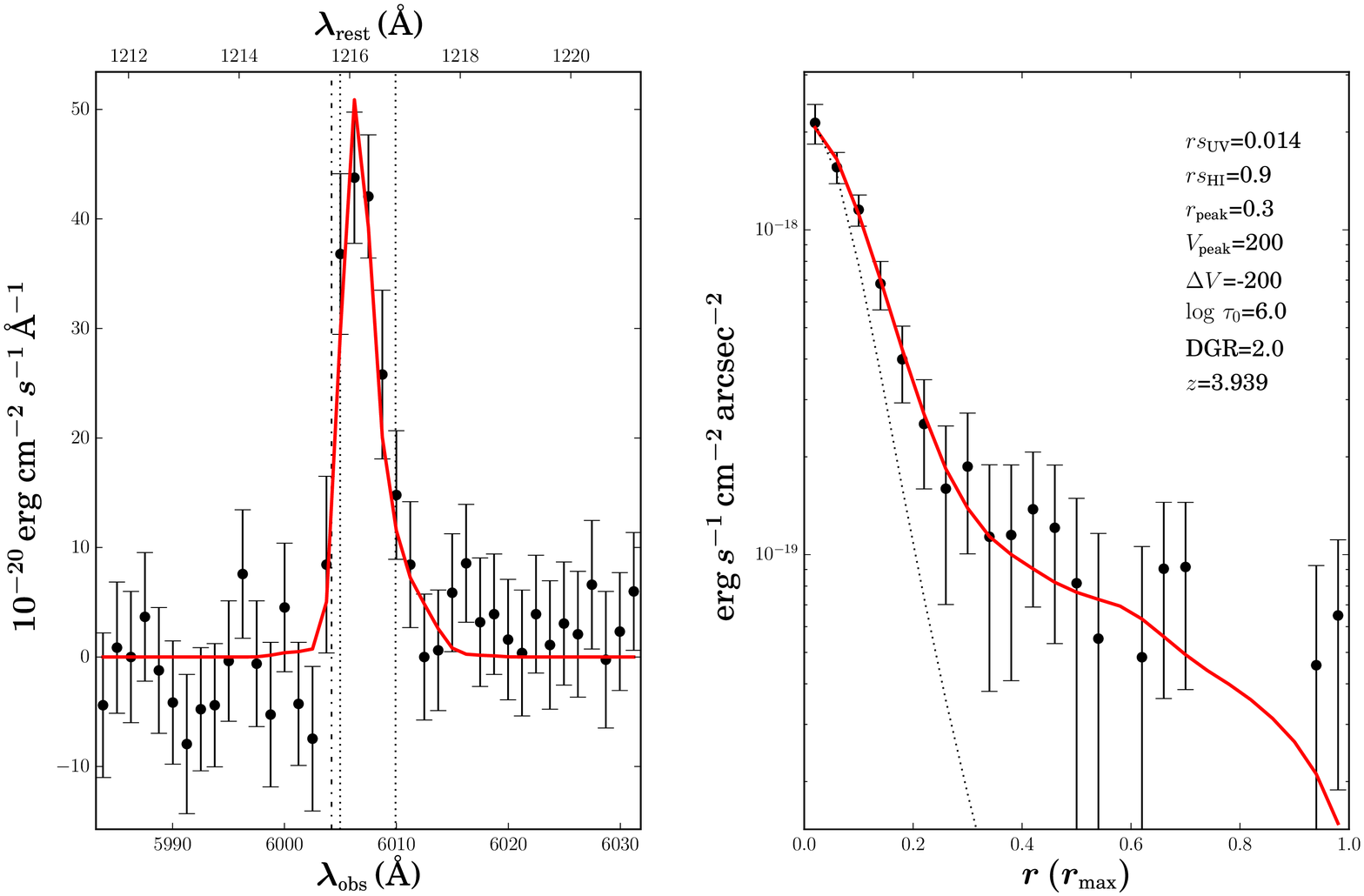}
	\caption{
		Similar to Figure \ref{fig-bestfit-tot-1185}, 
			but for MUSE \#364.
	}\label{fig-bestfit-tot-364}
\end{figure*}
%%%%%%%%%%%%%%%%%%%%%
%}}}

\section{\Lyaa Spectrum and Surface Brightness Profile Variation in Model Parameter Space (Continued)}\label{sec-parvar-rest}
%{{{
In this section, we present the impact of the rest of model parameters on
	the shapes of spectrum and surface brightness profile
	that are not fully presented in Section \ref{ssec-parvar}.

%%%%%%%%%%%%%%%%%%%%%
% Figure
\begin{figure*}
	\center
	\includegraphics[width=\textwidth]{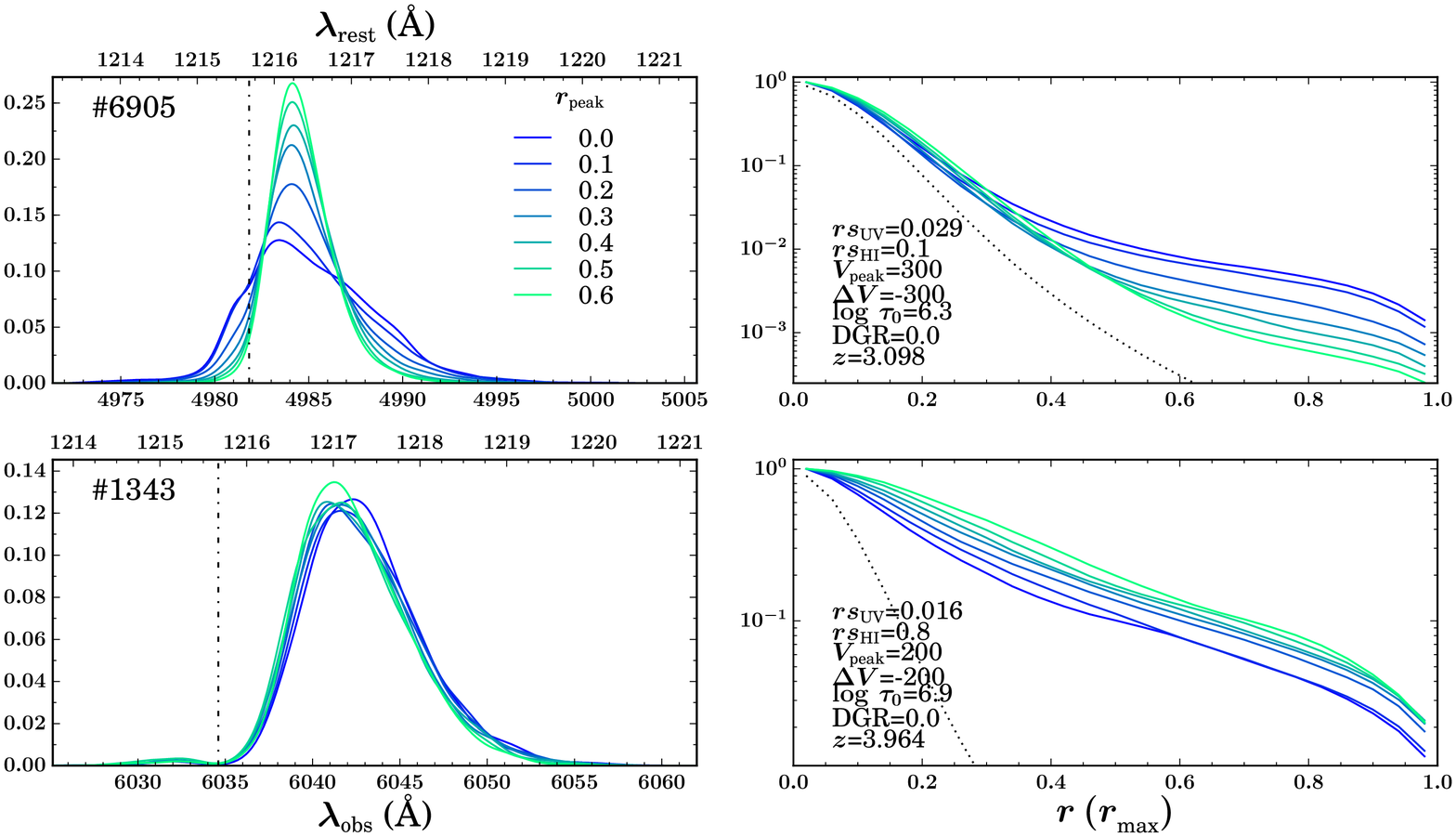}
	\caption{
		Similar to Figure \ref{fig-rscale}, but with varying \rpeak.
	}\label{fig-rpeak}
\end{figure*}
%%%%%%%%%%%%%%%%%%%%%
Figure \ref{fig-rpeak} shows the changes of spectrum and surface brightness profile with \rpeak.
The spectrum tends to be broader with higher intensities at the central wavelength
	and at the red wing as \rpeakk decreases (see the case of MUSE \#6905).
The higher intensity at the central wavelength is because 
	the inner medium becomes more transparent with a larger expanding velocity at a given radius 
	when \rpeakk is smaller,
	thus letting more photons at the central wavelength to escape from the system without scatterings.
The higher intensities at the red wing for smaller \rpeakk is because
	photons are more likely to be scattered by larger velocities,
	and thus are scattered into wavelengths farther away from the central wavelength.
A surface brightness profile tends to be flatter at small radii when \rpeakk is larger
	(because of more scatterings there),
	but its trend with \rpeakk at large radii seems to depend on other parameters.
When the outer part of the galaxy is optically thick
	(with large \rscale, small \delV, and large $\tau_0$),
	the change of spectrum with \rpeakk becomes negligible
	and the flattening of surface brightness profile by increasing \rpeakk
	extends to a large radius (e.g., MUSE \#1343).

%%%%%%%%%%%%%%%%%%%%%
% Figure
\begin{figure*}
	\center
	\includegraphics[width=\textwidth]{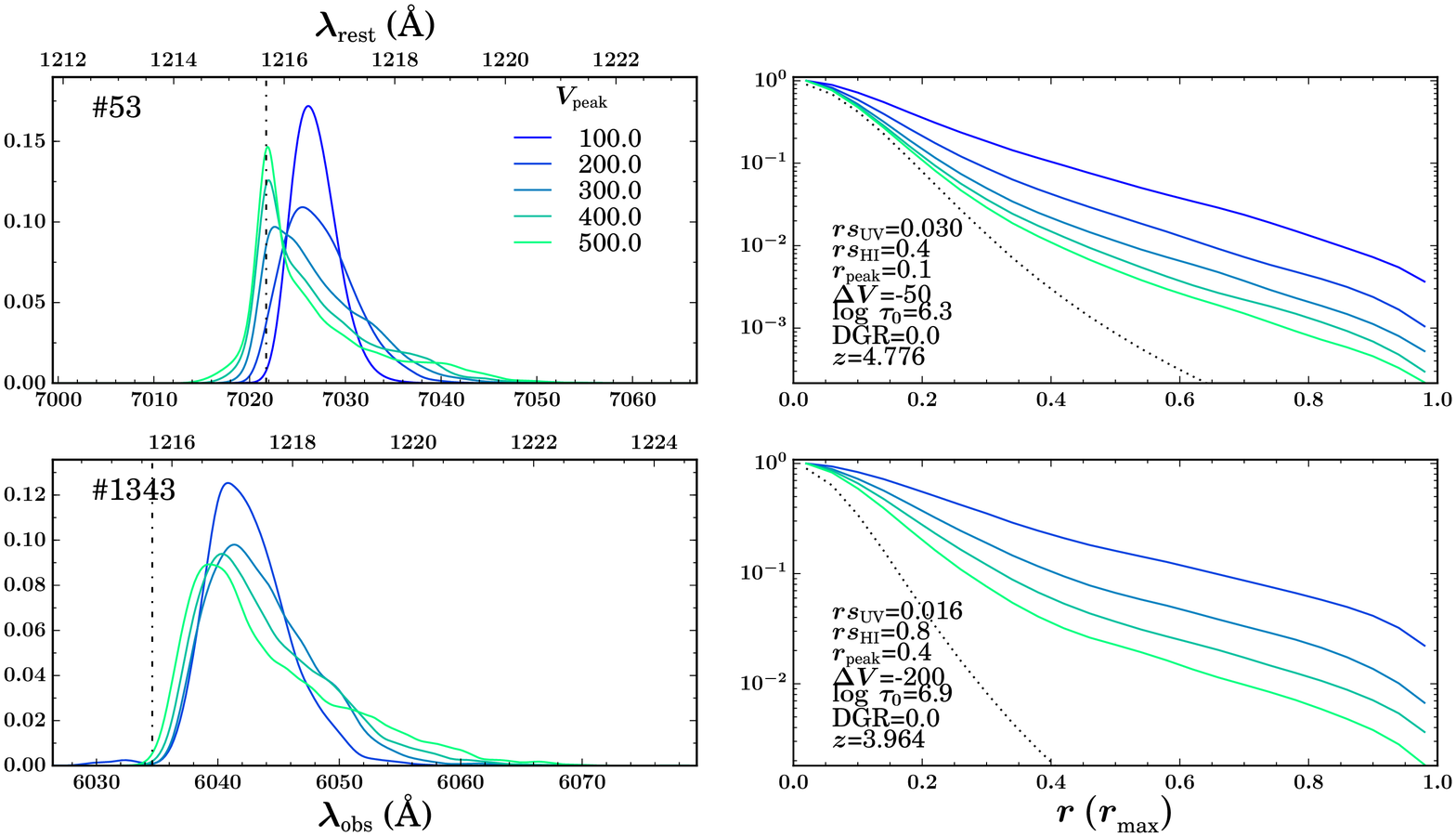}
	\caption{
		Similar to Figure \ref{fig-rscale}, but with varying \Vpeak.
	}\label{fig-Vpeak}
\end{figure*}
%%%%%%%%%%%%%%%%%%%%%
We can speculate the changes of spectrum and surface brightness profile with \Vpeakk
	on the analogy of those with \rpeak.
Increasing \Vpeakk has an effect similar to decreasing $r_\textrm{\cs peak}$,
	making the medium more transparent at a given radius.
Therefore, as $V_\textrm{\cs peak}$ increases,
	a spectrum will have a sharper peak that is closer to the central wavelength
	and a broader red tail.
A surface brightness profile becomes steeper with \Vpeak.
These trends are well shown in Figure \ref{fig-Vpeak} as expected.
Compared to the case of \rpeak,
	the impact of \Vpeakk on both spectrum and surface brightness profile is more significant.
It is because the maximum levels of medium transparency and the frequency change 
	that can be reached by varying \rpeakk are limited by a given \Vpeakk value.
Moreover, it is limited to $r<r_\textrm{\cs peak}$ 
	that the medium becomes more transparent as \rpeakk decreases.
The medium at $r>r_\textrm{\cs peak}$ becomes less transparent as \rpeakk decreases.
However, increasing \Vpeakk makes the whole medium more transparent.

%%%%%%%%%%%%%%%%%%%%%
% Figure
\begin{figure*}
	\center
	\includegraphics[width=\textwidth]{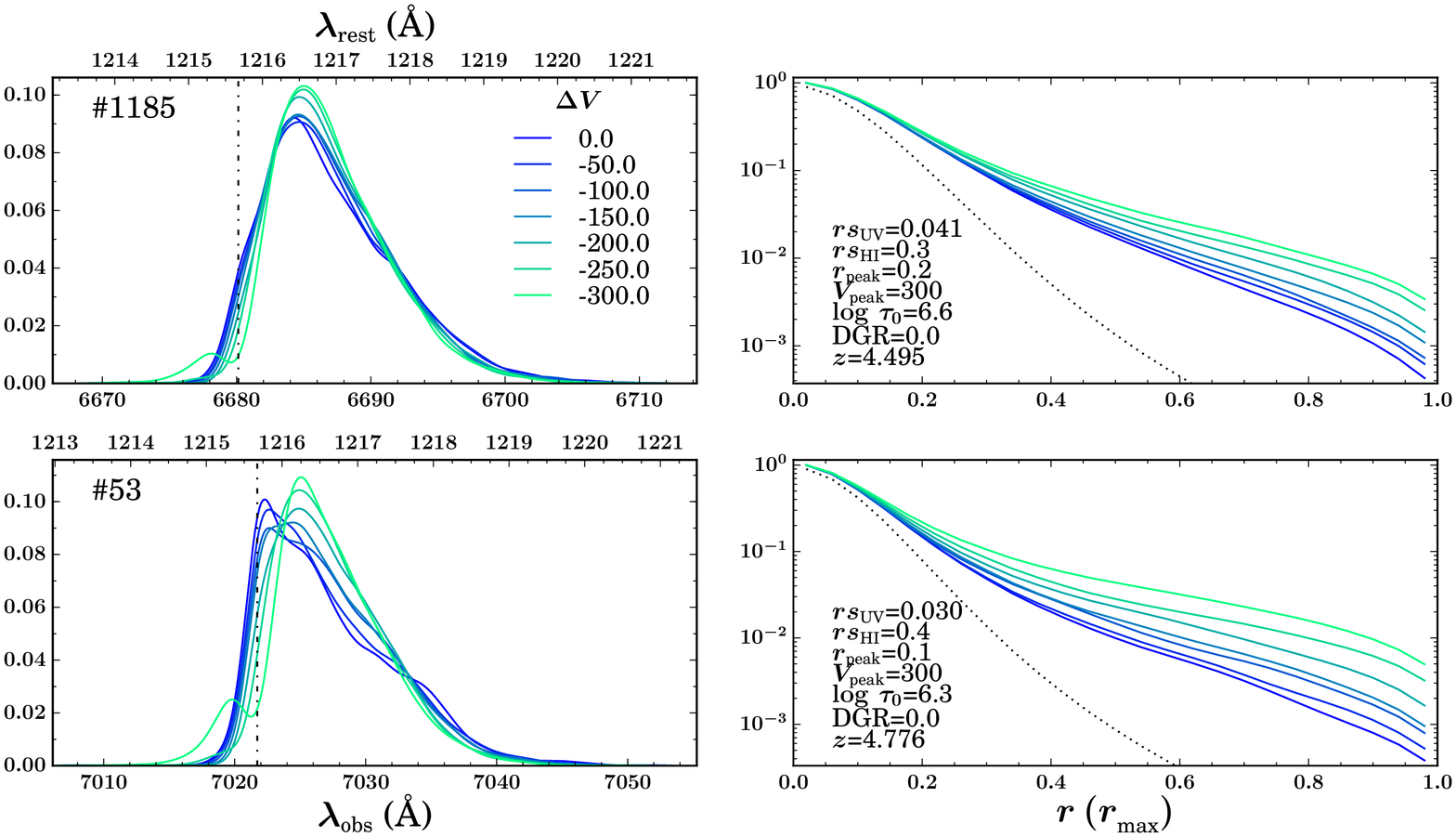}
	\caption{
		Similar to Figure \ref{fig-rscale}, but with varying \delV.
	}\label{fig-delV}
\end{figure*}
%%%%%%%%%%%%%%%%%%%%%
The impact of \delVV on spectrum and surface brightness profile is presented in Figure \ref{fig-delV}.
The change of surface brightness profile with \delVV is obvious;
	as \delVV decreases (i.e., as the velocity at the edge decreases closer to zero),
	the surface brightness profile becomes flatter.
This is because photons that were lucky enough not to experience many scatterings in the inner region
	are more likely to be scattered in the outer region
	as the medium there becomes more static (i.e., less transparent) with decreasing \delV.
The change of spectrum with $\Delta V$ appears less obvious,
	which can depend on other parameters.
Yet, the spectrum tends to have a higher intensity at the central wavelength 
	when \delVV is larger (i.e., larger velocity at the edge).
One interesting feature in the spectrum is a small blue bump
	that appears when the velocity at the edge is zero. 
This blue bump may consist of lucky photons 
	that have experienced almost no scattering until they reach the edge.
However, these photons are scattered by static medium at the edge
	to either redder or bluer wavelengths than the central wavelength,
	forming a (symmetric) double-peak spectrum.
Because there are not many such lucky photons,
	such a double-peak feature is subdominant;
	the red peak of the double-peak spectrum is hidden in the dominant main body
	and the blue peak appears as a small bump.

%%%%%%%%%%%%%%%%%%%%%
% Figure
\begin{figure*}
	\center
	\includegraphics[width=\textwidth]{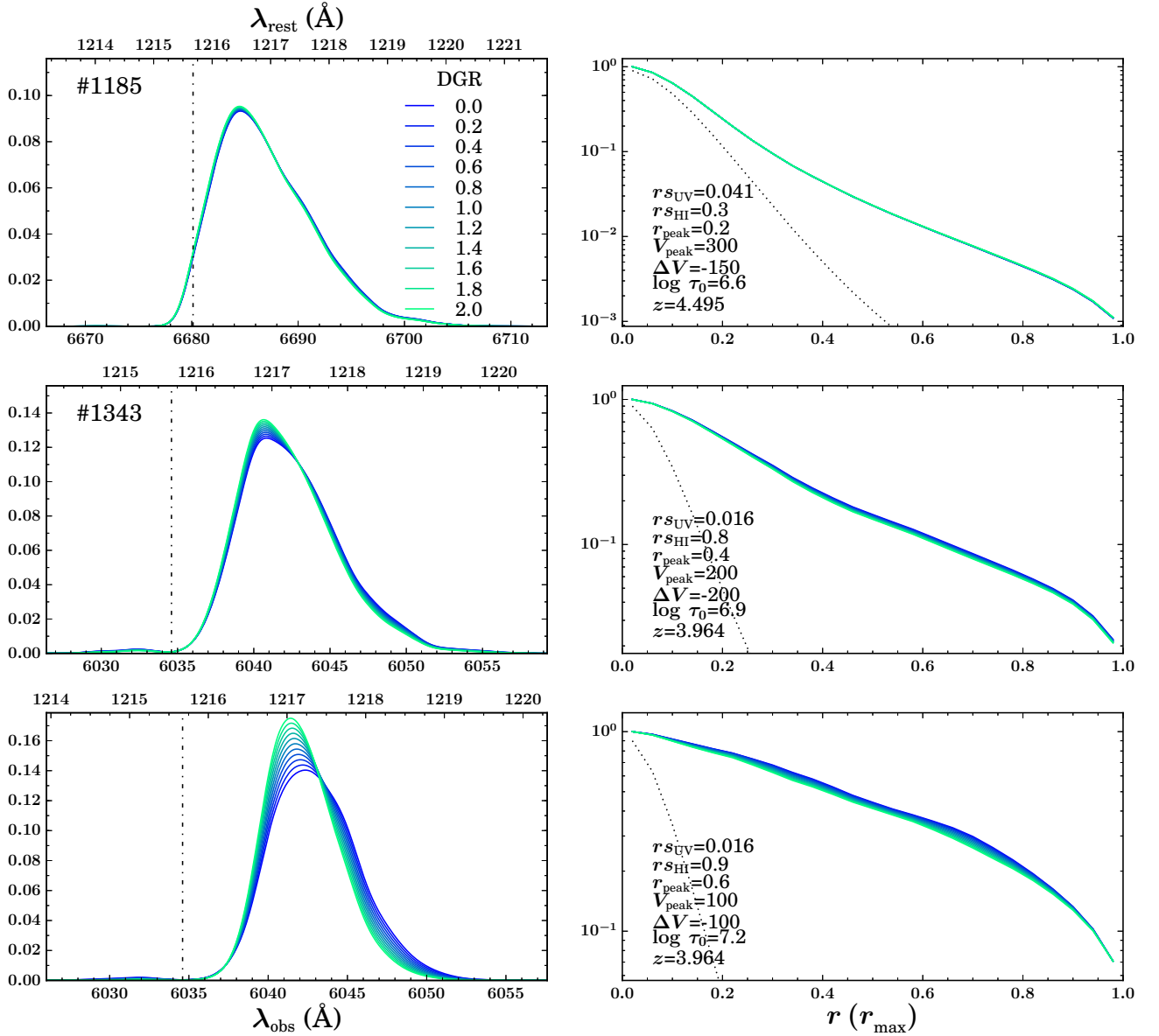}
	\caption{
		Panels in the upper two rows are similar to those in Figure \ref{fig-rscale}, but with varying DGR.
		Panels in the bottom row are with a parameter set that could maximize the number of scatterings
			(large $\tau_0$, large \rpeak, small \Vpeak, and small \delV),
			which is adopted to see the impact of dust more clearly.
	}\label{fig-DGR}
\end{figure*}
%%%%%%%%%%%%%%%%%%%%%
Because dust absorbs \Lyaa photons and re-emits photons at longer wavelengths 
	(i.e., at mid- or far-infrared wavelengths),
	the main effect of dust is the reduction in the amount of \Lyaa photons.
Meanwhile, the chance of a photon being destructed by dust
	is proportional to the number of scatterings that the photon has experienced until its escape.
Photons at wavelengths far away from the central wavelength
	are those that have been scattered a lot,
	and thus they are more likely to interact with dust.
Therefore, the intensity at a longer wavelength (in the redder side than the \Lyaa central wavelength) 
	might be reduced more than that at a shorter wavelength
	compared to the case without dust,
	implying that dust could have an impact on the shape of spectrum as well as the absolute intensity level.
However, the spectra and surface brightness profiles with different DGRs 
	show not much differences as seen in the panels of upper two rows of Figure \ref{fig-DGR}.
It could be because the best-fit parameter set of each target galaxy
	is not optimal to see the effect of dust on the shapes of spectrum and surface brightness profile.
To maximize the dust effect,
	we use a parameter set that is thought to maximize the number of scatterings 
	(e.g., large $\tau_0$, large \rpeak, small \Vpeak, and small \delV).
Now we can see some changes in the spectral shape with DGR in the bottom left panel of Figure \ref{fig-DGR};
	larger intensity reduction at long wavelengths when DGR is larger, 
	resulting in a sharper spectrum with its peak closer to the central wavelength.
It should be noted that each spectrum is normalized by the amount of escaping photons
	to compare the spectral shape not the intensity.
The overall intensity decreases with DGR,
	which will be discussed in the following section.
The effect of dust on the shape of surface brightness profile appears less significant.

\begin{figure*}
	\center
	\includegraphics[width=\textwidth]{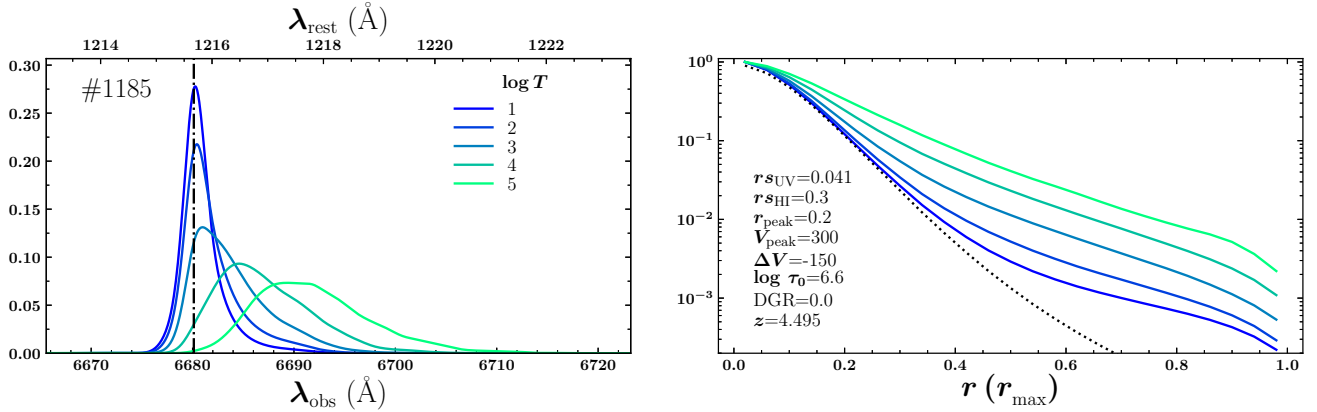}
	\caption{
			Similar to Figure \ref{fig-rscale}, but with varying medium temperature.
	}\label{fig-temp}
\end{figure*}

There is one parameter that we have not varied but may have a strong impact on \Lyaa transfer process,
	which is medium temperature.
We fix medium temperature at $10^4\textrm{K}$
	in order to focus on the impact of other parameters such as
	the scale radius of medium density and parameters for medium velocity profile
	that have not been studied much so far.
However, it is worth checking how medium temperature affects 
	the shapes of emerging \Lyaa spectrum and surface brightness profile briefly.
We perform four additional runs for medium temperature of $10^1$, $10^2$, $10^3$, and $10^5\textrm{K}$,
	fixing other parameters at their best-fit values to MUSE \#1185.
We note that hydrogen atoms are almost fully ionized at a temperature as high as $10^5\textrm{K}$.
In our simulation, the medium temperature of $10^5\textrm{K}$ should be regarded
	as an effective temperature that includes the effect of turbulent motion.
Figure \ref{fig-temp} shows that spectrum becomes broader with a larger peak shift 
	and surface brightness profile becomes shallower as medium temperature increases.
This is because hydrogen atoms have a broader velocity distribution when medium temperature is higher,
	and they scatter \Lyaa photons of a broader frequency range.
In the result, high medium temperature leads to more scatterings
	(under the assumption that the ionization state of hydrogen atoms
	does not change with temperature).
Similar to the case that we change the input spectrum shown in Section \ref{ssec-intsp}, 
	the best-fit parameter set for a galaxy will differ if we change medium temperature.
Nevertheless, our approach of fixing medium temperature is useful 
	for exploring the impact of other parameters more efficiently
	because we can reduce the dimension of the parameter space we are exploring
	and the resulting computation time.
It will be our future work to consider medium temperature together 
	with all relevant quantities that include density, pressure, and ionization state
	to be self consistent.
%%%%%%%%%%%%%%%%%%%%%
% Figure
%%%%%%%%%%%%%%%%%%%%%
%}}}

\section{Peak Shift of \Lyaa Spectrum and Peak Velocity of Medium}\label{sec-peak-Vmax}
%{{{
In this section, we examine
	the impact of the velocity of expanding medium 
	on the emerging spectrum in more detail,
	assuming an expanding velocity ranging from several tens to several hundreds km s$^{-1}$.
To focus on the impact of medium velocity only,
	we simplify the models
	by considering a central, monochromatic source 
	in an expanding \ion{H}{1} halo of uniform density (optical depth of $10^6$)
	and uniform temperature ($10^4$K).
We adopt three different types of velocity profile:
	(a) constant expansion, (b) Hubble-like expansion, 
	and (c) accelerated and then decelerated expansion (as in this paper).

\begin{figure*}
	\center
	\includegraphics[width=\textwidth]{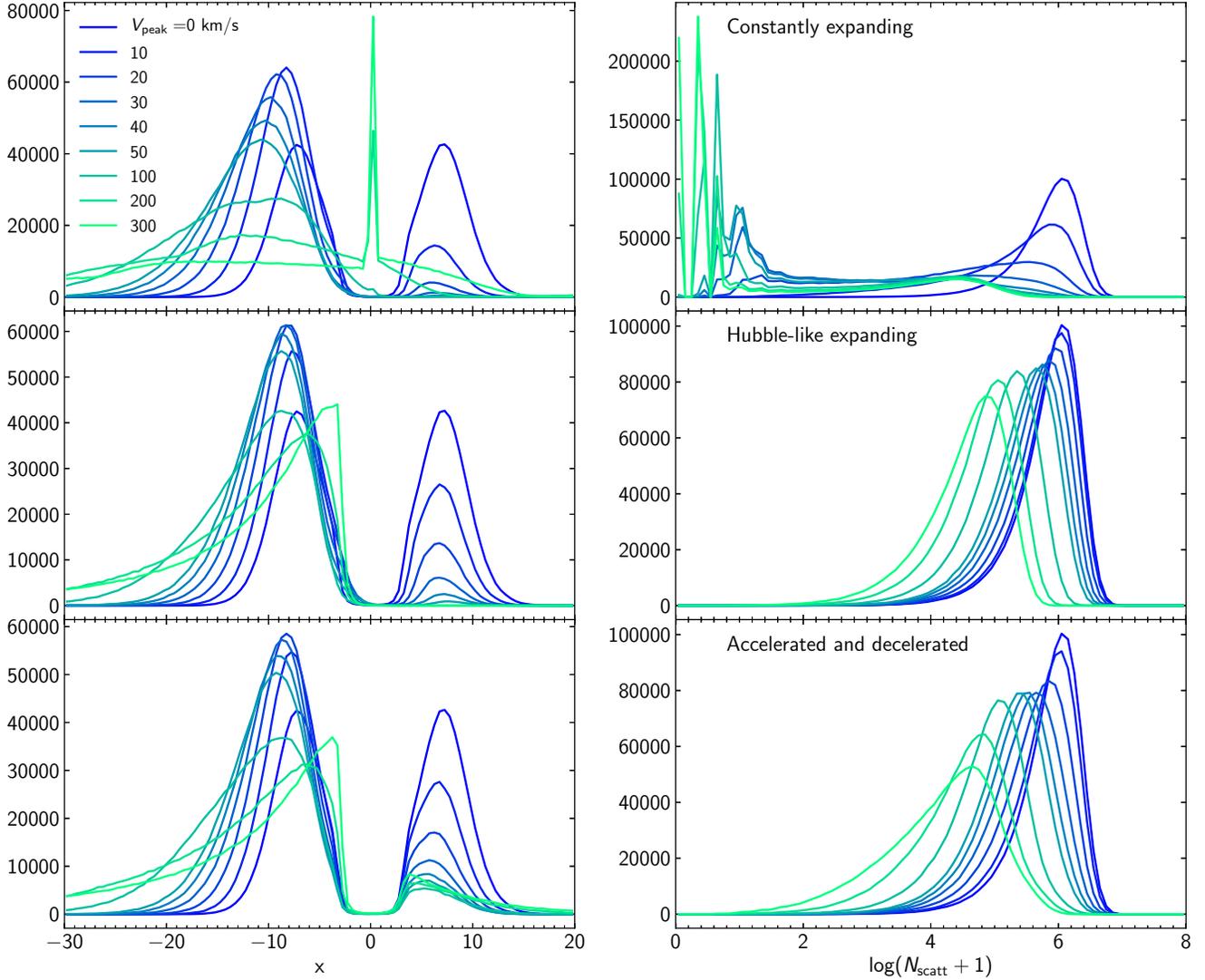}
	\caption{
		Spectra (left panels) and numbers of scatterings (right panels)
			of photons that emerge from a central monochromatic source
			in an expanding \ion{H}{1} halo of uniform density (optical depth of $10^6$)
			and uniform temperature ($10^4K$),
			for nine peak expanding velocities (denoted by different colors)
			for three different velocity types:
			constantly expanding (top), 
			Hubble-like expanding (middle), and
			accelerated and then decelerated expanding 
			(bottom, \rpeak$=0.5$ and \delV$=-$\Vpeak).
	}\label{fig-peakshift-Vpeak}
\end{figure*}

Figure \ref{fig-peakshift-Vpeak} shows (left) emerging spectra
	and (right) the distributions of numbers of scatterings ($N_\textrm{\cs scatt}$)
	for nine peak expanding velocities (denoted by different colors)
	for each velocity type (each row).
From $V_\textrm{\cs peak}=0$ to $\sim50\,\textrm{km s}^{-1}$
	the red peak of emerging spectrum appears to be progressively displaced redwards.
When $V_\textrm{\cs peak} \gtrsim 50\,\textrm{km s}^{-1}$, however,
	the peak appears closer to the central wavelength (i.e., $x=0$)
	as the velocity increases.
This trend is consistent with the result shown in Figure 7 of \citet{Verhamme_etal_2006}.
The number of scatterings, on the other hand, 
	consistently decreases as velocity increases,
	which is attributed to decreased effective optical-depth.
Therefore, as the medium velocity is too high,
	photons will escape without having chances to undergo a large number of scatterings;
	as a consequence, no significant frequency change is found in the emergent spectrum.

The behavior of spectral peak 
	when $V_\textrm{\cs peak} \lesssim 50\,\textrm{km s}^{-1}$
	can be understood with the help of 
	the redistribution function, $\mathcal{R}(x_\textrm{\cs out}|x_\textrm{\cs in})$,
	that describes the probability of the ``out" frequency ($x_\textrm{\cs out}$) after a single scattering 
	for a given ``in" frequency ($x_\textrm{\cs in}$, the frequency before scattering)\footnote{
		Refer to Section 7.3 of \citet{Dijkstra2017} for more details.}.
Figure 21 of \citet{Dijkstra2017} shows that
	when the input frequency is farther away from the central frequency,
	so is the output frequency.
As the velocity of the medium is larger,
	the input frequency is shifted farther away from the central frequency in the medium frame,
	which will end up with larger frequency change after a number of scatterings,
	despite decreased effective optical-depth.
It should be noted, however, that such an effect
	can happen for low medium velocities
	where enough number of scatterings can still occur;
	i.e., $V_\textrm{\cs peak} \lesssim 50\,\textrm{km s}^{-1}$ in our particular example.
This critical expanding-velocity will depend on 
	the optical depth, temperature of medium, and the input spectrum.
%%%%%%%%%%%%%%%%%%%%%
% Figure
%%%%%%%%%%%%%
%}}}

\section{Correlation between Extent of \Lyaa Source and \Lyaa Halo}\label{sec-rshalo-rscont}
\begin{figure*}
	\center
	\includegraphics[width=0.6\textwidth]{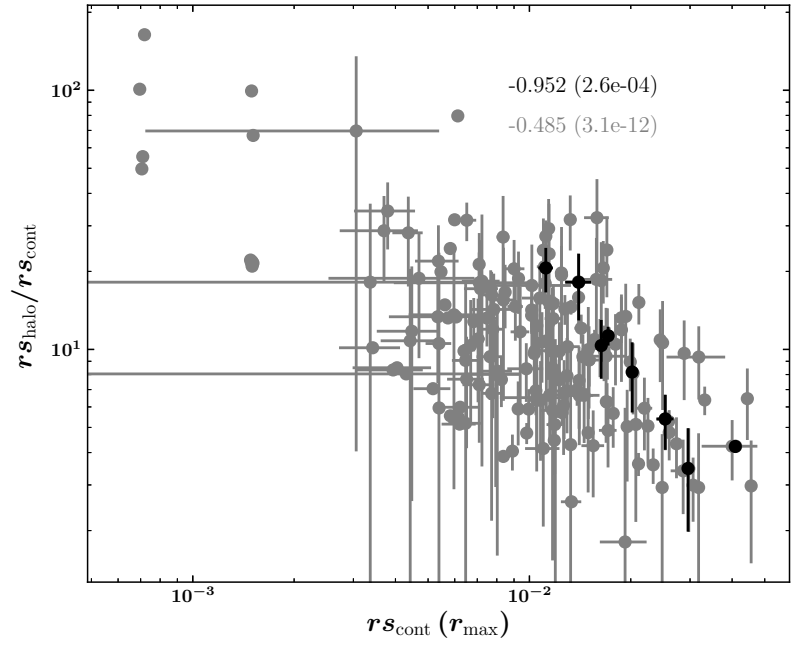}
	\caption{
		Anticorrelation between \rshalorscontt and \rscontt
			of eight target galaxies (filled black circles)
			and of the full sample of L17 (filled black circles and open gray circles).
		The Spearman rank-order correlation coefficients and p-values (in the parenthesis)
			for these two samples are given in the top right panel.
	}\label{fig-rshalo-rscont}
\end{figure*}

%{{{
Figure \ref{fig-rshalo-rscont} shows a correlation between
	\rshalorscontt and \rscontt for eight target galaxies of this study (filled black circles)
	and for the full sample of L17 (filled black circles and open gray circles altogether).
\rshalorscontt is strongly anticorrelated with \rscontt
	(refer to the Spearman rank-order correlation coefficients and $p$-values given in the figure).
It should be noted that this anticorrelation between \rshalorscontt and \rscontt
	is not derived from the simulation results, but from the observational results.
This anticorrelation is interesting
	because \rshaloo itself is positively correlated with \rscontt (see Figure 13 in L17).
The positive correlation between \rshaloo and \rscontt simply indicates that
	\Lyaa photons are generated where UV photons are (i.e., \ion{H}{2} regions)
	through photoionization followed by recombination.
%\textbf{On the other hand, the anticorrelation between \rshalorscontt and \rscontt 
%	indicates that the size of \Lyaa halo is predominantly determined by the scattering of \Lyaa photons.}
However, the anticorrelation between \rshalorscontt and \rscontt implies that
	the \rshalo--\rscontt correlation is not that strong but rather relatively weak.
This weakness in the \rshalo--\rscontt relationship suggests that
	other factors than \rscontt will be more critical in determining \rshalo;
	the factor that mostly matters the size of \Lyaa halo would be
	how significantly the \Lyaa photons diffuse out in space by resonance scattering,
	rather than the spatial size of \Lyaa source.
%%%%%%%%%%%%%%%%%%%%%
% Figure
%%%%%%%%%%%%%
%}}}

\bibliographystyle{apj}

\end{document}